\documentclass[twocolumn,english,showpacs,preprintnumbers,amsmath,amssymb,floatfix,nofootinbib]{revtex4-1}

\usepackage{ulem}
\usepackage{tikz,xcolor}
\usepackage[colorlinks = true,
linkcolor = blue,
urlcolor  = blue,
citecolor = blue,
anchorcolor = blue]{hyperref}
\definecolor{lime}{HTML}{A6CE39}
\DeclareRobustCommand{\orcidicon}{%
	\begin{tikzpicture}
		\draw[lime, fill=lime] (0,0)
		circle [radius=0.16]
		node[white] {{\fontfamily{qag}\selectfont \tiny ID}};
		\draw[white, fill=white] (-0.0625,0.095)
		circle [radius=0.007];
	\end{tikzpicture}
	\hspace{-2mm}
}
\foreach \x in {A, ..., Z}{%
	\expandafter\xdef\csname orcid\x\endcsname{\noexpand\href{https://orcid.org/\csname orcidauthor\x\endcsname}{\noexpand\orcidicon}}
}

\usepackage[T1]{fontenc}
\usepackage[latin9]{inputenc}
\usepackage{color}
\usepackage{array}
\usepackage{amstext}
\usepackage{graphicx}
\usepackage{esint}
\usepackage{rotating}
\usepackage[font=small,labelfont=bf]{caption}
\usepackage{xcolor}

\makeatletter

\@ifundefined{textcolor}{}
{%
	\definecolor{BLACK}{gray}{0}
	\definecolor{WHITE}{gray}{1}
	\definecolor{RED}{rgb}{1,0,0}
	\definecolor{GREEN}{rgb}{0,1,0}
	\definecolor{BLUE}{rgb}{0,0,1}
	\definecolor{CYAN}{cmyk}{1,0,0,0}
	\definecolor{MAGENTA}{cmyk}{0,1,0,0}
	\definecolor{YELLOW}{cmyk}{0,0,1,0}
}

\@ifundefined{definecolor}
{\usepackage{color}}{}
\@ifundefined{definecolor}
{\usepackage{color}}{}
\makeatother
\usepackage{babel}

\begin{document}



%
%
\title{Study of nuclear corrections on the charged hadron fragmentation functions in a Neural Network global QCD analysis} 
%
%

\author{Maryam Soleymaninia$^{1}$\orcidA{}}
\email{Maryam\_Soleymaninia@ipm.ir}

\author{Hadi Hashamipour$^{1}$\orcidB{}}
\email{H\_Hashamipour@ipm.ir}

\author{Hamzeh Khanpour$^{2,4,1}$\orcidE{}}
\email{Hamzeh.Khanpour@cern.ch}

\author{Samira Shoeibi$^{1}$}
\email{Samira.Shoeibi@ipm.ir}

\author{Alireza Mohamaditabar$^{1}$}
\email{A.Mohamaditabar@ipm.ir}

\affiliation {
$^{1}$School of Particles and Accelerators, Institute for Research in Fundamental Sciences (IPM), P.O.Box 19395-5531, Tehran, Iran. \\
$^{2}$AGH University, Faculty of Physics and Applied Computer Science, Al. Mickiewicza 30, 30-055 Kraków, Poland. \\
$^{4}$Department of Physics, University of Science and Technology of Mazandaran, P.O.Box 48518-78195, Behshahr, Iran. 
}

\date{\today}

%
\begin{abstract}

In this work, we present a new global QCD analyses, 
referred to as {\tt PKHFF.23}, for charged pion, kaon, 
and unidentified light hadrons. 
We utilize a Neural Network to fit the high-energy 
lepton-lepton and lepton-hadron scattering data, 
enabling us to determine parton-to-hadron fragmentation 
functions (FFs) at next-to-leading-order (NLO) accuracy.
The analyses include all available single-inclusive $e^+e^-$ 
annihilation (SIA) and semi-inclusive deep-inelastic scattering 
(SIDIS) data from the {\tt COMPASS} Collaboration for charged pions, 
kaons, and unidentified light hadrons. 
Taking into account the most recent nuclear parton distribution 
functions (nuclear PDFs) available in the literature, we evaluate 
the effect of nuclear corrections on the determination of light hadrons FFs.
The Neural Network parametrization, enriched with the Monte 
Carlo methodology for uncertainty estimations, is employed for all 
sources of experimental uncertainties and the proton PDFs.
Our results indicate that incorporating nuclear corrections has a marginal impact on 
the central values of FFs and their corresponding uncertainty bands. 
The inclusion of such corrections does not significantly affect the fit quality of the 
data as well. 
The study suggests that while nuclear corrections are a consideration, 
their impact in such QCD analysis is limited.

\end{abstract}
%


\maketitle
\tableofcontents{}

%
\section{Introduction}\label{sec:introduction}
%

Fragmentation Functions (FFs) play a crucial role in investigating 
the non-perturbative parton structure of hadrons at 
the Large Hadron Collider (LHC). 
They are also of great significance 
for future high-precision hadron production measurements at 
the Electron Ion Collider (EIC)~\cite{Aguilar:2019teb,AbdulKhalek:2021gbh} and 
the Future Circular Collider (FCC)~\cite{FCC:2018byv,FCC:2018vvp}.

Hence, methodological, experimental, and theoretical 
developments need to be considered to calculate well-established FF sets as well.
To this end, most recently, several global QCD analyses have 
been performed to determine the FFs of charged pion and 
kaon~\cite{AbdulKhalek:2022laj} and unidentified light 
hadrons~\cite{Soleymaninia:2022alt} by including all available data 
from single-inclusive electron-positron annihilation (SIA) and 
semi-inclusive deep-inelastic scattering (SIDIS) processes as well.

The Neural Network (NN) is considered as a modern parametrization 
technique to minimize the bias. 
To obtain the probability density distribution of extracted 
FF used for uncertainty estimation in the fitting 
analysis, the Monte Carlo method has been employed.
These analyses have been performed using the publicly 
available \texttt{MontBlanc} package~\cite{MontBlanc}.
The analysis for FFs of unidentified light charged hadrons 
has been extracted up to next-to-leading order (NLO), and 
the FFs of pion and kaon have been determined up to 
next-to-next-to-leading order (NNLO) accuracy.

The perturbative corrections for single-inclusive 
electron-positron annihilation (SIA) up to NNLO are 
available~\cite{Rijken:1996ns, Almasy:2011eq}. 
Recently, the derivation of approximate NNLO corrections 
to semi-inclusive deep-inelastic scattering (SIDIS) has 
been completed by expanding the resumed expressions~\cite{Abele:2021nyo}.
The authors of Ref.~\cite{AbdulKhalek:2022laj} have also 
extended the theoretical accuracy up to NNLO corrections.
Consequently, the \texttt{MontBlanc} package takes advantage of these 
developments and can calculate the FFs for different hadrons up 
to NNLO by including SIA and SIDIS measurements.

Although the recently reported FF sets~\cite{AbdulKhalek:2022laj,Soleymaninia:2022alt} 
take advantage of the model bias reduction in FF 
parametrization and consider the propagation of experimental 
data and proton PDF uncertainties, the nuclear corrections have 
not been taken into account so far.
The cross-section in the 
SIDIS process has been calculated considering only the 
proton PDFs set at NLO and NNLO accuracy, and the analyses 
have been performed without considering the nuclear 
corrections on the PDFs in SIDIS.

In this work, our primary focus is to revisit the analyses mentioned 
above for charged pions ($\pi^{\pm}$), kaons ($K^\pm$), and unidentified 
light hadrons ($H^\pm$) in order to assess the impact of 
various nuclear PDF sets on the determination of FFs.
To achieve this objective, we begin by conducting global 
QCD analyses for FFs of charged pions, kaons, and unidentified 
light hadrons separately. We consider three proton PDF sets as 
baseline fits in the case of SIDIS for NLO accuracy.
Afterwards, we repeat the analyses, but this time, 
we use the most recent nuclear PDF sets to investigate 
the effects of applying the nuclear corrections on the 
determination of FFs. By comparing the results obtained 
with and without the nuclear corrections, we gain insight into how these corrections impact the analysis and 
the precision of the final FF determinations for charged 
pions, kaons, and unidentified light hadrons.

We generate three FF sets using the {\tt nNNPDF3.0}~\cite{AbdulKhalek:2022fyi}, 
{\tt EPPS21}~\cite{Eskola:2021nhw}, and {\tt nCTEQ15WZ}~\cite{Kusina:2020lyz} nuclear 
PDF sets at NLO accuracy. 
Our analysis examines the impact of nuclear corrections on 
the determination of FFs in global NLO QCD analyses. 
We present and discuss our primary findings, showing that nuclear corrections have a 
marginal impact on both the shape and uncertainty bands of FFs and do not 
significantly affect the fit quality. This suggests that while nuclear 
corrections are a consideration, their overall impact on the FF QCD analysis is limited.

The rest of the paper is organized as follows.
In Sec.~\ref{sec:Theoretical}, we provide a summary of the 
theoretical formalism concerning the SIDIS and SIA processes.
The parametrization of FFs and the flavor decomposition for charged pion, kaon and 
unidentified light hadrons in terms of Neural Network will be detailed
in this section.
The SIA and SIDIS experimental data sets analyzed in this work 
and the kinematical cuts applied  
will be summarized in Sec.~\ref{sec:data}.
In Sec.~\ref{sec:fit} we present the fitting methodology as 
well as the Monte Carlo method to 
calculate the uncertainties of FFs.
Sec.~\ref{sec:results} includes the detailed discussion of 
the numerical results in 
the presence of nuclear corrections.
Finally, in Sec.~\ref{sec:Summary}, we will provide 
the summary and conclusion of the study.

%
\section{Theoretical Setup}\label{sec:Theoretical}
%

According to the standard collinear factorization, 
the QCD cross sections can be separated into the perturbative partonic 
cross sections convoluted with the partonic and hadronic distribution 
functions as non-perturbative objects~\cite{esw-book}. 
Therefore, the cross sections for SIA and SIDIS can be written as,

\begin{eqnarray}
\label{eq:SIA-cross}
\sigma ^{\rm {SIA}}&=&
\hat{\sigma} 
\otimes \rm {FF},\nonumber\\
\sigma ^{\rm {SIDIS}}&=&
\hat{\sigma} 
\otimes \rm {PDF} 
\otimes \rm {FF}\,.
\end{eqnarray}

The $\hat{\sigma}$ indicates partonic cross section as a process dependent and
perturbative part of the cross sections.
The universal and non-perturbative parts 
of the cross sections are indicated by PDFs and FFs.
The details of the computation of the SIA and SIDIS cross sections can be found  
in the literature, and we refer the 
reader to Refs.~\cite{Soleymaninia:2022alt,Khalek:2021gxf,Moffat:2021dji} for a clear review.

The NLO and NNLO corrections to the time-like DGLAP evolution equations and 
to the SIA coefficient functions carried out in the 
literature~\cite{Abdolmaleki:2021yjf,Bertone:2017tyb}.
The NNLO corrections to SIDIS coefficient functions can be taken from
Ref.~\cite{Abele:2021nyo} in which the authors have calculated 
the NNLO perturbative corrections by 
applying the threshold resummation formalism up to 
next-to-next-to-leading logarithmic (NNLL) order. 
Then the approximate NNLO corrections to 
unpolarized and polarized SIDIS are achievable.

Accordingly, in the present paper, we investigate the impact of nuclear corrections 
on the FFs of charged hadrons at NLO and NNLO accuracy. 
It is important to highlight that we thoroughly examined the inclusion 
of NNLO QCD corrections but found no improvement in terms of both data/theory 
agreement and the uncertainty bands. Consequently, we decided to restrict our 
analysis to report the results of the NLO analysis in this article.

In the following we present the parameterization and flavor decomposition for 
charged pion, kaon and light hadrons at the initial scale $\mu _0 =5$ GeV. 
Since the SIA data along with SIDIS measurements are included in our analyses, 
the direct constrain on the FFs of light quarks and antiquarks could be provided.
In the case of pions and kaons, we follow the flavor 
decomposition of Ref.~\cite{AbdulKhalek:2022laj}    

\begin{eqnarray}
\label{pionkaon_decomp}
&&D^{\pi^+}_{g},~~~~~~~~~~~~~~~~~~~~~~~D^{K^+}_{g}, \nonumber\\
&&D^{\pi^+}_{u},~~~~~~~~~~~~~~~~~~~~~~~D^{K^+}_{u}, \nonumber\\
&&D^{\pi^+}_{\bar{d}},~~~~~~~~~~~~~~~~~~~~~~~D^{K^+}_{\bar{s}} \nonumber\\
&&D^{\pi^+}_{d}=D^{\pi^+}_{\bar{u}},~~~~~~~~~~~~~D^{K^+}_{s}=D^{K^+}_{\bar{u}}, \nonumber\\
&&D^{\pi^+}_{s}=D^{\pi^+}_{\bar{s}},~~~~~~~~~~~~~D^{K^+}_{d}=D^{K^+}_{\bar{d}} \nonumber\\
&&D^{\pi^+}_{c}=D^{\pi^+}_{\bar{c}},~~~~~~~~~~~~~D^{K^+}_{c}=D^{K^+}_{\bar{c}}, \nonumber\\
&&D^{\pi^+}_{b}=D^{\pi^+}_{\bar{b}},~~~~~~~~~~~~~D^{K^+}_{b}=D^{K^+}_{\bar{b}}\,.
\end{eqnarray}

In the case of unidentified light charged hadrons, 
we follow the flavor decomposition as our previous work~\cite{Soleymaninia:2022alt}, 

\begin{eqnarray}
\label{hadron_decomp}
&&D^{H^+}_{g}, \nonumber\\
&&D^{H^+}_{u}, \nonumber\\
&&D^{H^+}_{\bar{u}} \nonumber\\
&&D^{H^+}_{d}=D^{H^+}_{s}, \nonumber\\
&&D^{H^+}_{\bar{d}}=D^{H^+}_{\bar{s}} \nonumber\\
&&D^{H^+}_{c}=D^{H^+}_{\bar{c}}, \nonumber\\
&&D^{H^+}_{b}=D^{H^+}_{\bar{b}}\,.
\end{eqnarray}

These FFs are obtained in the framework of Neural Networks.
Corresponding to the momentum fraction $z$, we use one input 
node in a one-layered feed-forward Neural Network, $20$ intermediate nodes in a 
one hidden layer and a sigmoid activation function. 
Since the number of independent FFs of the positive pion, kaon and unidentified light 
hadrons are seven, we use $7$ nodes in the output with  a 
linear activation function. The architecture of the Neural Network is ${1-20-7}$ 
and the number of Monte Carlo replicas is $200$. Indeed, the uncertainty 
propagation has been estimated by using the Monte Carlo method.

As mentioned before, the main objective of this paper is to investigate 
the impact of applying nuclear corrections to calculate the SIDIS observables.
Since the PDFs does not have role in the SIA process, the nuclear corrections are
irrelevant to electron-positron annihilation process.

To achieve this goal, in the first step, we utilize the 
proton PDF sets {\tt nNNPDF3.0-p}\cite{AbdulKhalek:2022fyi}, 
{\tt CT18A}~\cite{Hou:2019efy}, and {\tt nCTEQ15WZ-1-1}~\cite{Kusina:2020lyz} as 
input to calculate the SIDIS cross sections. These 
proton PDF sets serve as the baseline for {\tt nNNPDF3.0}~\cite{AbdulKhalek:2022fyi}, 
{\tt EPPS21}~\cite{Eskola:2021nhw}, and {\tt nCTEQ15WZ}~\cite{Kusina:2020lyz} nuclear PDFs, respectively.

In the second step, we incorporate the nuclear PDF sets to 
account for the nuclear corrections.
The computation of FFs at NLO accuracy has been 
performed by means of various nuclear PDF sets, namely 
the {\tt nNNPDF3.0}~\cite{AbdulKhalek:2022fyi}, 
{\tt EPPS21}~\cite{Eskola:2021nhw} and 
{\tt nCTEQ15WZ}~\cite{Kusina:2020lyz}.

We determine the charged pion, kaon and light hadron FFs in 
the Zero-Mass Variable-Flavor-Number Scheme (ZM-VFNS), in which 
the active flavors are considered to be massless. 
In our analysis, the masses of charm and bottom quarks 
are considered to be fixed at $m_c = 1.51$ GeV and $m_b = 4.92$ GeV, respectively. 
Finally, we should mention that we utilize the {\tt MontBlanc} public package for our global QCD 
analyses in this project.

%
\section{Experimental Data Setup}\label{sec:data}
%

The experimental data sets used in this work include 
both the SIA and SIDIS measurements for charged pion, kaon and 
unidentified light hadrons productions. 

In the case of SIA, the measured differential cross section is 
normalized to the total cross section and reported as 
sum of the production of positively and negatively charged hadrons. 
The data analyzed in this work include different experiments 
performed by {\tt TASSO}~\cite{TASSO:1980dyh,TASSO:1982bkc,TASSO:1988jma,TASSO:1990cdg} 
experiment at DESY, {\tt TPC}~\cite{Aihara:1988su}, {\tt BABAR}~\cite{BaBar:2013yrg} and 
{\tt SLD}~\cite{Abe:2003iy} experiment at SLAC, {\tt BELLE}~\cite{Belle:2013lfg} and 
{\tt TOPAZ}~\cite{TOPAZ:1994voc} at KEK, and
{\tt ALEPH}~\cite{Buskulic:1995aw,ALEPH:1994cbg}, {\tt DELPHI}~\cite{Abreu:1998vq} and 
{\tt OPAL}~\cite{OPAL:1994zan,Ackerstaff:1998hz} experiments at CERN.

In the case of SIDIS observables, the experimental data correspond to 
the SIDIS cross section which is normalized to the 
inclusive DIS cross section. 
The SIDIS measurements have been reported both for the production of 
positive and negative charged hadrons. 
The {\tt COMPASS} experiment  at CERN~\cite{COMPASS:2016crr,COMPASS:2016xvm} and 
{\tt HERMES} experiment at DESY~\cite{HERMES:2012uyd,HERMES:2009uge,HERMES:2007plz} are two main experiments which have 
reported the experimental data for SIDIS processes for pion, 
kaon and unidentified light hadrons productions.
It is worth emphasizing that the SIDIS data not only have played a 
crucial role in determining of FFs but also the medium-modified FFs, as discussed 
in Ref.~\cite{Sassot:2009sh,Zurita:2021kli}

Considering the impact of nuclear PDF sets in the SIDIS case, 
one should take into account the nuclear PDF 
sets relevant to the nuclear target.
The {\tt COMPASS} experiment used a muon beam, and the target 
consisted of Li-6 deuteride immersed in a mixture of He-3 and He-4.
As a theoretical aspect, we should choose one of them as the 
isoscalar target material. 
Since $^6$LiD has been considered as an isoscalar 
target in the measurement by the {\tt COMPASS} 
Collaboration, we have chosen Li-6 as the main target.

The data are 
collected by the {\tt HERMES} experiment at the HERA 
storage ring using electron and positron beams 
on a hydrogen or deuterium gas target.
Since {\tt MontBlanc} unable to use different PDF sets at the 
same time to calculate the SIDIS observables, 
we limit our study to the {\tt COMPASS} data sets which reported 314 points 
for pion, kaon and unidentified light hadrons. 
The published {\tt HERMES} data only include the 4 points and neglected in our study.
The SIA and SIDIS experimental data for light charged hadron 
production used in our analysis follow those of our previous work~\cite{Soleymaninia:2022alt}. 
We include all the charged pion and kaon data of SIA 
and SIDIS experiments, which are applied in recent 
analyses of {\tt MAPFF1.0}~\cite{AbdulKhalek:2022laj}.  
As we mentioned earlier the {\tt HERMES} data is excluded from our analysis. 
More detailed study on the data selection can be 
found in Refs.~\cite{AbdulKhalek:2022laj, Soleymaninia:2022alt}.
The proton PDF set of {\tt NNPDF3.0} has been used for 
determination of SIDIS observables as our baseline fit. Hence, in order to 
study the nuclear corrections one need 
to include the nuclear PDFs for $^{6}$LiD.

All measured observables analyzed in this study along with their published 
references and the number of data points are presented in 
Tables.~\ref{tab:datasets_kaon_NLO} for kaon, and in Tables.~\ref{tab:datasets_hadron_NLO} 
for the unidentified light charge hadrons, 
and in Tables.~\ref{tab:datasets-pion_NLO} for pion. 
The values of individual and total $\chi^2$ are presented for different PDF sets.
In the case of SIA experimental data sets, the experimental data  
have reported as total inclusive, light-, charm-, and bottom-tagged 
cross sections measurements.

One needs to impose kinematical cuts on small and large values of $z$ 
in which the perturbative fixed-order predictions should be reliable.
We follow Refs.~\cite{AbdulKhalek:2022laj, Soleymaninia:2022alt} to apply 
the kinematic cuts on the SIA and SIDIS data sets. 
For the SIA data points, we apply the kinematic cuts on $z$ based on 
the value of a center-of-mass (CM) energy:  
for data points with the scale of energy less than $M_Z$ we 
use $0.075\leq z\leq 0.9$, and for the data points with the 
scale of energy equal to $M_Z$ we consider $0.02\leq z\leq 0.9$.

For SIDIS, we consider the kinematical cut on the small $Q$ 
and retain the experimental data points in region $Q \geq 2$ GeV.
The same kinematical cuts are applied for charged pion, kaon and 
unidentified light hadrons analyses.
In addition, since the small-$z$ corrections at energy scale of $B$ factories 
for kaon production occur at higher value of $z$ in comparison with pion, 
we adopt $0.2\leq z\leq 0.9$ for the BABAR experiments for kaon production.

%
\section{Analysis setup and fitting procedure}\label{sec:fit}
%

After briefly presenting the flavor decomposition of FFs 
and related to observables of interest, 
we are in a position to discuss the analysis setup and fitting procedure. 

We will state other theoretical choices and 
consider their effect on the extraction of FFs. 
The architecture of Neural Network used for parametrization is 
(1-20-7) which is quite simple but it proves to 
be sufficient~\cite{Soleymaninia:2022alt} for the FFs analysis, 
it has 187 free parameters (weights and biases) need to be fixed from the data. 
To force the distributions to vanish at $z=1$ and to be positive, 
value of NN at $z=1$ (i.e. $\mathcal{N}(1;\boldsymbol{\theta})$) 
is subtracted from the output of 
the Neural Network and then the outcome is squared, 
so the parametrization can be written as,

\begin{eqnarray}
z D(z,\mu_0) = \left(\mathcal{N}(z;\boldsymbol{\theta})-\mathcal{N}(1;\boldsymbol{\theta})\right)^2.
\end{eqnarray}

In the equation above, the $\mathcal{N}$ represents a Neural Network with 
set of parameters $\boldsymbol{\theta}$, 
and the parametrization is assumed to be at $\mu_0=5$~GeV, 
above bottom quark mass threshold \footnote{Mass thersholds considered to 
be, $1.51$~GeV for charm and $4.92$~GeV for bottom quark.}. 
This parametrization is used for NLO calculation with 
strong coupling constant $\alpha_S=0.118$ at Z boson mass scale ($91.1876$~GeV)  
together with appropriate evolution.

As mentioned, the analysis made in this paper utilizes \texttt{MontBlanc} package, 
a \texttt{C++} package dedicated to determination of collinear FFs~\cite{MontBlanc}. 
In order to perform perturbative QCD analysis 
\texttt{MontBlanc} package uses the following packages; 
\texttt{GSL} or GNU scientific library for numerical utilities~\cite{gsl}, 
\texttt{yaml-cpp}~\cite{yaml-cpp} for reading data 
and writing the results, \texttt{LHAPDF} library for reading 
input PDFs~\cite{Buckley:2014ana}, \texttt{Ceres Solver}~\cite{ceres-solver} 
for minimizing the $\chi^2$ function, \texttt{NNAD}~\cite{nnad} for defining Neural Network 
and their derivatives, \texttt{APFEL++}~\cite{Bertone:2017gds} for evolution of FFs 
and \texttt{NangaParbat}~\cite{nanga} for calculation of observables. 
The $\chi^2$ function minimized during analysis is defined as,

\begin{eqnarray}
	\chi^{2}_{(k)} \equiv \left(\boldsymbol {\mathrm{T}}\left(\boldsymbol{\theta}_{(k)}\right) - \boldsymbol x_{(k)}\right)^T \cdot \boldsymbol {\mathrm{C}^{-1}} \cdot \left(\boldsymbol {\mathrm{T}}\left(\boldsymbol{\theta}_{(k)}\right) - \boldsymbol x_{(k)}\right)\,,\nonumber \\
	\label{chi2}
\end{eqnarray}
where $\boldsymbol {\mathrm{C}^{-1}}$ is inverse of 
covariance matrix which includes all statistical and systematic uncertainties. 
The symbol $\boldsymbol x_{(k)}$ in represents $k$-th replica or 
pseudo-data set, analogously $\boldsymbol{\theta}_{(k)}$ is 
parameters of $k$-th Neural Network replica. 
Here $\boldsymbol {\mathrm{T}}$ represents the theoretical prediction 
calculated using mentioned Neural Network parameters. 
The symbol $T$ indicates to the matrix transposing as usual. 
So far we have introduced the different parts of Eq.~\ref{chi2}, 
now we are in a position to explain what we mean by replica. 
The method of Monte Carlo is used in this analysis to propagate 
experimental uncertainty to the resulting FFs. 
In this method a suitable ensemble of copies of data is chosen, 
after performing $N_{\rm rep.}$ fits over 
individual (pseudo-)data sets in the ensemble, 
produced FFs are such that their simple statistical parameters as 
average and standard deviation constitute their central value and uncertainty, 
for more details see e.g.~\cite{Soleymaninia:2022alt} and references therein.

To consider the SIDIS data, (nuclear) PDFs are needed to calculate the cross section. In the 
\texttt{MontBlanc} approach, a random replica of the input 
PDF is chosen in every fit to incorporate 
PDF uncertainty in the determination of FFs. Despite the fact that this source of 
uncertainty is shown to be minor~\cite{Khalek:2021gxf}, we opt to include it for the sake 
of completeness. Most of the available nuclear PDFs present the uncertainty using 
the Hessian method, and one needs to convert the Hessian error sets into Monte Carlo 
sets using the \texttt{MCGEN} code~\cite{Hou:2016sho,valerio} and then implement them 
in the analysis. The \texttt{MCGEN} creates Monte Carlo replicas using random 
displacements of the Hessian sets and ensures that important properties 
of the Hessian PDFs are preserved.

%
\section{Numerical Results}\label{sec:results}
%

In this section, we will present the main findings of this study and provide a 
comparison between the computed FFs with and without nuclear corrections.
We will begin by presenting the impact of nuclear corrections on our QCD 
analysis of pion FFs. We will provide a detailed discussion regarding the 
effect on both the central values and the error bands.
Then, we will examine the effect of nuclear corrections on the extracted kaon FFs
and unidentified charged hadron FFs.
We will also provide a comprehensive analysis of the impact resulting from 
different nuclear correction on the individual 
and total $\chi^2$ values for various hadrons.

%
\subsection{Nuclear effects of the pion FFs}\label{sec:pion_result}
%

In this section, we will present the results for the pion FFs 
when incorporating nuclear corrections.
To this aim and in order to investigate the 
impact arising from the inclusion 
of the nuclear effects at NLO accuracy, 
we consider three different sets of nuclear PDFs 
available in the literature, namely 
the {\tt nNNPDF3.0}~\cite{AbdulKhalek:2022fyi}, 
{\tt EPPS21}~\cite{Eskola:2021nhw}, 
and {\tt nCTEQ15WZ}~\cite{Kusina:2020lyz}.

It is worth noting that the global QCD analyses 
reported by the {\tt nNNPDF3.0}, {\tt EPPS21}, and {\tt nCTEQ15WZ} 
PDF sets are limited to NLO accuracy. This means that the 
nuclear PDFs provided by these sets are derived within the 
framework of NLO perturbative QCD. 
In the following, we present a short summary of these nuclear PDFs.	

The {\tt nNNPDF3.0}~\cite{AbdulKhalek:2022fyi} analysis presented an 
updated determination of nuclear PDFs from global datasets, including the 
neutral-current (NC) nuclear fixed-target DIS measurements, the charged-current (CC) 
neutrino-nucleus DIS data, pPb collision at LHC for boson production, 
data for CMS dijet~\cite{CMS:2018jpl}, LHCb D0-meson production~\cite{LHCb:2017yua}, 
isolated photon production~\cite{ATLAS:2019ery}, and fixed-target Drell-Yan (DY) measurements. 
As a final note, it is worth mentioning that the proton PDF 
set used as a baseline for {\tt nNNPDF3.0} is {\tt nNNPDF3.0-p}.

The EPPS collaboration~\cite{Eskola:2021nhw} has presented an 
updated {\tt EPPS21} nuclear PDF sets by new experimental data,  
the LHC pPb data for dijets (Run-I), D-mesons 
(Run-I) and $W^\pm$ bosons (Run-II). 
In addition, they have included Jefferson Lab measurements 
of DIS which probe nuclear PDFs. 
In  {\tt EPPS21} the inclusive pion production at RHIC has been 
included and hence they have used a 
FF sets of pion in their analysis.
The proton PDF set {\tt CT18A}~\cite{Hou:2019efy}  has been 
reported as the proton PDF baseline of {\tt EPPS21}.

Recently, the nCTEQ Collaboration has expanded the {\tt nCTEQ15} dataset by 
incorporating the $W$ and $Z$ data obtained 
from pPb collisions at the LHC~\cite{Kusina:2020lyz}. 
In addition to the nuclear DIS, DY lepton pair production, 
and RHIC pion data had been employed in the earlier nCTEQ15 analysis. 
They referred to it as {\tt nCTEQ15WZ} and the accuracy is up to NLO.
The Proton PDF baseline of this nuclear PDF is free 
proton PDF set {\tt nCTEQ15WZ-1-1}~\cite{Kusina:2020lyz}. 

Considering the above mentioned nuclear PDFs, 
the total and individual values for the $\chi^2$ per data point 
are presented in Tables.~\ref{tab:datasets-pion_NLO}
for the NLO analyses.

As shown in Table.~\ref{tab:datasets-pion_NLO}, the incorporation of 
nuclear PDFs from {\tt nNNPDF3.0} leads to a significant 
reduction in the $\chi^2$ per data point value for 
the {\tt COMPASS} $\pi^+$ dataset, resulting in a 
reduction in the total $\chi^2$ value from 0.879 to 0.809.

However, the use of nuclear PDFs {\tt EPPS21} and {\tt nCTEQ15WZ} does not 
result in improvements in the individual $\chi^2$ values per 
data point, particularly for the {\tt COMPASS} dataset.
The total $\chi^2$ value increases from 0.757 to 0.765 and 
from 0.742 to 0.761, respectively, when considering these two 
nuclear PDF sets. 
The $\chi^2$ values for both {\tt nCTEQ15WZ} and {\tt EPPS21} 
are closely aligned with our baseline fits as well.

\begin{table*}[htb]
	\renewcommand{\arraystretch}{2}
	\centering 	\scriptsize
	\begin{tabular}{|l|c|cc|cc|cr|}				\hline
	
	       ~        &  ~ &  \multicolumn{6}{c|}{$\frac{\chi^2}{N_{\rm dat}}$:}
	        \\
	    Experiment & $N_{\rm dat}$ &~  {\tt nNNPDF3.0-p}    ~&~   {\tt nNNPDF3.0} ~&~  {\tt CT18A} ~&~ {\tt EPPS21} ~&~ {\tt nCTEQ15WZ-1-1} ~&~ {\tt nCTEQ15WZ} 
		\rule[-3mm]{0mm}{5mm}
		\\
		\hline \hline
		{\tt COMPASS} $\pi ^+$\cite{COMPASS:2016xvm}   & 157  & 1.010& 0.731&0.570&0.598&0.538& 0.593 \\
		{\tt COMPASS} $\pi ^-$~\cite{COMPASS:2016xvm} &   157 & 0.412 & 0.422&0.437&0.424&0.414&0.430 \\
		{\tt BELLE}\cite{Belle:2013lfg}   & 70 & 0.094& 0.092&0.092&0.092&0.094& 0.093 \\
		{\tt BABAR}~\cite{BaBar:2013yrg} &   39 & 1.309 & 1.300&1.173&1.242&1.421&1.375 \\
		{\tt TASSO12}~\cite{TASSO:1980dyh}  & 4 & 0.975 &0.975&0.976&0.976&0.978&0.977 \\
		{\tt TASSO14}~\cite{TASSO:1982bkc}  & 9 & 1.385& 1.389 &1.384&1.379&1.373&1.374\\       					
		{\tt TASSO22}~\cite{TASSO:1982bkc}   & 8 & 1.931& 1.915&1.870&1.858&1.817&1.838\\
		{\tt TPC}~\cite{Aihara:1988su}  & 13 & 0.232 &0.229&0.222&0.223&0.230& 0.223\\
		{\tt TASSO30}~\cite{TASSO:1980dyh}   & 2 &0.353 &0.347&0.341&0.339&0.322&0.336\\
		{\tt TASSO34}~\cite{TASSO:1988jma} & 9 & 1.378&  1.353&1.241&1.216&1.133&1.165\\                                          				        
		{\tt TASSO44}~\cite{TASSO:1988jma}& 6 & 1.260&1.247&1.198&1.183&1.117&1.164\\
		{\tt TOPAZ}~\cite{TOPAZ:1994voc}& 5 & 0.326 &0.319&0.284&0.273 &0.237&0.260\\
		{\tt ALEPH}~\cite{ALEPH:1994cbg}  & 23 & 1.121 &1.139&1.216&1.189&1.159&1.164\\
		{\tt DELPHI} (incl.)~\cite{Abreu:1998vq}   & 21 & 1.330 &1.311&1.294&1.299&1.296&1.305\\
		{\tt DELPHI} ($uds$ tag)~\cite{Abreu:1998vq}  & 21 & 2.541 &2.573&2.561&2.559&2.512&2.517 \\
		{\tt DELPHI} ($b$ tag)~\cite{Abreu:1998vq}  &21 & 1.734  & 1.785 &1.754& 1.806&1.741&1.756\\
		{\tt OPAL} (incl.)~\cite{OPAL:1994zan} & 24& 1.646& 1.674&1.688&1.681&1.659&1.661\\                                                    				
		{\tt SLD} (incl.)~\cite{Abe:2003iy}   & 34 & 1.060&1.096&1.125&1.108&1.082&1.087\\
		{\tt SLD} ($uds$ tag)~\cite{Abe:2003iy}   &34 & 1.679 & 1.694&1.501& 1.571&1.499&1.463 \\                                             				
		{\tt SLD} ($b$ tag)~\cite{Abe:2003iy} &34 &0.595 &0.604&0.599&0.605&0.596&0.598\\ 				
		\hline \hline
		Total $\chi^2/{N_{\rm dat}}$ & 691 &0.879 &0.809&0.757&0.765 &0.742& 0.761  \\
		\hline \hline	
	\end{tabular}
	\caption{ \small 
		The list of input data sets for the pion production included in our pion FFs analysis. 
		For each data set, we have indicated 
		the experiments, corresponding published reference and the number of data points. 
		In the last four columns, we show the value of $\chi^2/{N_{\rm dat}}$ resulting from the 
		FF fit at NLO order by considering proton PDF sets from {\tt nNNPDF3.0-p}~\cite{ AbdulKhalek:2022fyi}, {\tt CT18A} ~\cite{Hou:2019efy}  and {\tt nCTEQ15WZ-1-1} \cite{Kusina:2020lyz}, and 
		nuclear PDF sets available in the literature, {\tt nNNPDF3.0}~\cite{AbdulKhalek:2022fyi}, 
		{\tt EPPS21}~\cite{Eskola:2021nhw}, 
		and {\tt nCTEQ15WZ}~\cite{Kusina:2020lyz}. 
		The total value of the total $\chi^2/{N_{\rm dat}}$ also is shown at the bottom of the table.} 
	\label{tab:datasets-pion_NLO}
\end{table*}

\begin{figure*}[htb]
\vspace{0.50cm}
	\resizebox{0.45\textwidth}{!}{\includegraphics{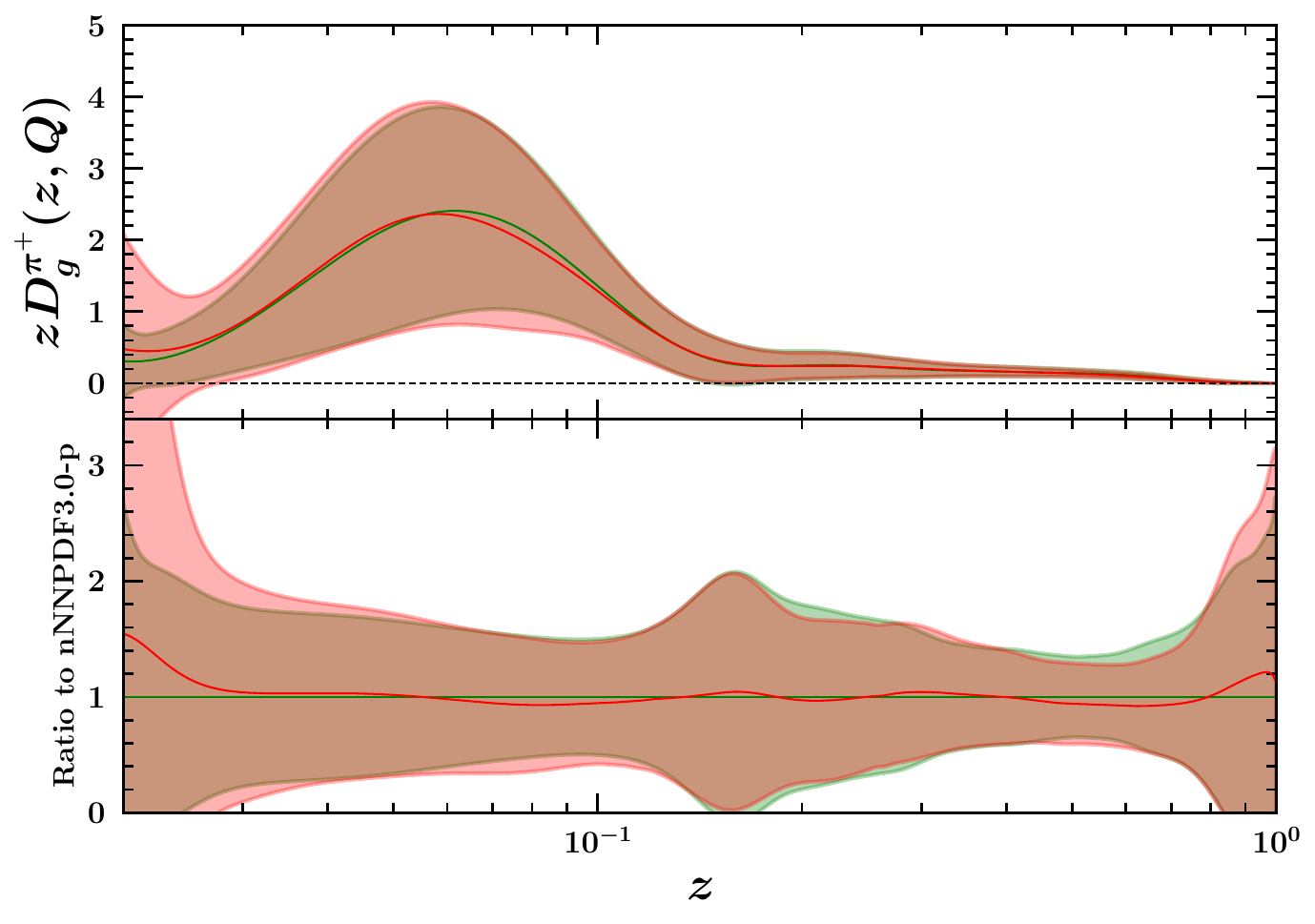}} 	
	\resizebox{0.45\textwidth}{!}{\includegraphics{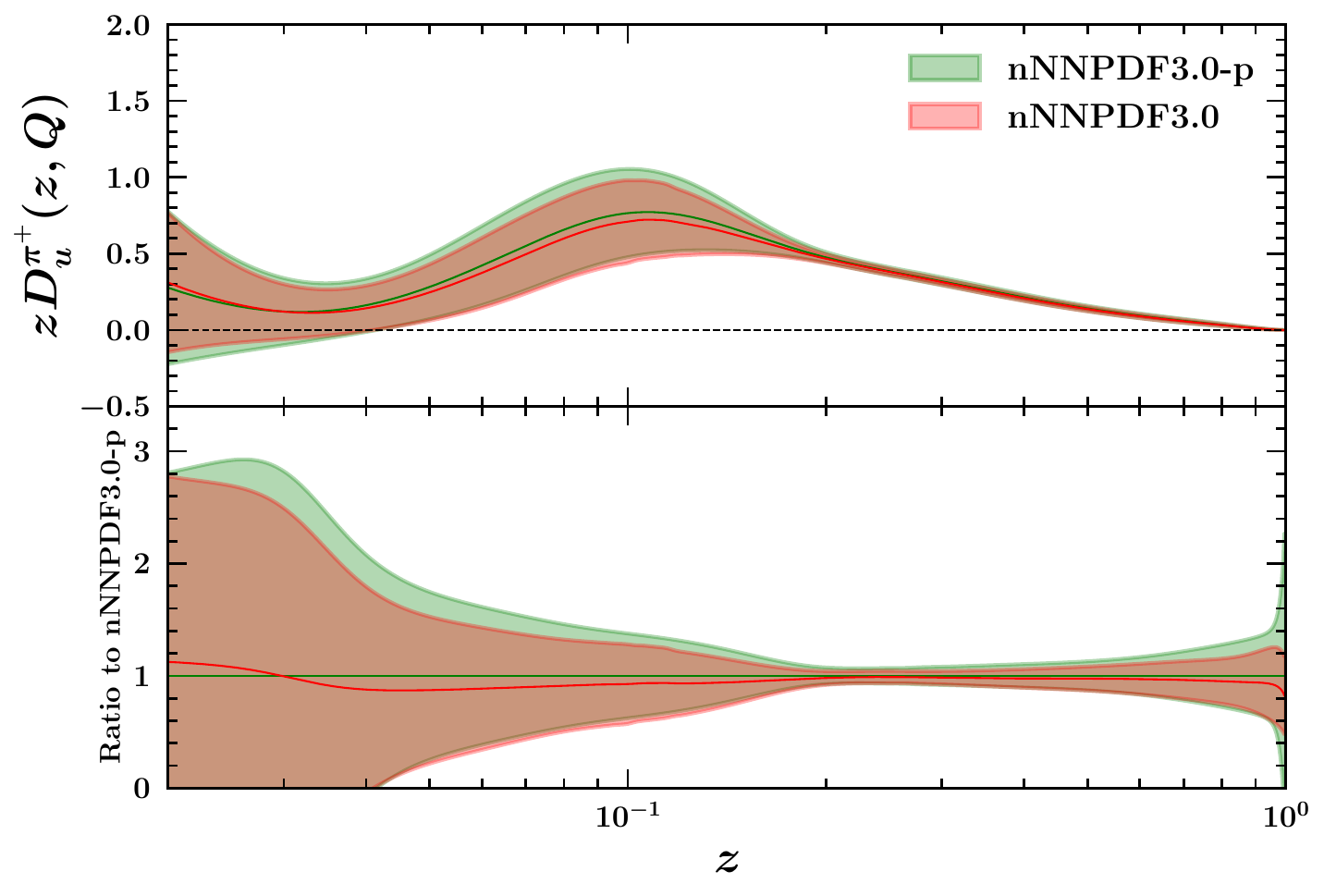}}   
	\resizebox{0.45\textwidth}{!}{\includegraphics{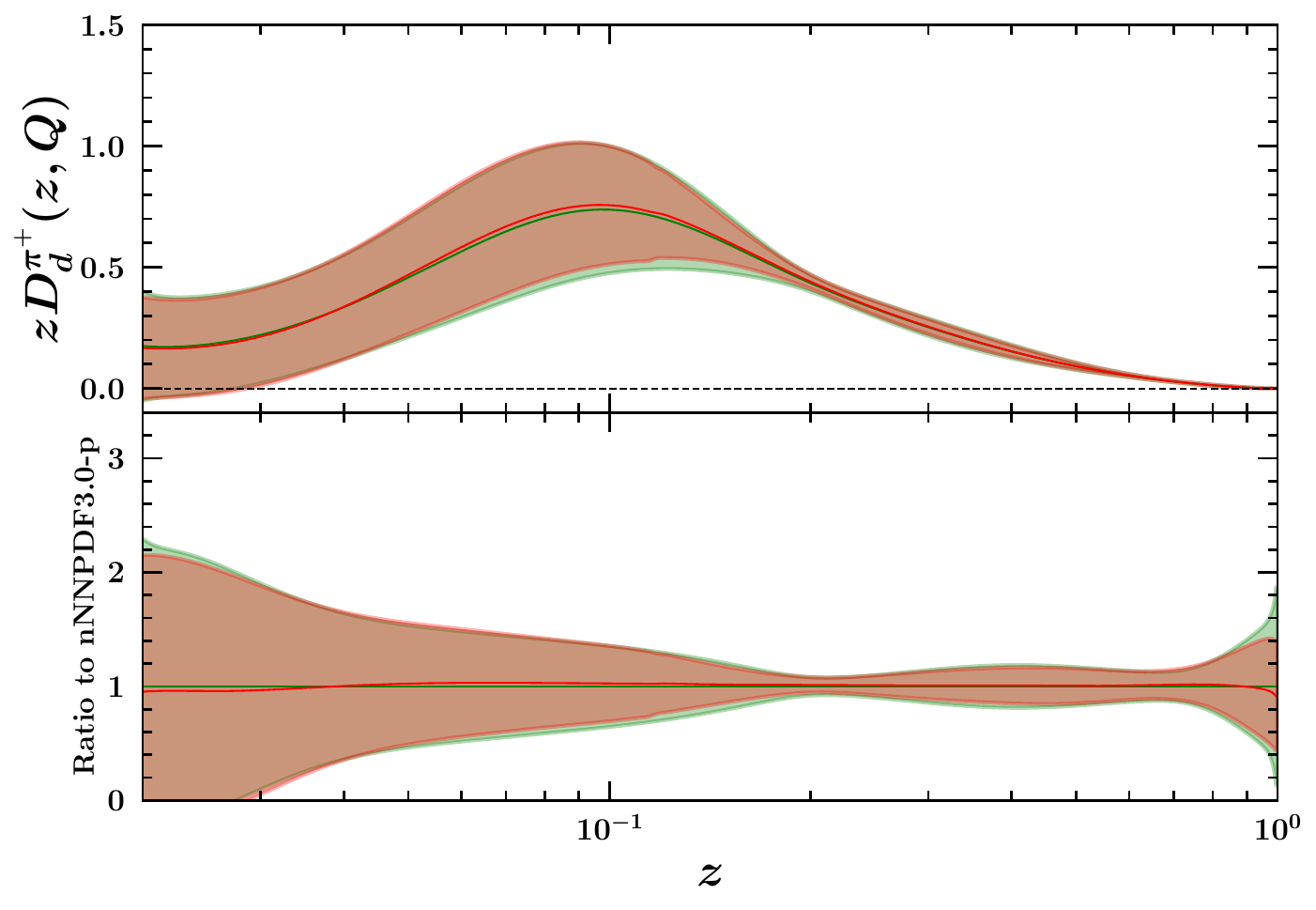}}	
	\resizebox{0.45\textwidth}{!}{\includegraphics{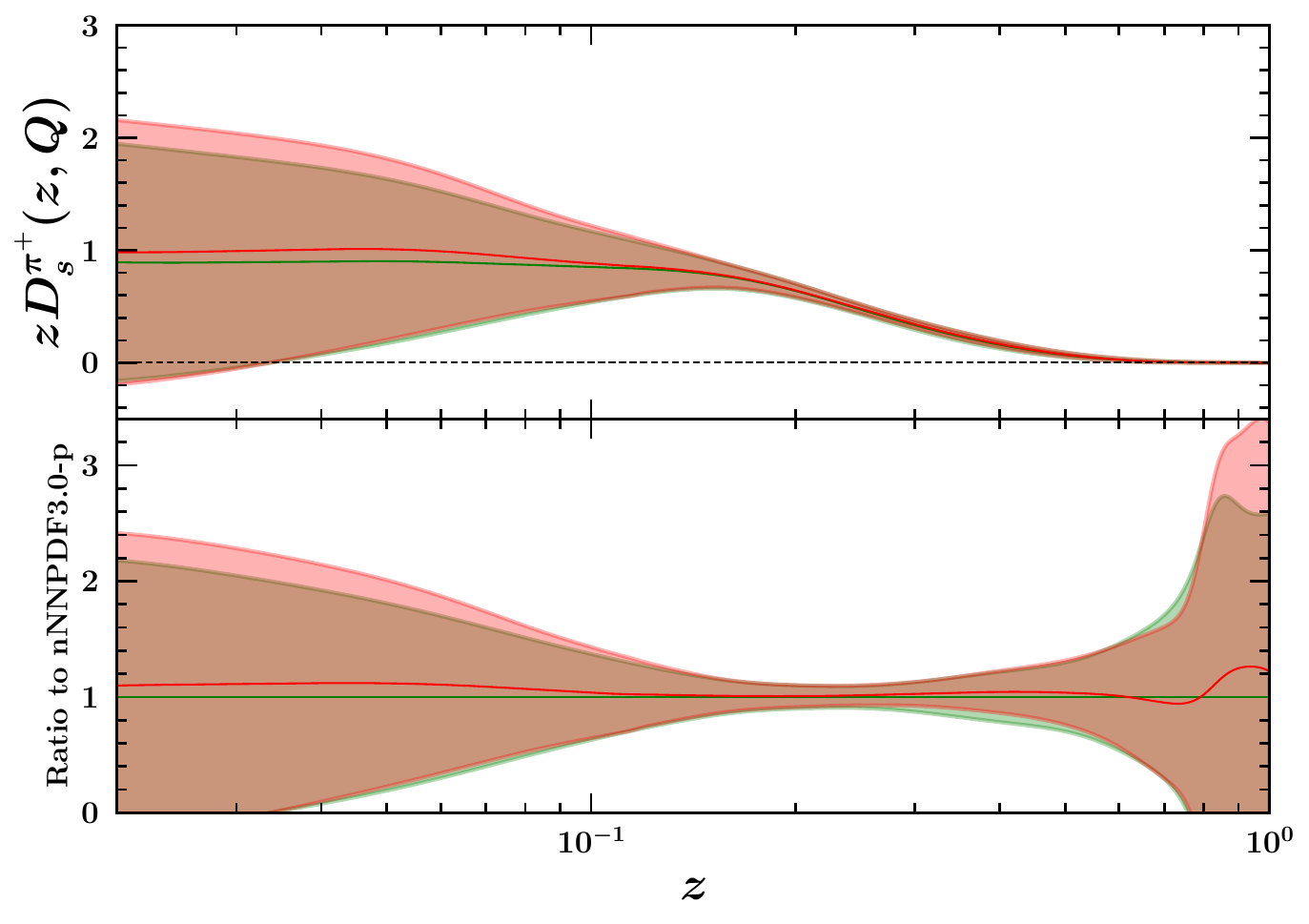}} 	
	\resizebox{0.45\textwidth}{!}{\includegraphics{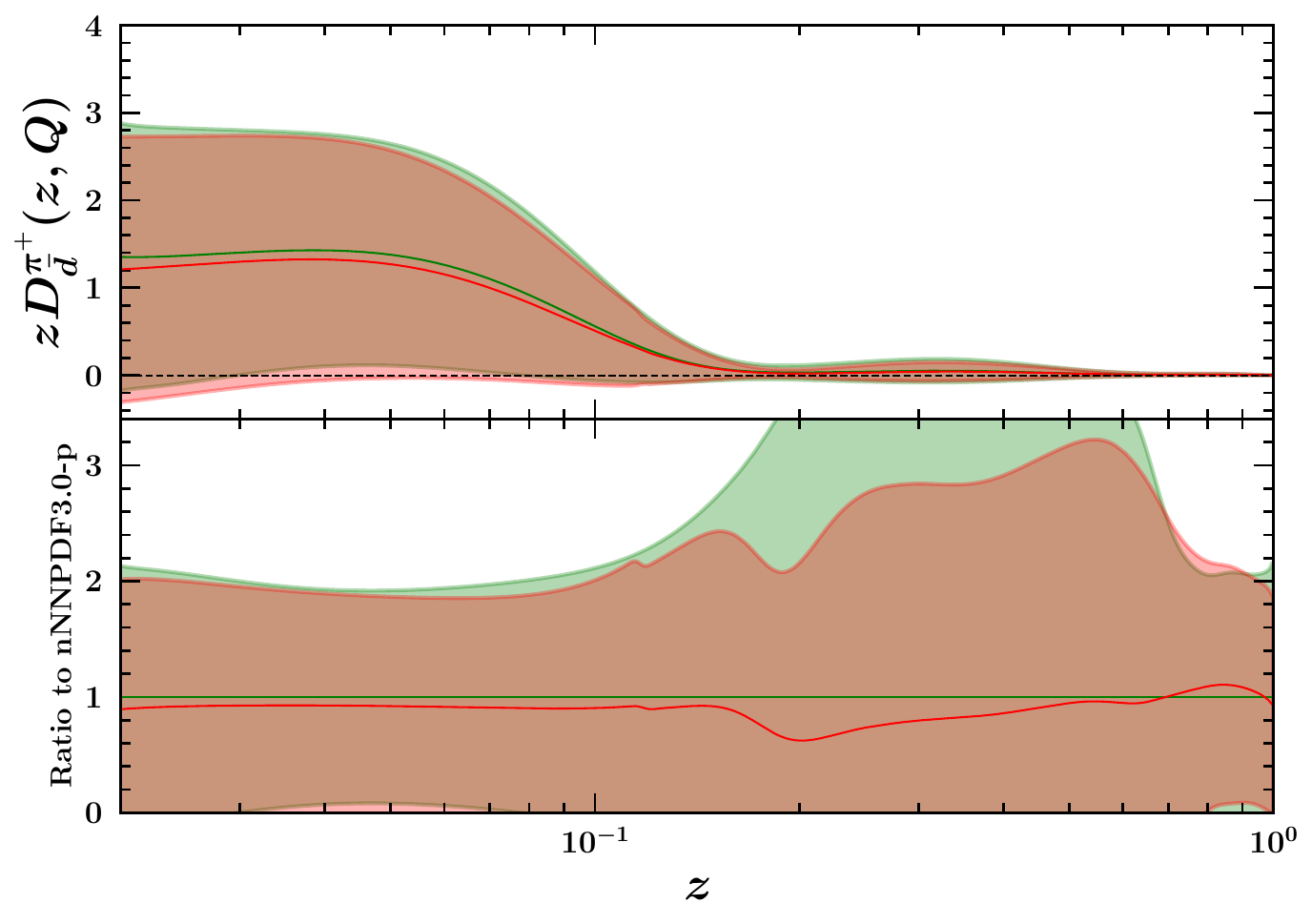}}
	\resizebox{0.45\textwidth}{!}{\includegraphics{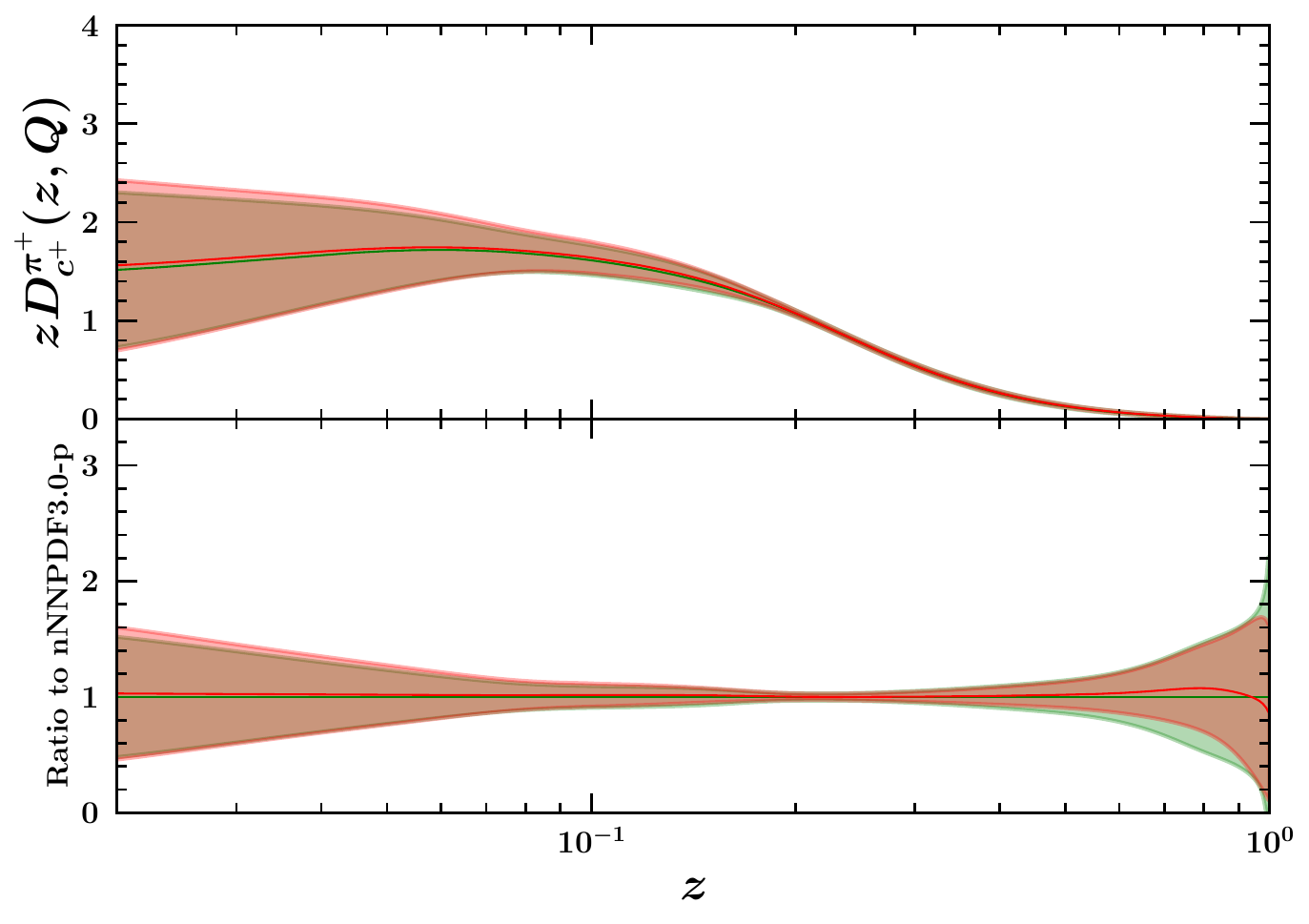}} 
	\resizebox{0.45\textwidth}{!}{\includegraphics{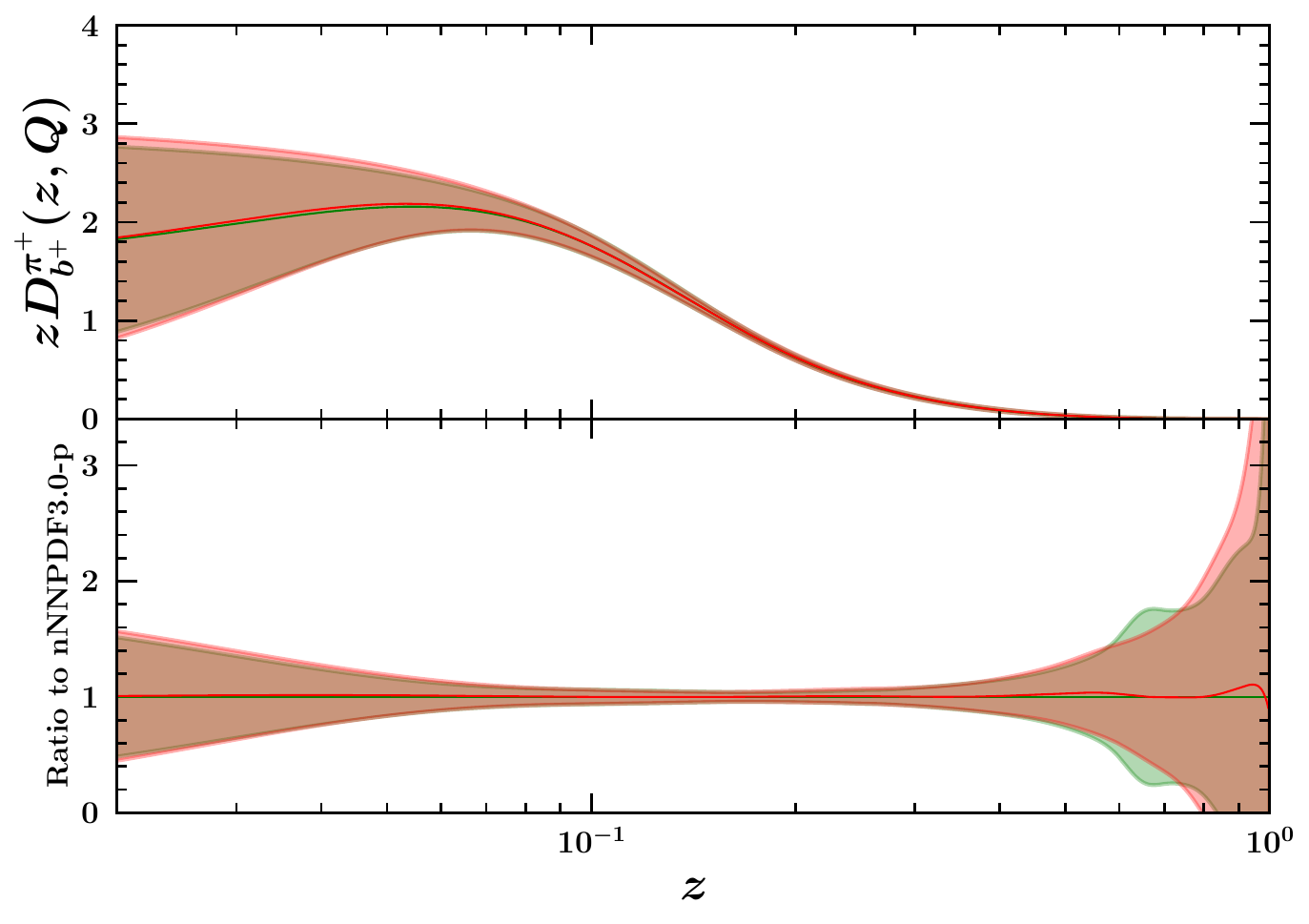}}   	 	
\begin{center}
\caption{ 
\small 
Comparison of pion FFs extracted from {\tt nNNPDF3.0-p}
proton PDF sets as our baseline, 
and the nuclear PDF set from {\tt nNNPDF3.0}. 
We present both the absolute values and the ratio 
to the {\tt nNNPDF3.0-p} proton PDFs baseline for the pion 
FFs extracted using the {\tt nNNPDF3.0} nuclear PDF set. 
The results are presented at a scale of $Q=5$ GeV. }
\label{fig:pion-FFs_NLO_nNNPDF}
\end{center}
\end{figure*}

\begin{figure*}[htb]
	\vspace{0.50cm}
	\resizebox{0.45\textwidth}{!}{\includegraphics{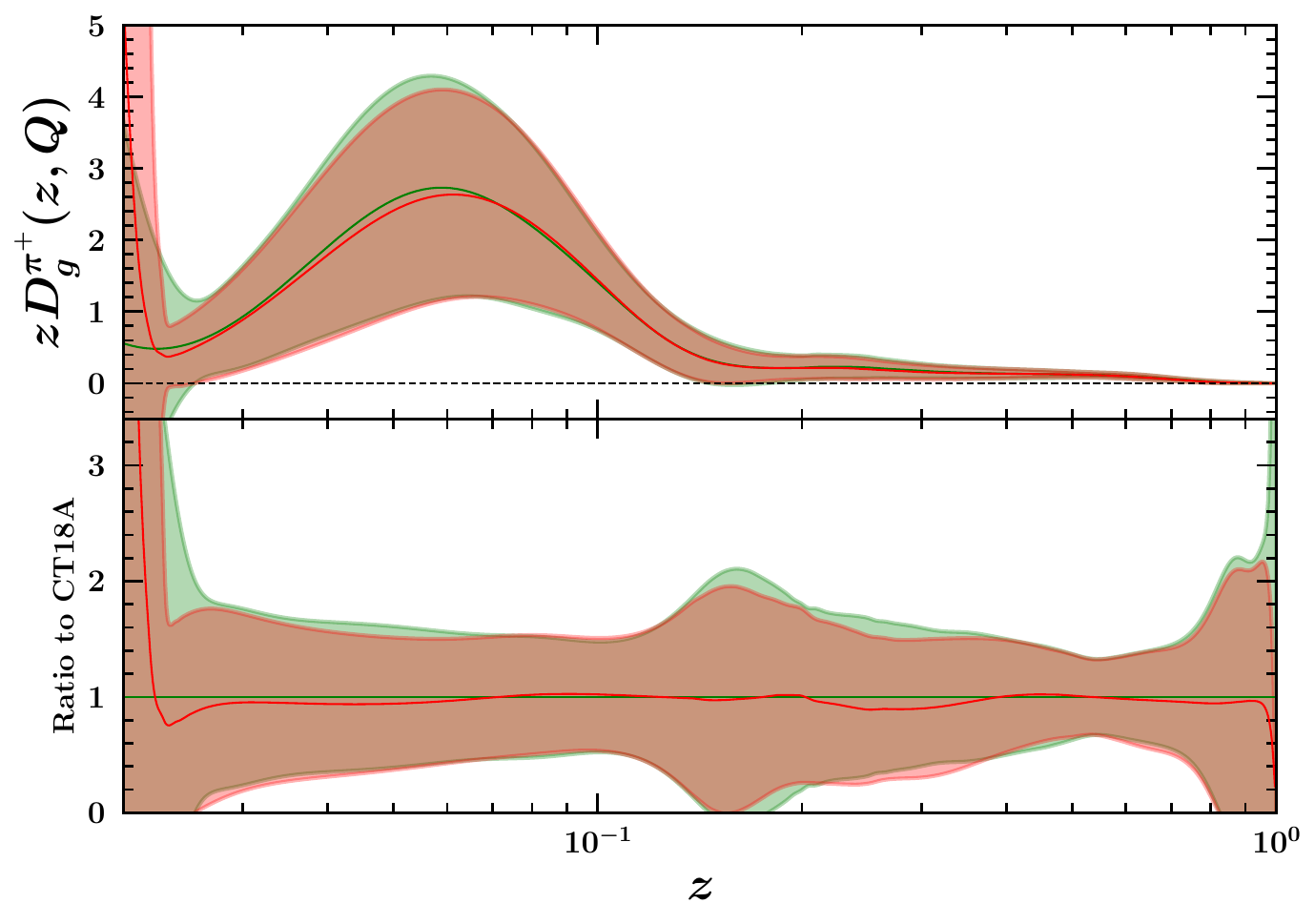}} 	
	\resizebox{0.45\textwidth}{!}{\includegraphics{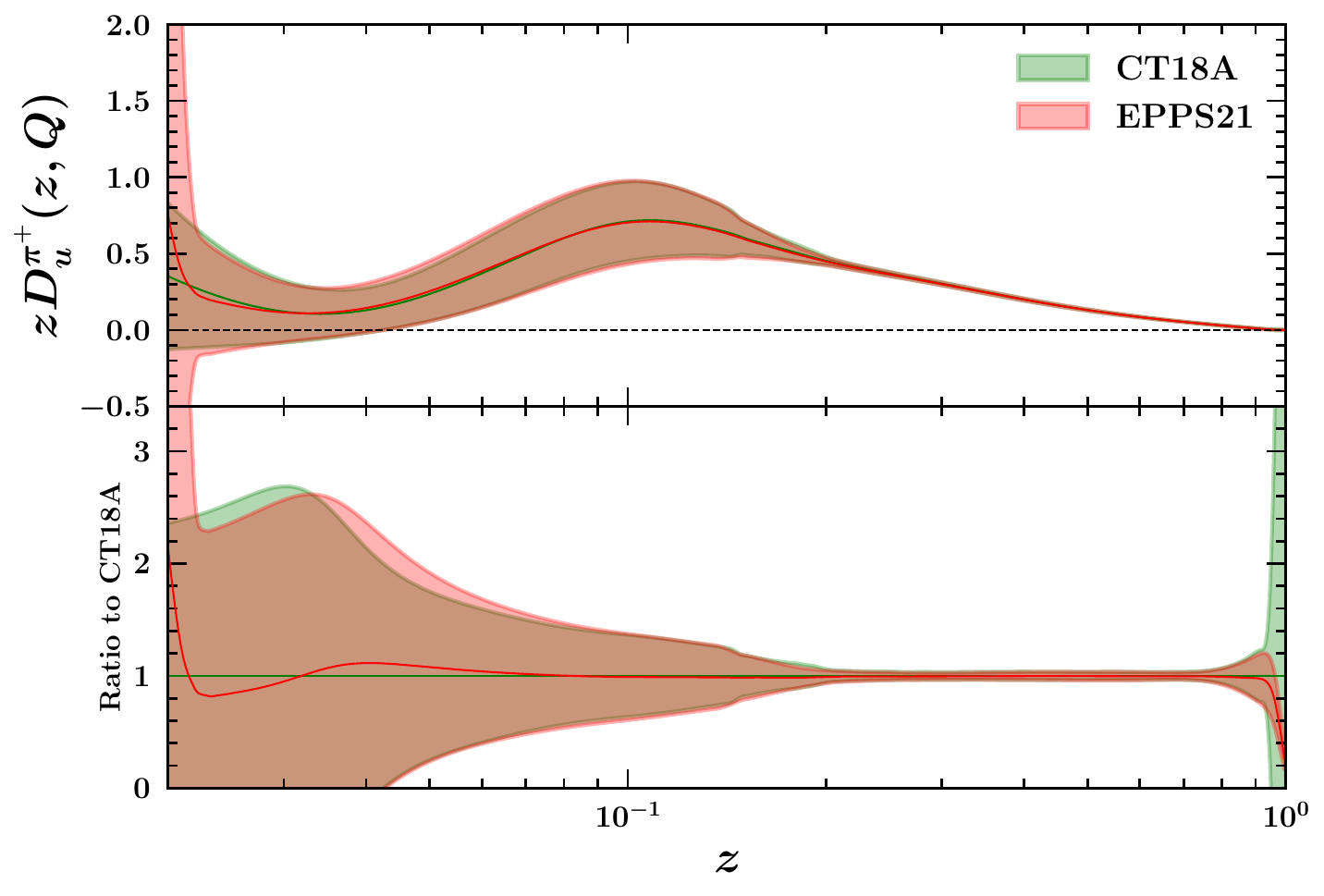}}   
	\resizebox{0.45\textwidth}{!}{\includegraphics{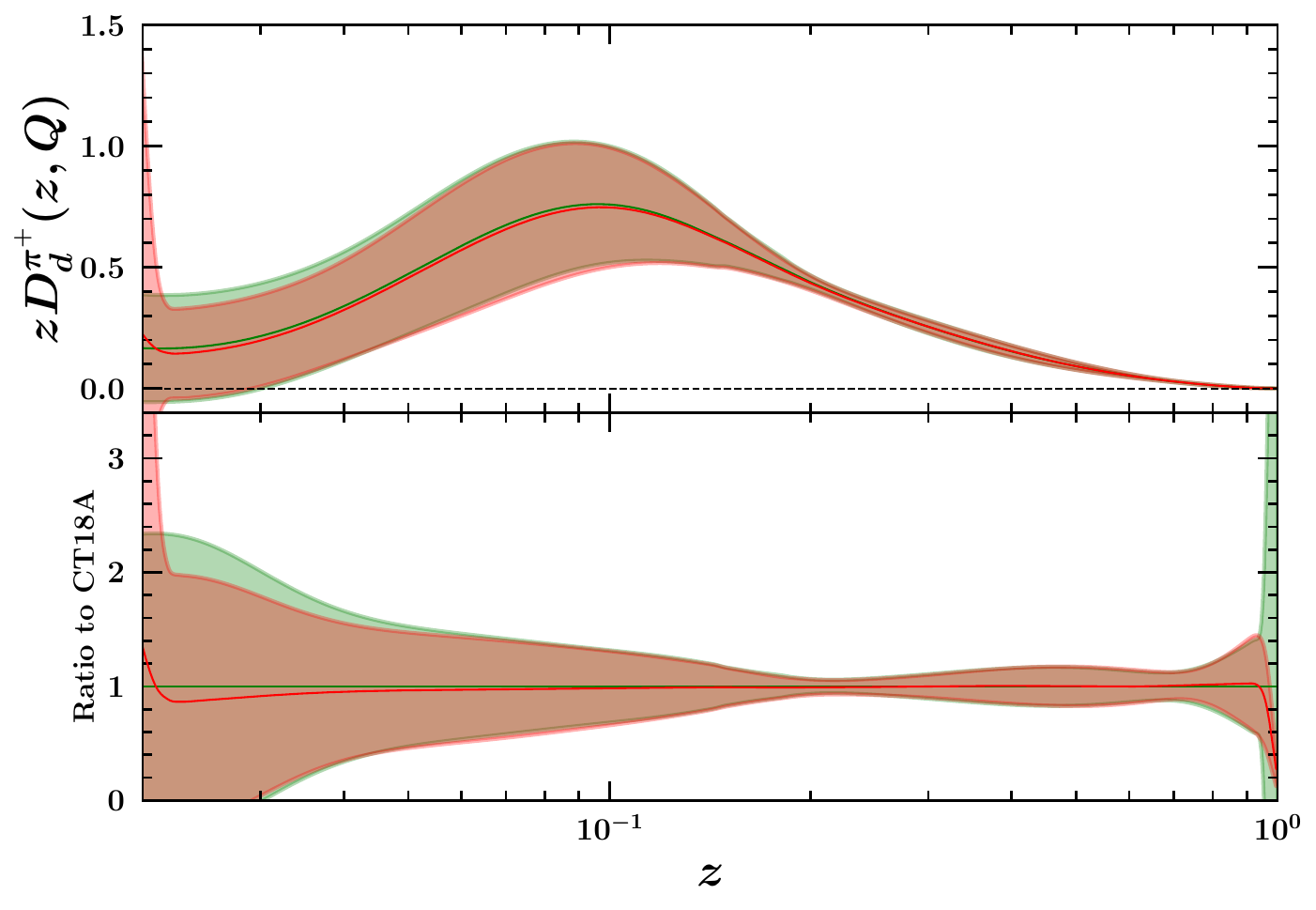}}	
	\resizebox{0.45\textwidth}{!}{\includegraphics{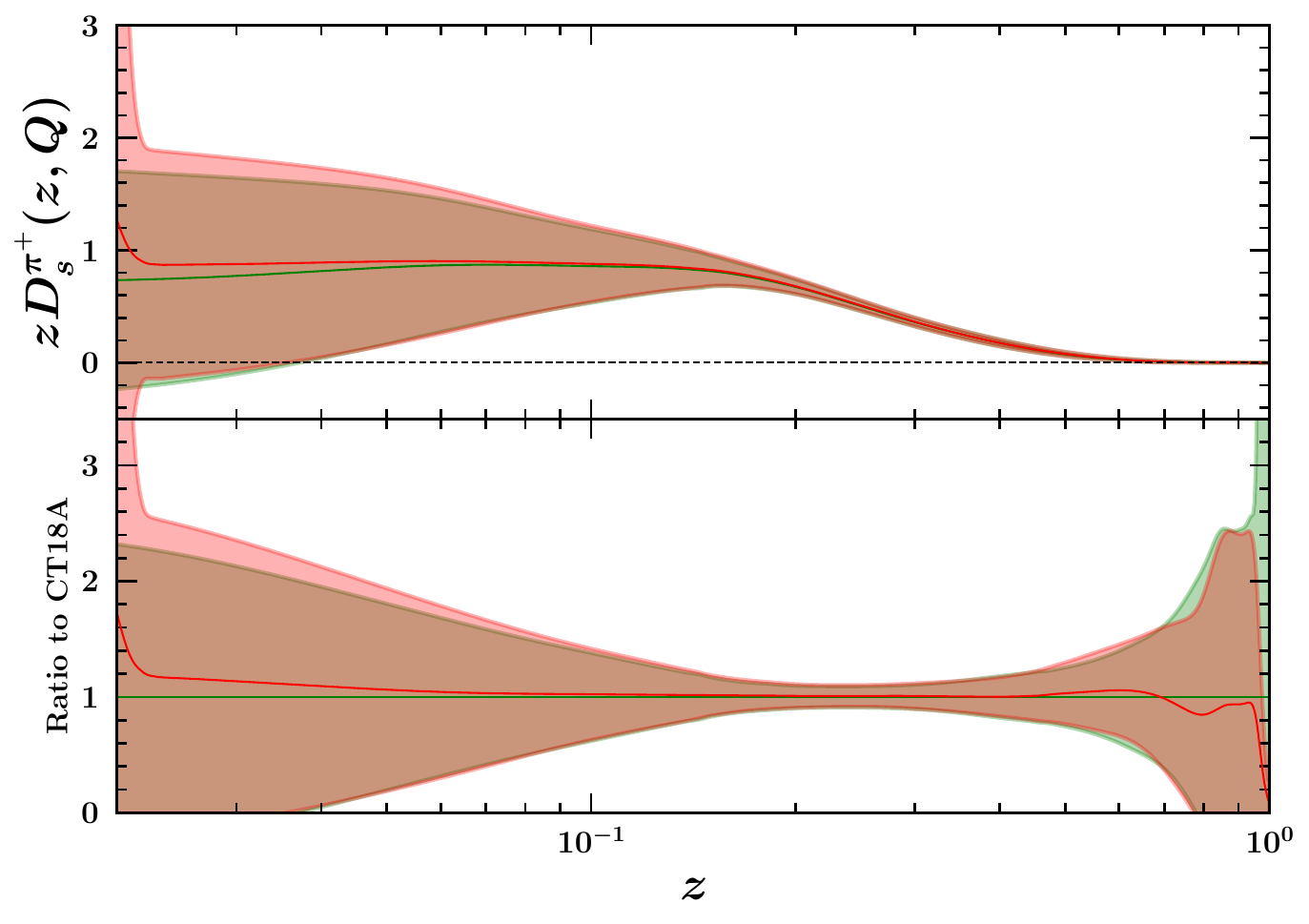}} 	
	\resizebox{0.45\textwidth}{!}{\includegraphics{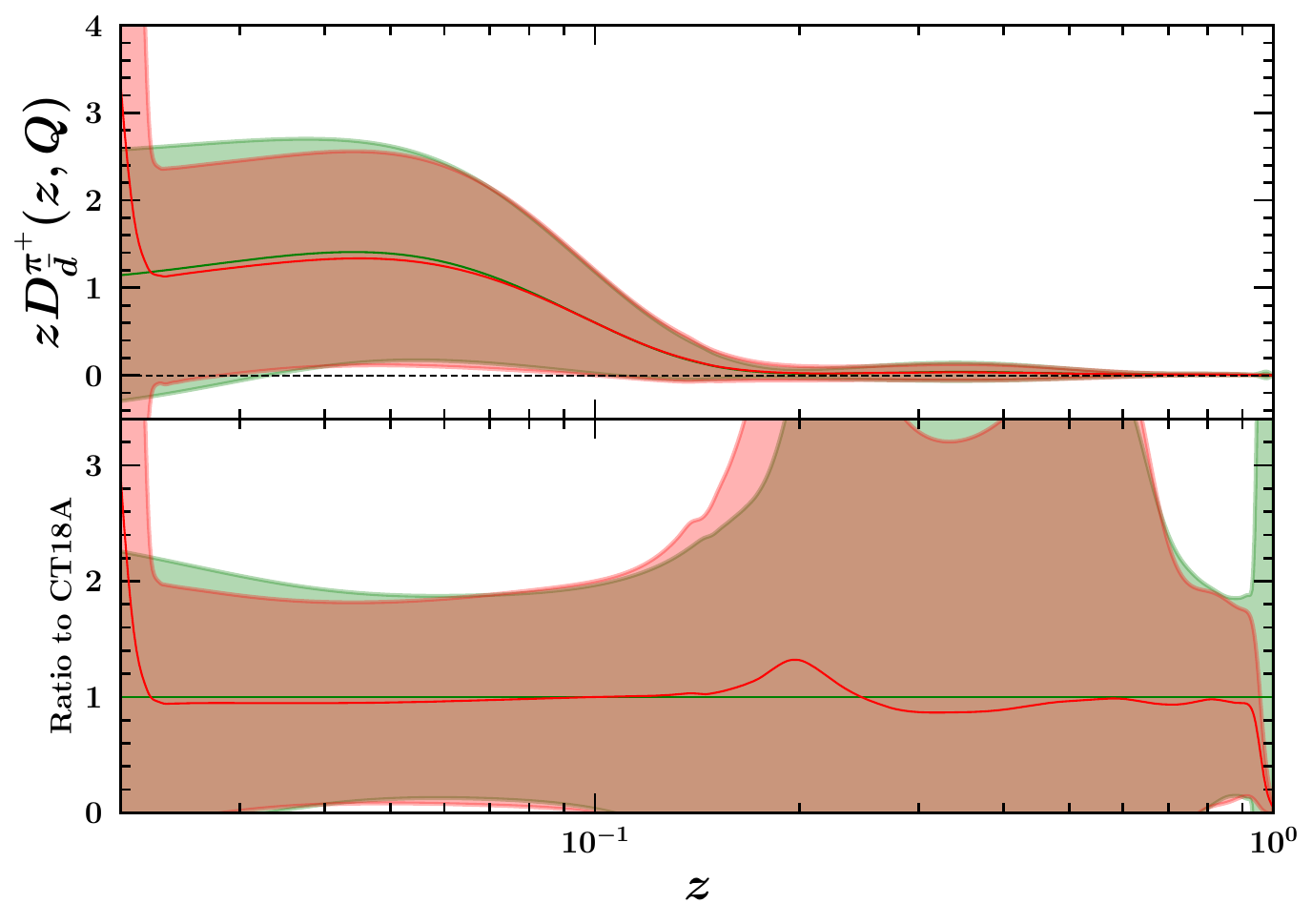}}
	\resizebox{0.45\textwidth}{!}{\includegraphics{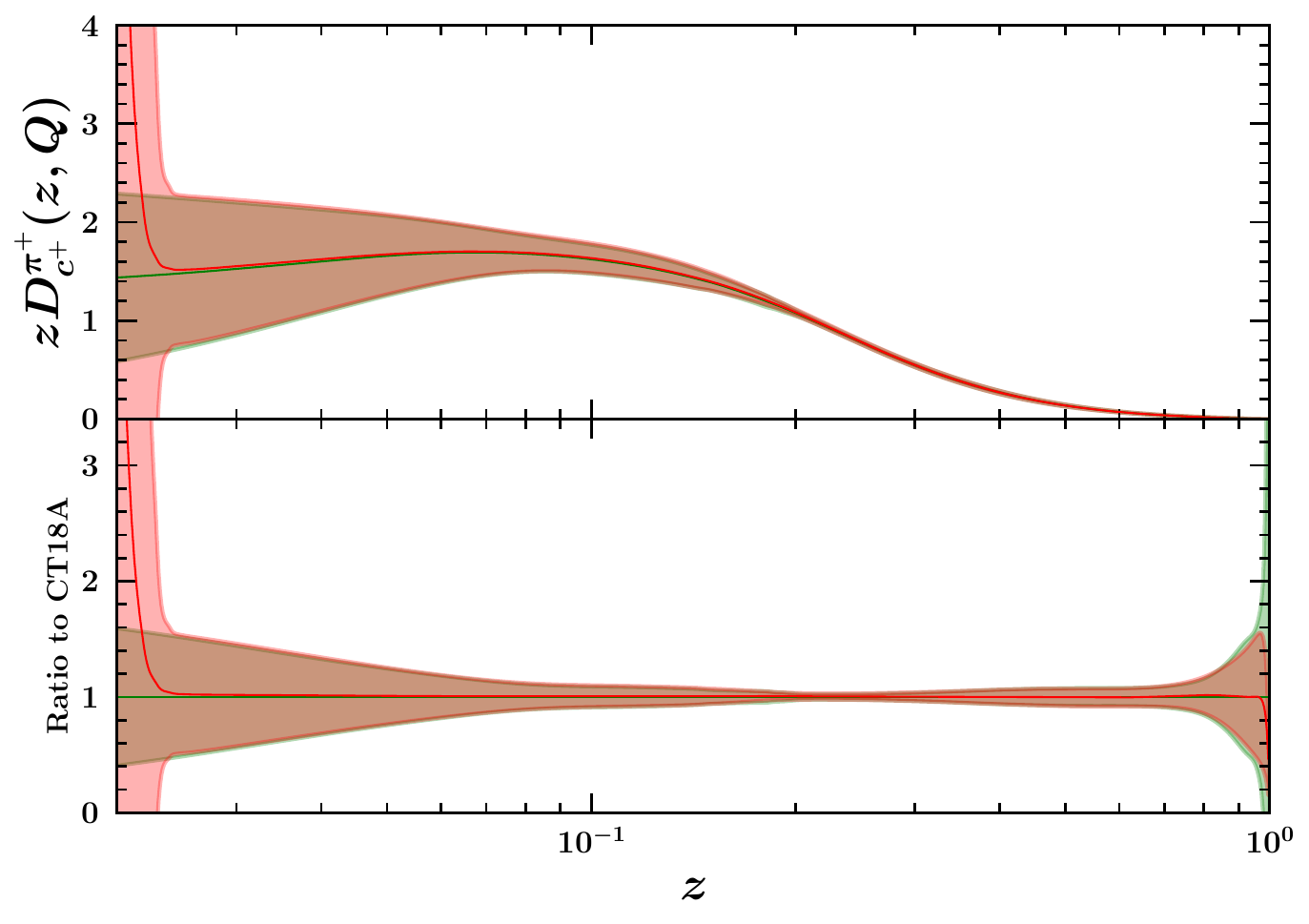}} 
	\resizebox{0.45\textwidth}{!}{\includegraphics{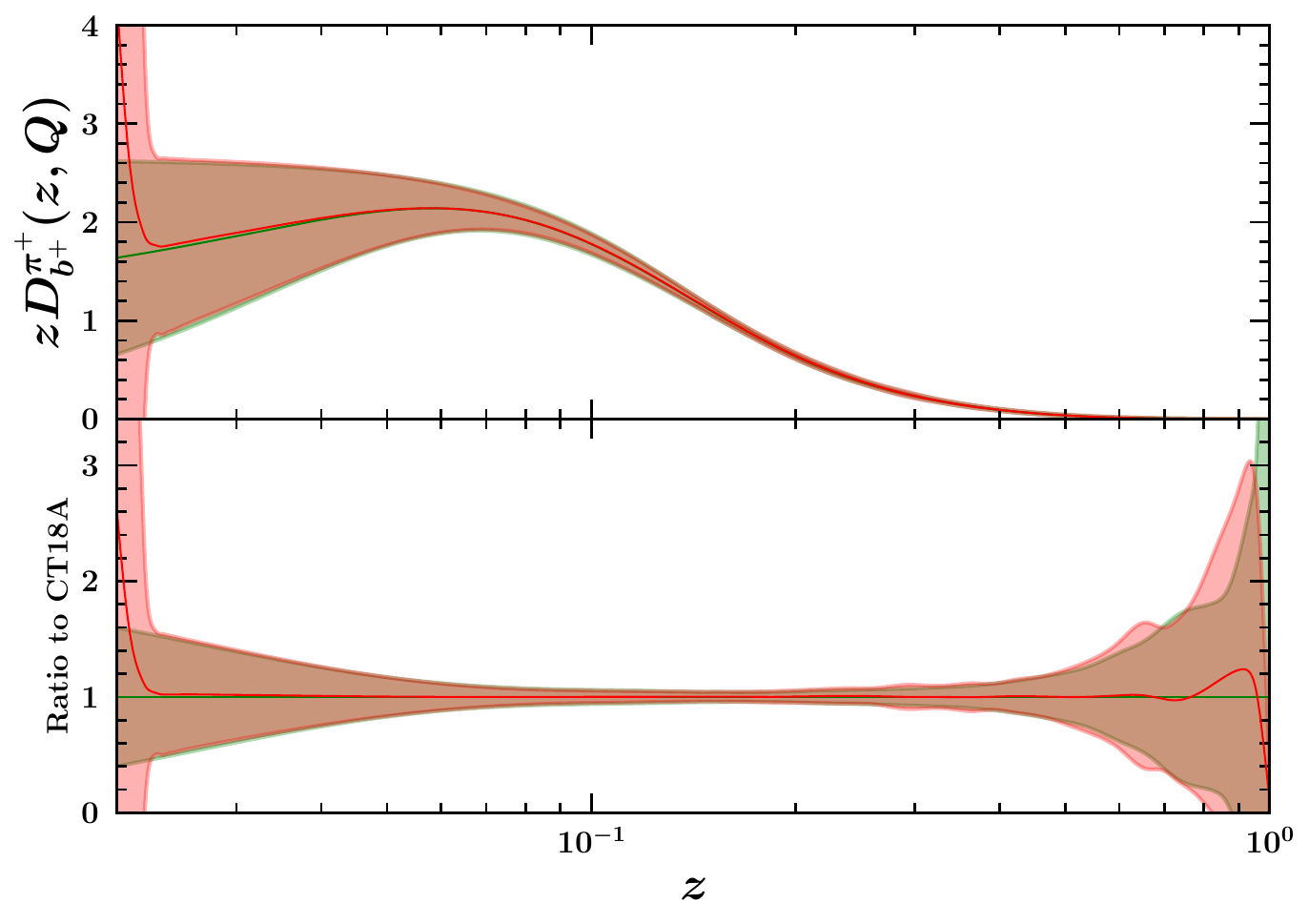}}   	 	
	\begin{center}
		\caption{ 
			\small 
Comparison of pion FFs extracted from {\tt CT18A} proton PDF sets as our baseline, 
			and the nuclear PDF set from {\tt EPPS21}. 
			We present both the absolute values, and the ratio to the 
			{\tt CT18A} proton PDFs baseline as well.
			The results presented at  $Q=5$ GeV.}
		\label{fig:pion-FFs_NLO_EPPS}
	\end{center}
\end{figure*}

\begin{figure*}[htb]
	\vspace{0.50cm}
	\resizebox{0.45\textwidth}{!}{\includegraphics{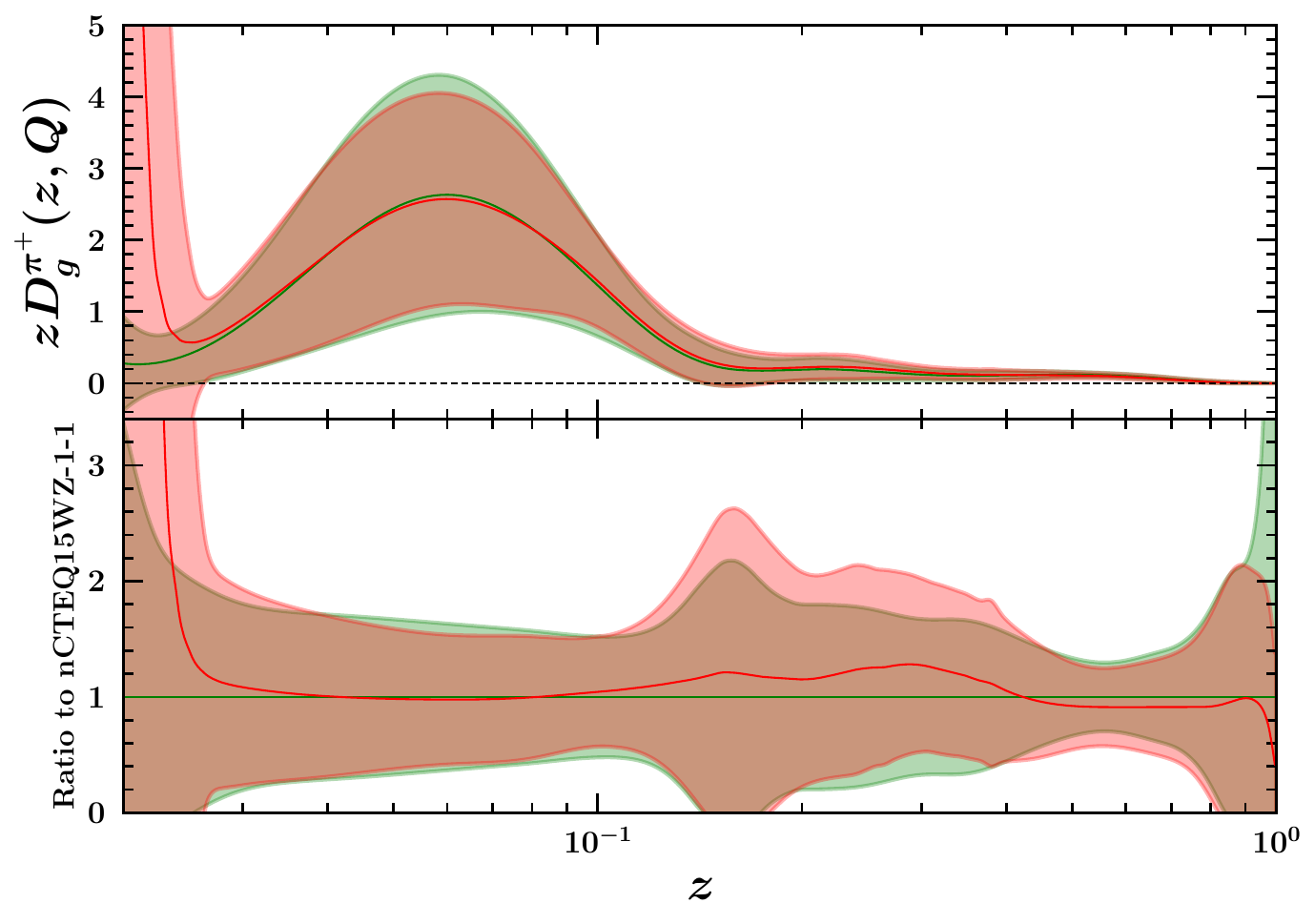}} 	
	\resizebox{0.45\textwidth}{!}{\includegraphics{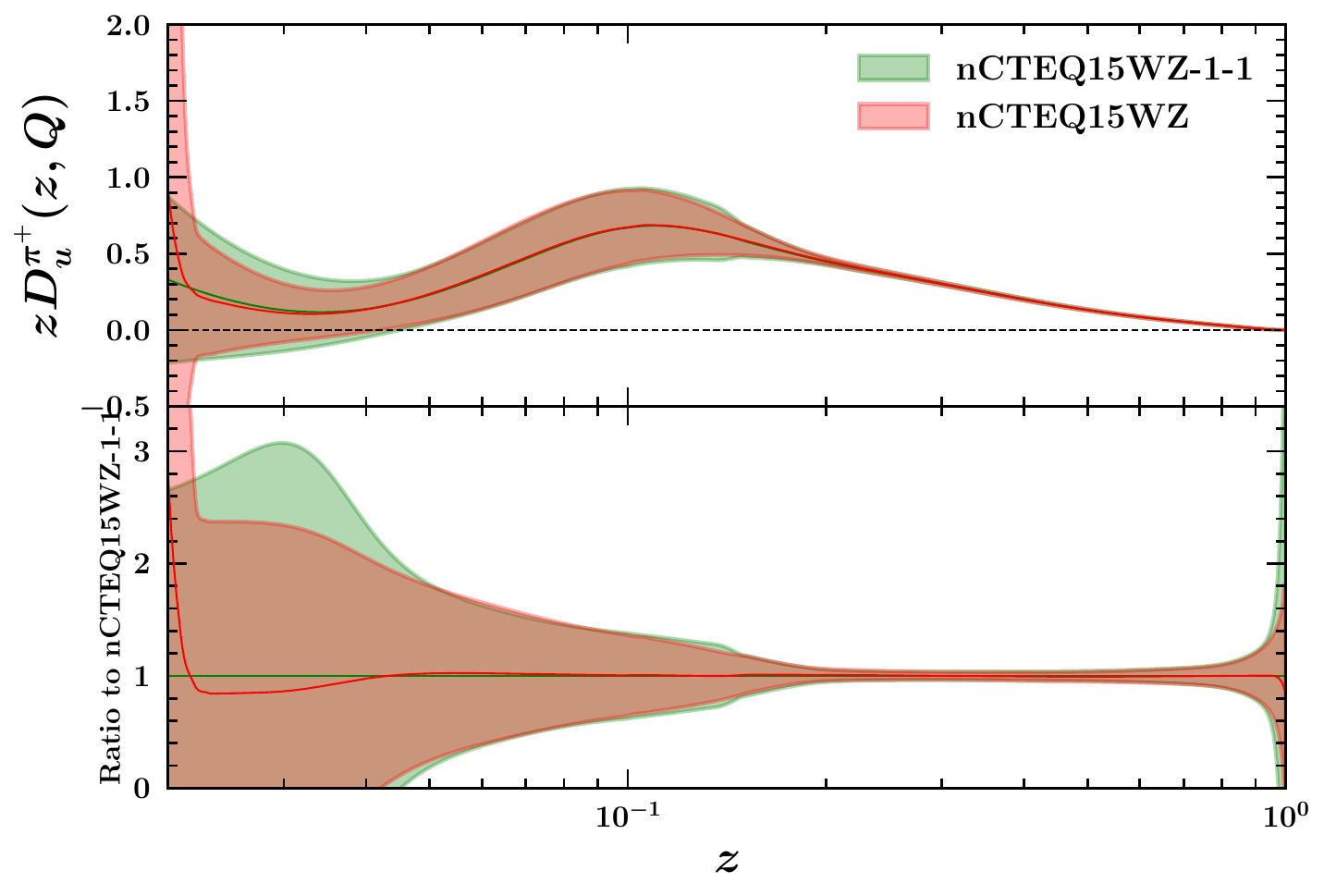}}   
	\resizebox{0.45\textwidth}{!}{\includegraphics{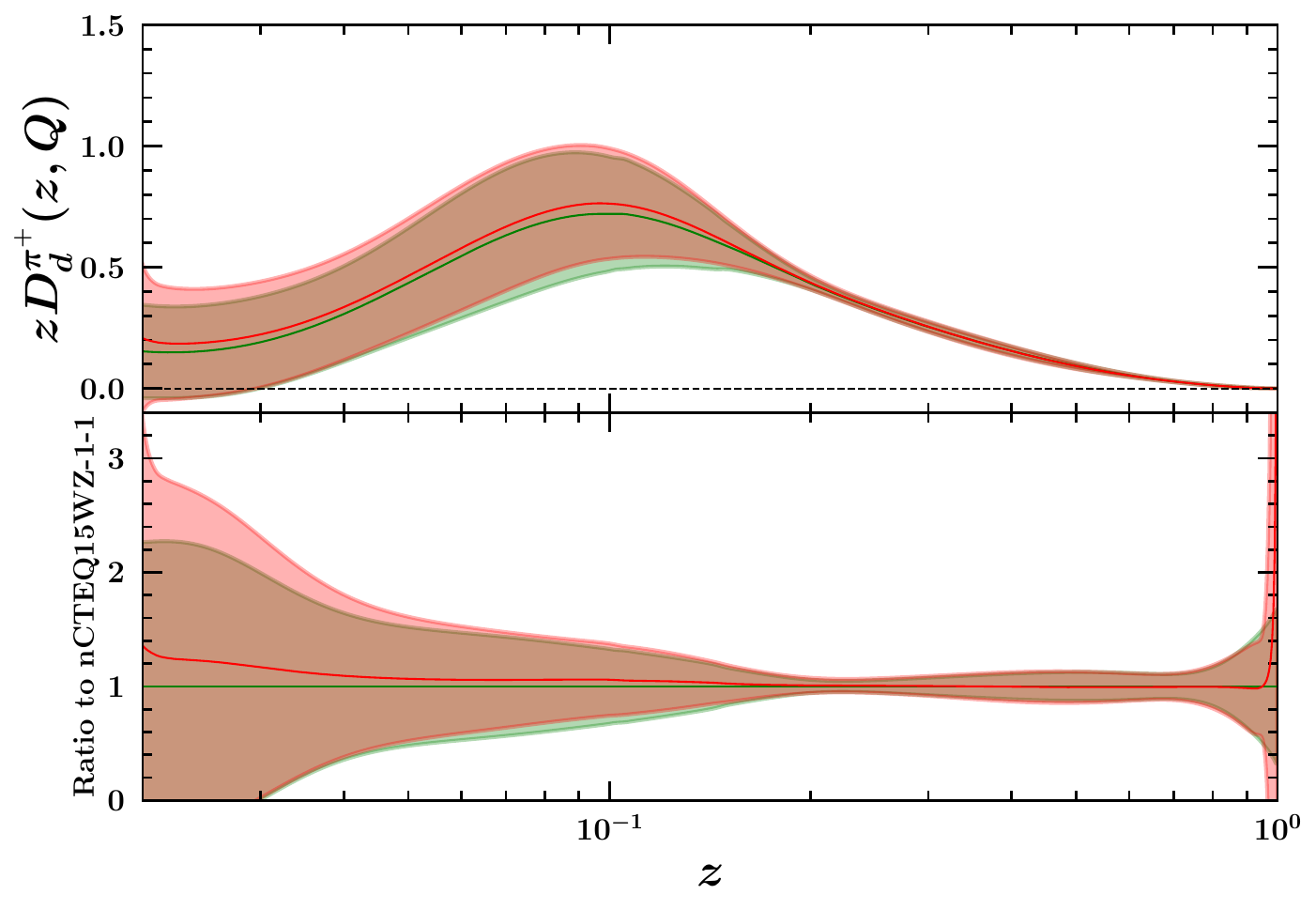}}	
	\resizebox{0.45\textwidth}{!}{\includegraphics{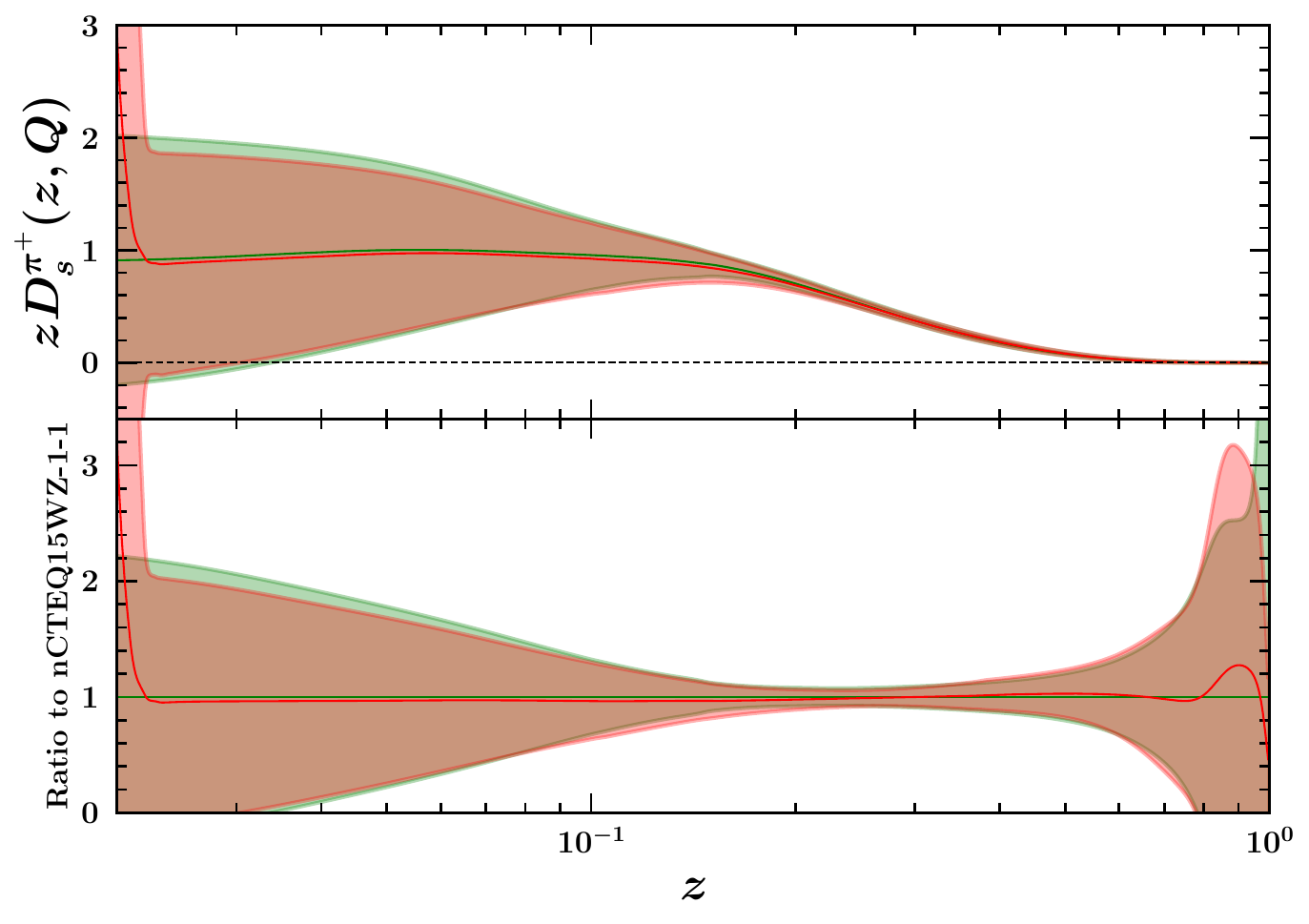}} 	
	\resizebox{0.45\textwidth}{!}{\includegraphics{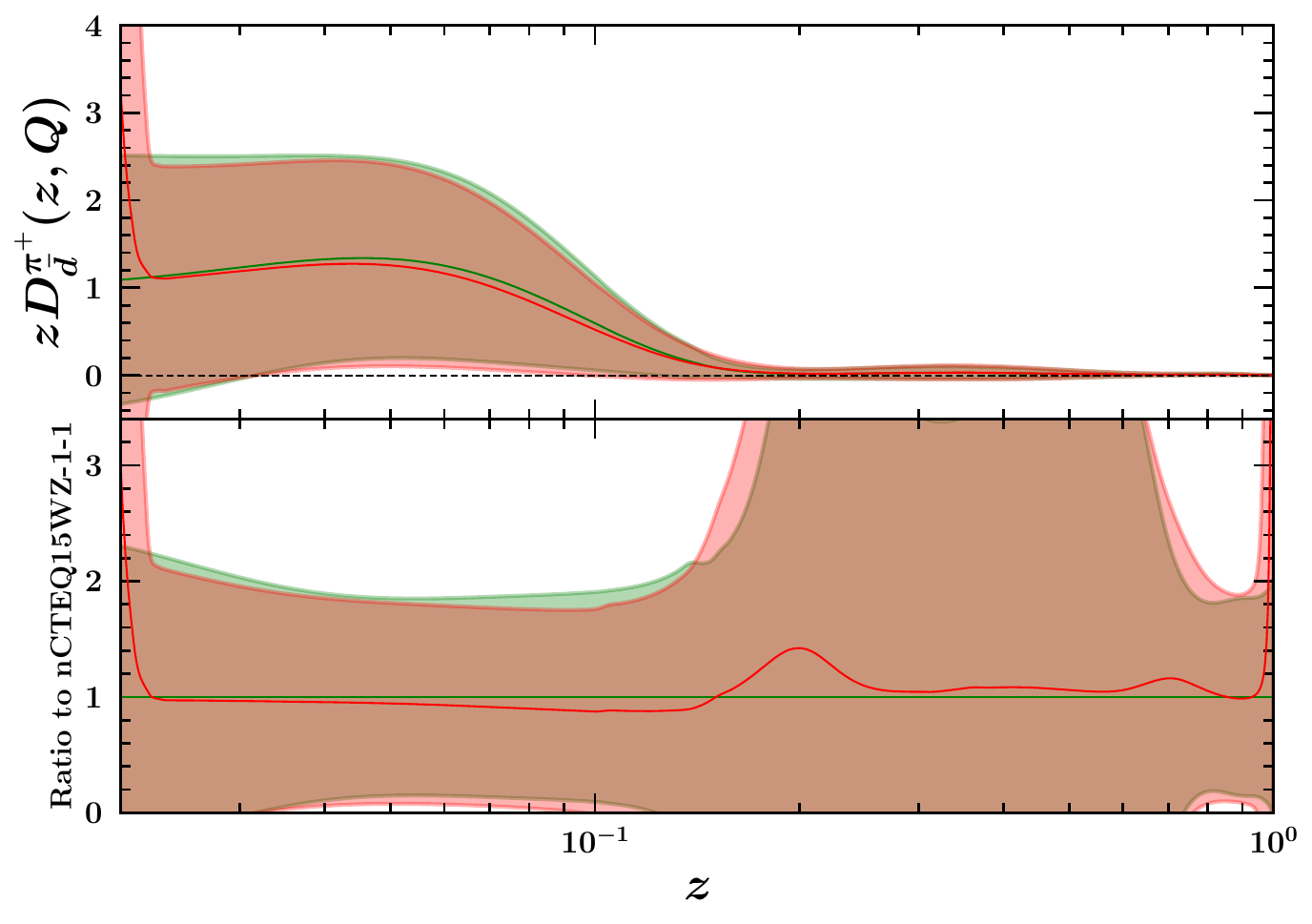}}
	\resizebox{0.45\textwidth}{!}{\includegraphics{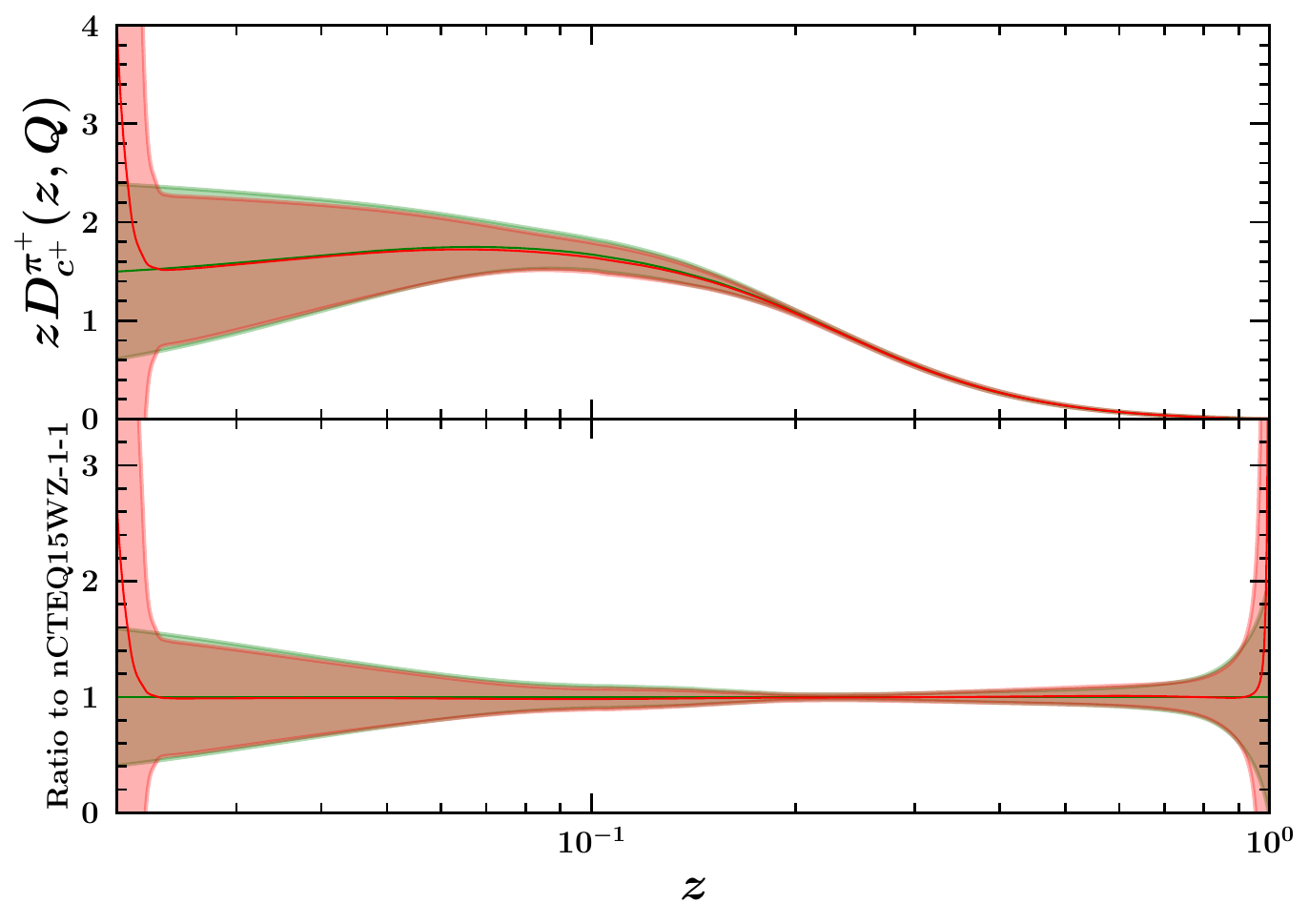}} 
	\resizebox{0.45\textwidth}{!}{\includegraphics{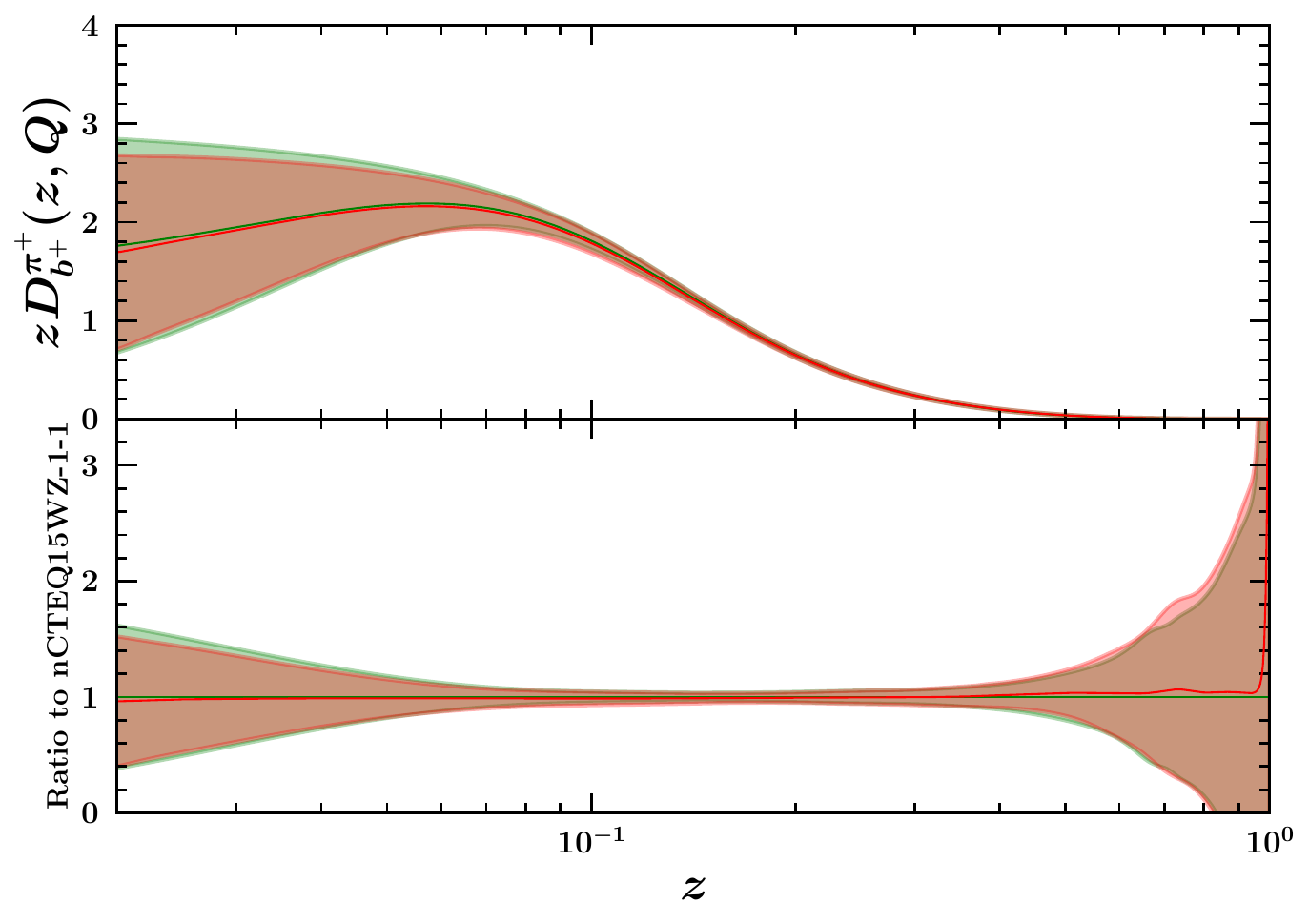}}   	 	
	\begin{center}
		\caption{ 
			\small 
 Comparison of pion FFs extracted from {\tt nCTEQ15WZ-1-1} 
	proton PDF sets as our baseline, 
			and the nuclear PDF set from {\tt nCTEQ15WZ}. 
			We present both the absolute values, and the ratio to the 
			{\tt nCTEQ15WZ-1-1} proton PDFs baseline as well.
			The results presented at  $Q=5$ GeV.} 
		\label{fig:pion-FFs_NLO_nCTEQ}
	\end{center}
\end{figure*}

In Figs.~\ref{fig:pion-FFs_NLO_nNNPDF}, \ref{fig:pion-FFs_NLO_EPPS}, 
and \ref{fig:pion-FFs_NLO_nCTEQ}, we present the QCD fits in 
the presence of the {\tt nNNPDF3.0-p}, {\tt CT18A}, and {\tt nCTEQ15WZ-1-1} 
proton PDFs as our baseline, along with corresponding three 
different sets of nuclear PDF sets.
Considering the results presented in Figs.~\ref{fig:pion-FFs_NLO_nNNPDF}, 
\ref{fig:pion-FFs_NLO_EPPS} and \ref{fig:pion-FFs_NLO_nCTEQ}, several comments are in order.
As one can see, including the nuclear PDFs not only changes the shape of the central value of pion FFs 
but also affects the error bands. However, these changes are rather small.

In terms of the individual flavor components of pion FFs the following conclusions can be made for 
the {\tt nNNPDF3.0}, {\tt EPPS21}, and {\tt nCTEQ15WZ} analyzed in this study.
Starting from the {\tt nNNPDF3.0} presented in Fig.~\ref{fig:pion-FFs_NLO_nNNPDF}, 
one can observe that the biggest changes in the central values 
of the pion FFs occur for the $\bar{d}$ FFs in the region $0.1 \leq z \leq 1$. 
Additionally, some fluctuations are present in gluon distribution, more pronounced changes can be seen for the gluon component 
in high and low $z$ regions.
One can also observe slight changes in the $s$ and $u$ FFs, 
specifically in the small region of $z$, when 
considering the nuclear corrections.
In the case of $c^+$ and $b^+$ FFs, the nuclear corrections 
primarily affect the large $z$ region. 
For the $d$, $\bar{d}$ and $u$ 
components of pion FFs, improvements in the error bands 
from medium to large values of $z$ can be 
observed by applying the nuclear correction from the {\tt nNNPDF3.0} PDF set. 
In the case of
$u$-quark FF, the reduction in the error bands is 
primarily observed in the small values of $z$.

For the case of {\tt EPPS21} presented in Fig.~\ref{fig:pion-FFs_NLO_EPPS}, 
changes can be observed in both the central values and error bands of 
the pion FFs, particularly in the medium to small values of $z$. 
We can discern minor fluctuations in the gluon distribution across the entire $z$ range. 
As for the $u$, $d$, and $s$ distributions, the variations become noticeable in 
the transition from medium to small values of $z$. However, the most substantial 
alterations in the central value of the $\bar{d}$ distribution are evident in the region 
extending from medium to large values of $z$.
Indeed, while for certain cases like the $\bar{d}$ FFs, the inclusion 
of nuclear corrections can lead to a reduction in the uncertainty bands, 
it may not necessarily improve the bands for other cases such 
as the $u$ and $s$ FFs.

In the context of nCTEQ15WZ, as depicted in Fig.~\ref{fig:pion-FFs_NLO_nCTEQ}, 
the most substantial alterations in both the central values and the uncertainty bands 
are notable within the gluon component across the entire $z$ range. These changes 
manifest as both increases and decreases in the width of the uncertainty bands at various $z$ regions.
Furthermore, discernible modifications and improvements in the uncertainty bands are 
observable for the $d$, $u$, and $s$ FFs, particularly within the realm of medium and 
small $z$ values. In contrast, for the remaining extracted FFs, both the central values
and the corresponding error bands remain largely consistent and almost identical across the entire $z$ range.

The results obtained for the {\tt nNNPDF3.0}, {\tt EPPS21}, and {\tt nCTEQ15WZ} 
nuclear PDF sets, as discussed so far, are consistent with the 
total $\chi^2$ values reported in Table.~\ref{tab:datasets-pion_NLO}. 
This indicates that the inclusion of these nuclear PDF sets in the pion FFs 
analysis is in agreement with the overall 
goodness-of-fit measures obtained from the experimental data.

Overall, in the case of pion, the largest changes in 
the central values and uncertainty bands 
can be observed for the gluon, $u$ and $\bar{d}$ components, 
 showing slight effects from the inclusion of nuclear PDFs.

\clearpage

%
\subsection{Nuclear effects of the  kaon FFs}\label{sec:kaon_result}
%

In this section, our main focus is on presenting the results for 
the kaon FFs considering the effects of nuclear corrections. 
We aim to analyze the impact of nuclear PDFs on the determination 
of kaon FFs and explore any changes or 
improvements observed in the presence of nuclear effects.

In Table.~\ref{tab:datasets_kaon_NLO}, we present 
the individual $\chi^2/{N_{\rm dat}}$ values 
obtained from the NLO analysis of the kaon FF using the nuclear PDF 
sets from {\tt nNNPDF3.0}, {\tt EPPS21}, and {\tt nCTEQ15WZ}, along with 
their respective proton PDF baselines.
The total $\chi^2/{N_{\rm dat}}$ values are displayed at 
the bottom of this table, as well.
Based on the numbers presented in the table
for the kaon FFs, several comments can be made.

An interesting observation in our kaon FFs fit is the decrease in the total $\chi^2/{N_{\rm dat}}$ 
value when incorporating the nuclear PDFs from {\tt nCTEQ15WZ}.
When utilizing the {\tt nCTEQ15WZ-1-1} proton PDFs, the $\chi^2/{N_{\rm dat}}$ 
value is 0.757. However, by incorporating the {\tt nCTEQ15WZ}  
nuclear PDFs, the $\chi^2/{N_{\rm dat}}$ value decreases 
to 0.704. The reduction in the individual $\chi^2$ per data point 
is particularly notable for the {\tt COMPASS} $K^+$ and {\tt COMPASS} $K^-$ data sets.

In the case of {\tt nNNPDF3.0-p}, the incorporation of this nuclear PDF set 
leads to a slight increase in the $\chi^2/{N_{\rm dat}}$, rising from 0.774 to 0.785. 
However, when examining the individual $\chi^2$ values per data point for 
the {\tt COMPASS} $K^+$ dataset, there is a reduction, while a minor increase is observed 
for the {\tt COMPASS} $K^-$ dataset.

For the {\tt EPPS21}, the inclusion of this nuclear PDF set results in a notable rise in 
the $\chi^2/{N_{\rm dat}}$, increasing it from 0.691 to 0.749. 
The overall upturn in 
the $\chi^2/{N_{\rm dat}}$ for {\tt EPPS21} is primarily 
affected by the {\tt COMPASS} $K^+$ 
and {\tt COMPASS} $K^-$ datasets, which contribute to the elevated 
individual $\chi^2$ values per data point.

\begin{table*}[htb]
	\renewcommand{\arraystretch}{2}
	\centering 	\scriptsize
	\begin{tabular}{|l|c|cc|cc|cr|}				\hline
	
	        ~        &  ~ & \multicolumn{6}{|c|}{$\frac{\chi^2}{N_{\rm dat}}$:}
	        \\
	        
		      Experiment & $N_{\rm dat}$&~  {\tt nNNPDF3.0-p}    ~&~   {\tt nNNPDF3.0} ~&~  {\tt CT18A} ~&~ {\tt EPPS21} ~&~ {\tt nCTEQ15WZ-1-1} ~&~ {\tt nCTEQ15WZ} 
		\rule[-3mm]{0mm}{5mm}
		\\
		\hline \hline
		{\tt COMPASS} $K^+$~\cite{COMPASS:2016crr}   & 156  & 0.685& 0.674&0.621&0.632&0.654& 0.632 \\
		{\tt COMPASS} $K^-$~\cite{COMPASS:2016crr} &   156 & 0.580 & 0.582&0.587&0.618&0.671&0.596 \\
		{\tt BELLE}~\cite{Belle:2013lfg}   & 70 & 0.540& 0.539&0.529&0.532&0.530& 0.525 \\
		{\tt BABAR}~\cite{BaBar:2013yrg} &   28 & 1.093 & 1.131&0.832&0.965&0.918&0.838 \\
		{\tt TASSO12}~\cite{TASSO:1980dyh}  & 3 & 0.781 &0.779&0.790&0.785&0.787&0.791 \\
		{\tt TASSO14}~\cite{TASSO:1982bkc}  & 9 & 1.407& 1.422 &1.344&1.386&1.378&1.344\\       					
		{\tt TASSO22}~\cite{TASSO:1982bkc}   & 6 & 0.652& 0.636&0.707&0.665&0.673&0.712\\
		{\tt TPC}~\cite{Aihara:1988su}  & 13 & 0.712 &0.730&0.641&0.686&0.664& 0.639\\
		{\tt TASSO34}~\cite{TASSO:1988jma} & 5 & 0.037&  0.035&0.040&0.037&0.038&0.041\\
		{\tt TOPAZ}~\cite{TOPAZ:1994voc}& 3 & 0.147 &0.147&0.142&0.146 &0.138&0.143\\
		{\tt ALEPH}~\cite{ALEPH:1994cbg}  & 18 & 0.464 &0.470&0.479&0.475&0.467&0.474\\
		{\tt DELPHI} (incl.)~\cite{Abreu:1998vq}   & 23 & 0.637 &0.639&0.655&0.652&0.629&0.650\\
		{\tt DELPHI} ($uds$ tag)~\cite{Abreu:1998vq}  & 23 & 0.260 &0.260&0.280&0.276&0.281&0.293 \\
		{\tt DELPHI} ($b$ tag)~\cite{Abreu:1998vq}  &23 & 0.866  & 0.918 &0.594& 0.770&0.719&0.610\\
		{\tt OPAL} (incl.)~\cite{OPAL:1994zan} & 10& 0.485& 0.496&0.473&0.487&0.487&0.474\\                                           				         
		{\tt SLD} (incl.)~\cite{Abe:2003iy}   & 35 & 1.753&1.908&1.296&1.683&1.579&1.360\\
		{\tt SLD} ($uds$ tag)~\cite{Abe:2003iy}   &35 & 1.882 & 1.877&1.738& 1.784&1.732&1.674 \\                                             				
		{\tt SLD} ($b$ tag)~\cite{Abe:2003iy} &35 & 2.630 &2.768&1.971&2.429&2.289&2.043\\ 				
		\hline \hline
		Total $\chi^2/{N_{\rm dat}}$ & 651 &0.774 &0.785&0.691&0.749& 0.757& 0.704  \\
		\hline \hline	
	\end{tabular}
	\caption{ \small 
 		The list of input data sets for the kaon production included in our kaon FFs analysis. 
		For each data set, we have indicated 
		the experiments, corresponding published reference and the number of data points. 
		In the last six columns, we show the value of $\chi^2/{N_{\rm dat}}$ resulting from the 
		FF fit  by considering the proton PDF sets from {\tt nNNPDF3.0-p}~\cite{AbdulKhalek:2022fyi}, {\tt CT18A} ~\cite{Hou:2019efy}  and {\tt nCTEQ15WZ-1-1} \cite{Kusina:2020lyz} ,and 
		nuclear PDF sets available in the literature, namely the {\tt nNNPDF3.0}~\cite{AbdulKhalek:2022fyi}, 
		{\tt EPPS21}~\cite{Eskola:2021nhw}, and {\tt nCTEQ15WZ}~\cite{Kusina:2020lyz}. 
		The total value of $\chi^2/{N_{\rm dat}}$ also is shown at the bottom of this table.  } 
	\label{tab:datasets_kaon_NLO}
\end{table*}

\begin{figure*}[htb]
	\vspace{0.50cm}
	\resizebox{0.45\textwidth}{!}{\includegraphics{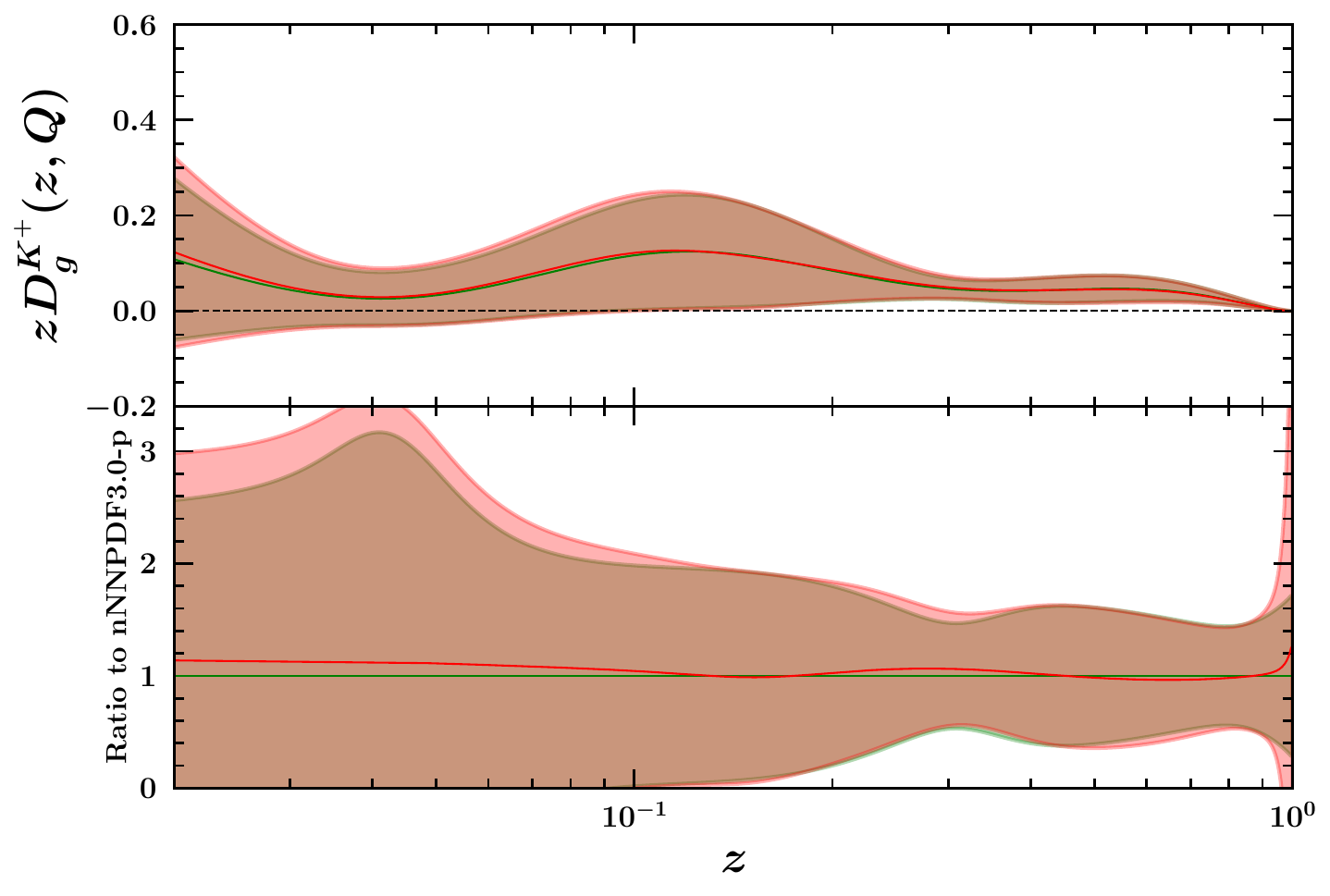}} 	
	\resizebox{0.45\textwidth}{!}{\includegraphics{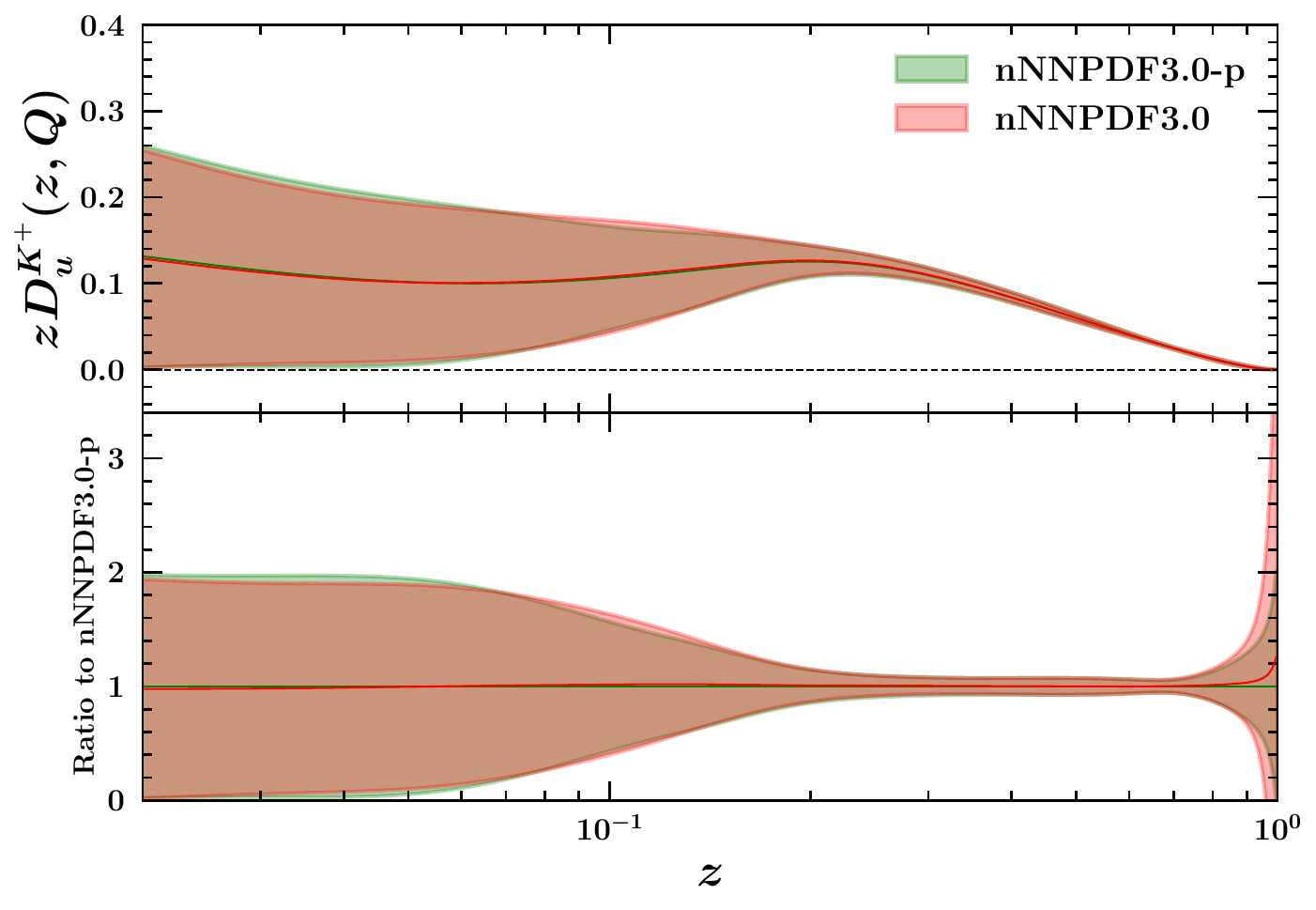}}  	
	\resizebox{0.45\textwidth}{!}{\includegraphics{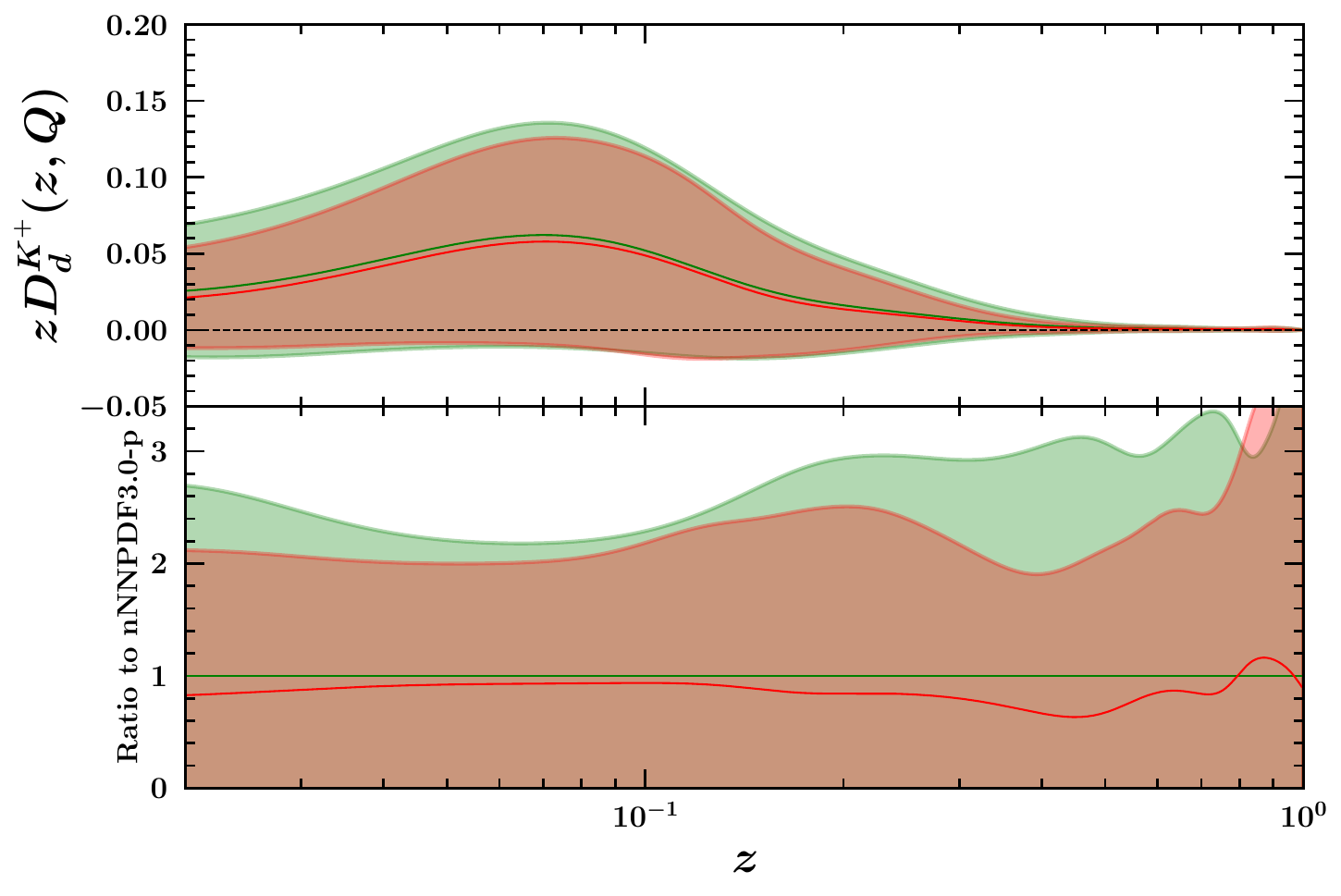}}
	\resizebox{0.45\textwidth}{!}{\includegraphics{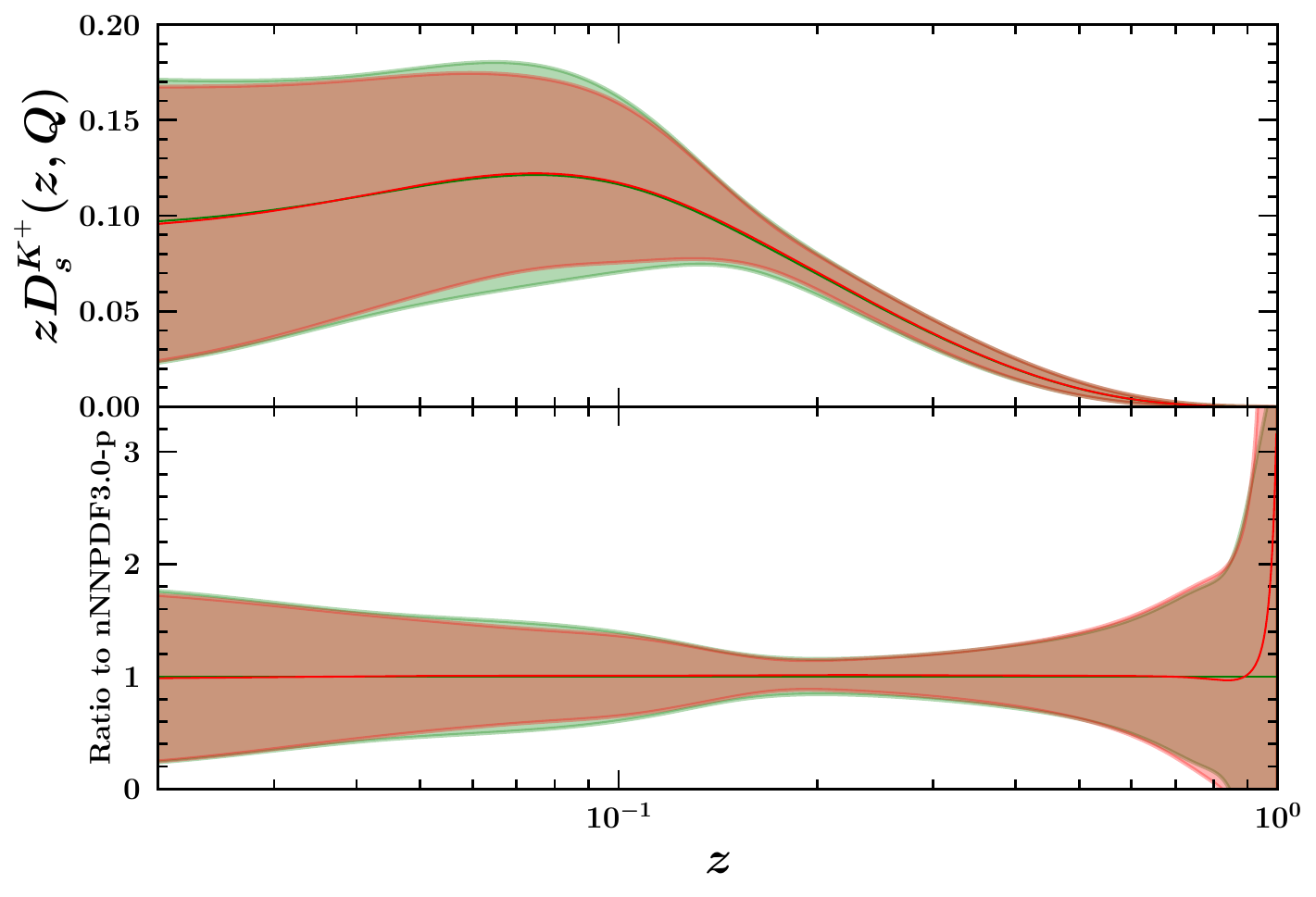}} 
	\resizebox{0.45\textwidth}{!}{\includegraphics{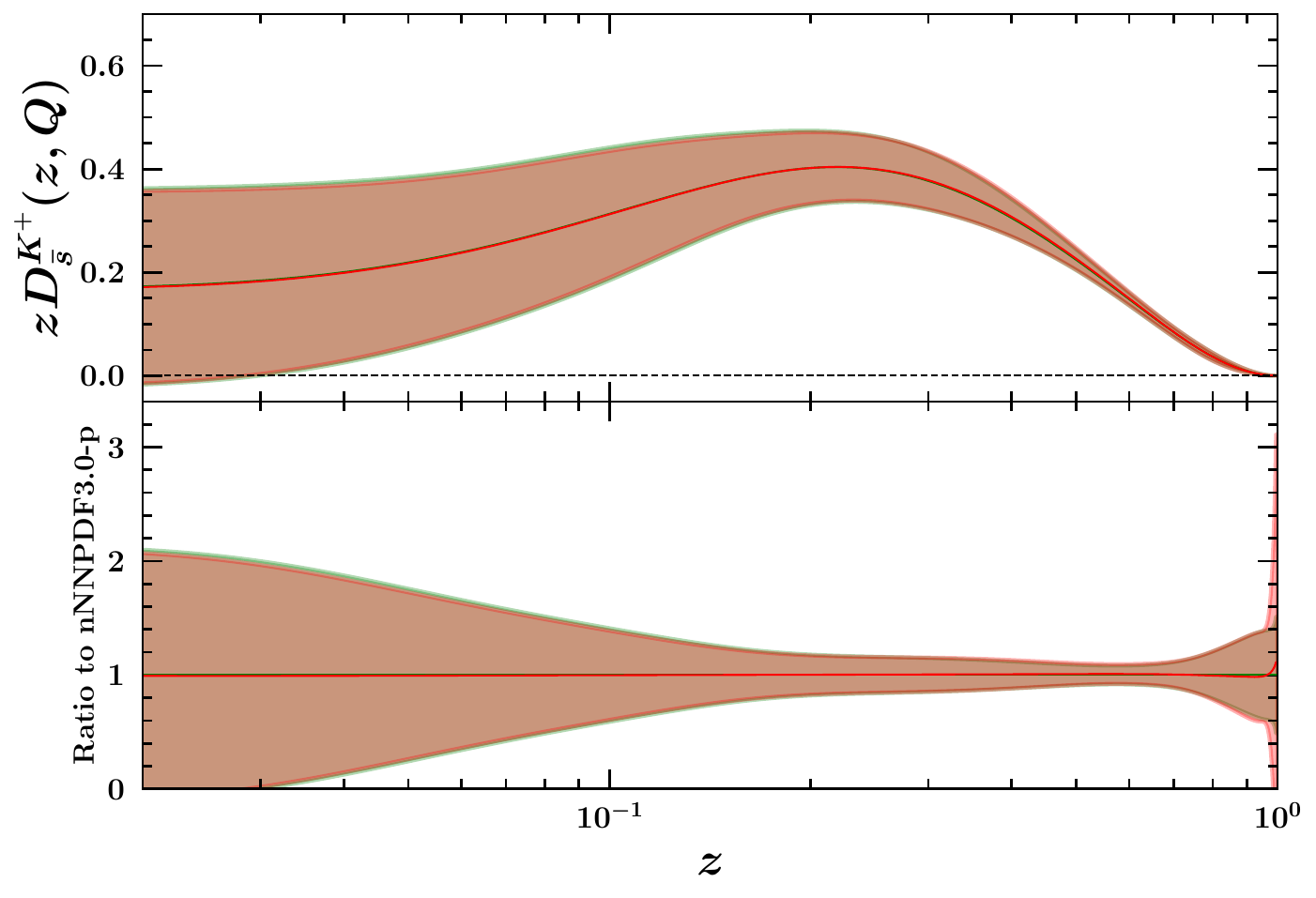}} 
	\resizebox{0.45\textwidth}{!}{\includegraphics{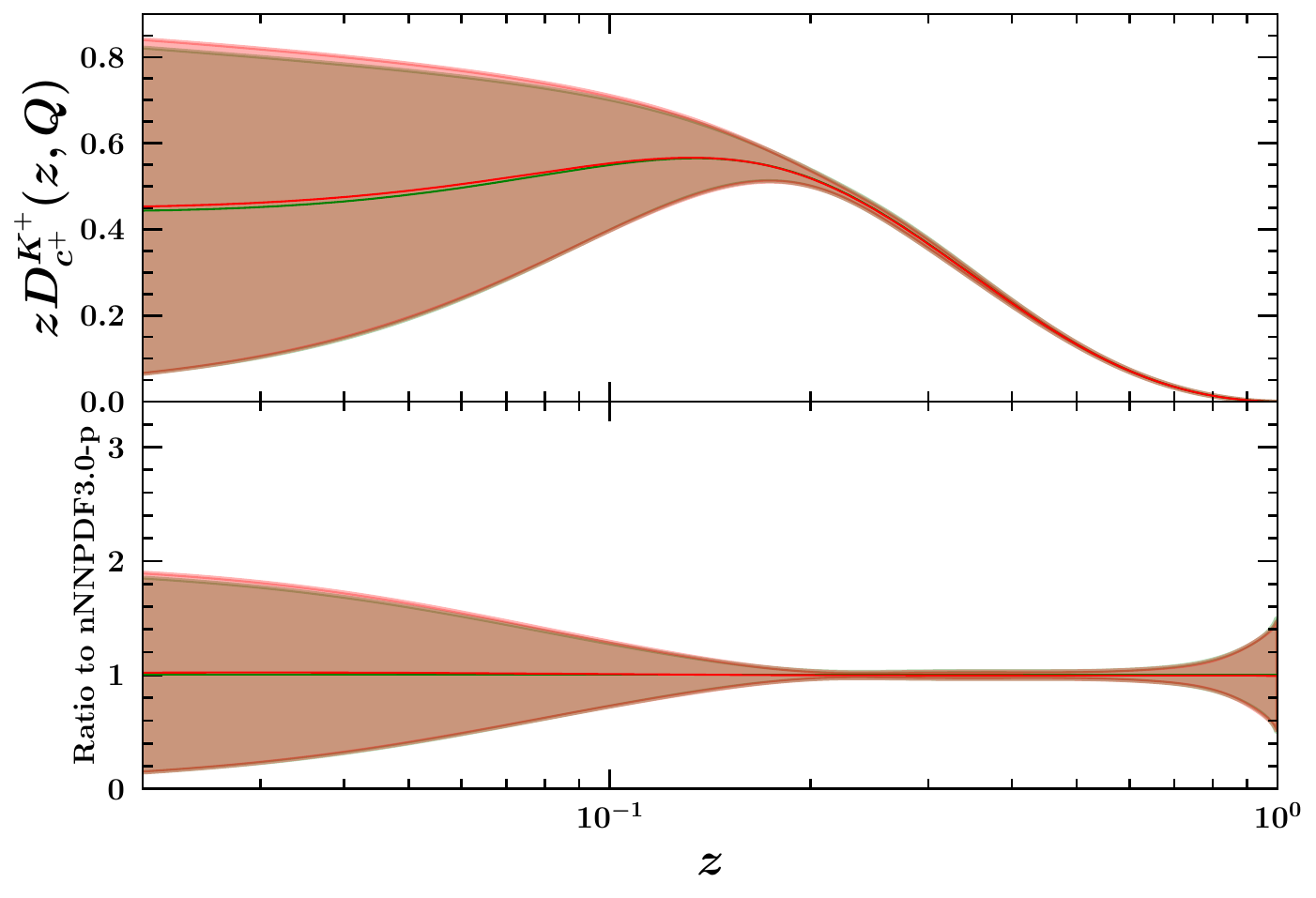}} 
	\resizebox{0.45\textwidth}{!}{\includegraphics{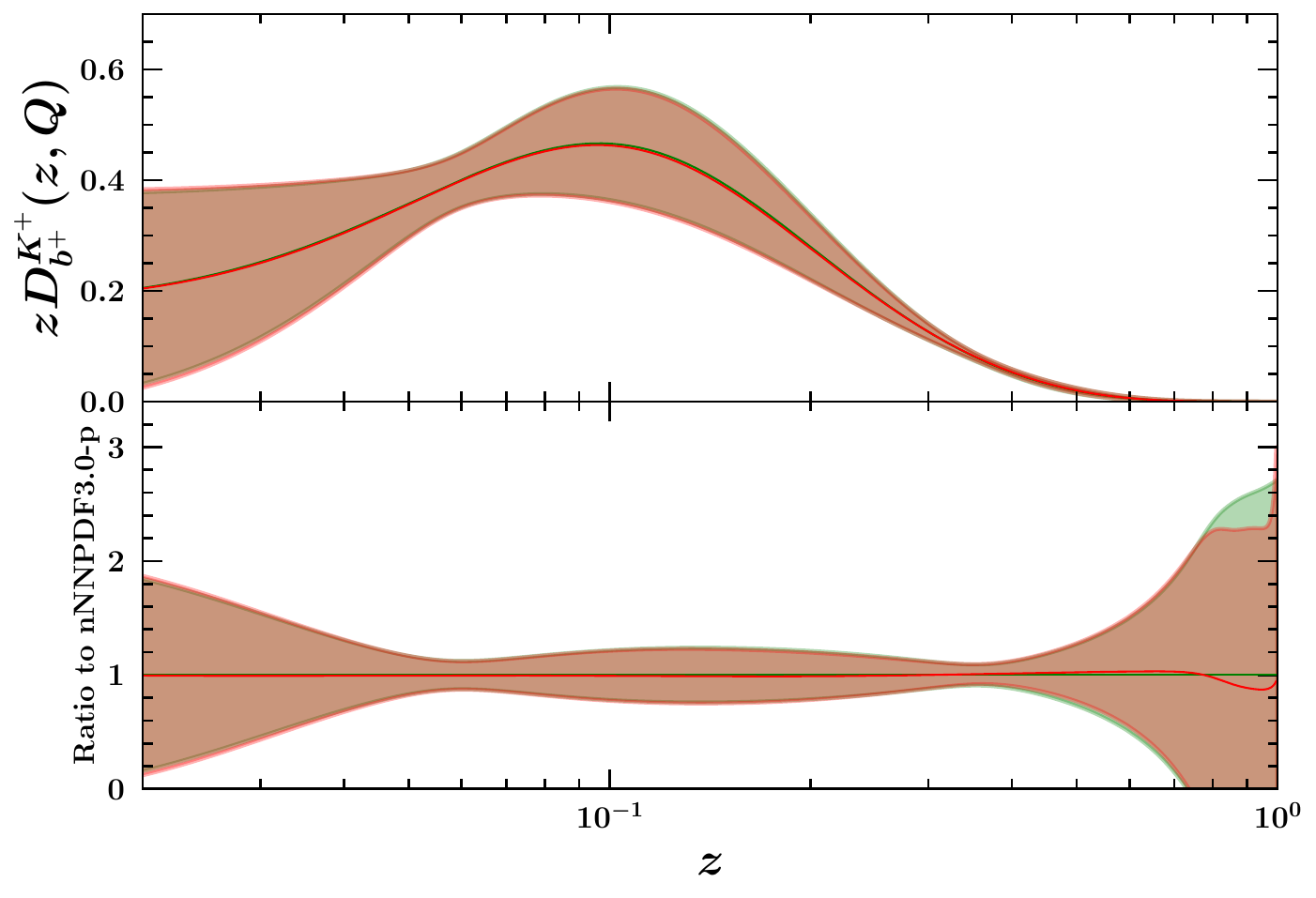}}   	 	
	\begin{center}
		\caption{ 
			\small 
	 		Comparison of kaon FFs extracted from {\tt nNNPDF3.0-p} proton PDF sets as our baseline, 
			and the nuclear PDF sets from {\tt nNNPDF3.0}. 
			We present both the absolute values, and the ratio to the 
			{\tt nNNPDF3.0-p} proton PDFs baseline as well.
			The results presented  at  $Q=5$ GeV.} 
		\label{fig:kaon-FFs_NLO_nNNPDF}
	\end{center}
\end{figure*}

\begin{figure*}[htb]
	\vspace{0.50cm}
	\resizebox{0.45\textwidth}{!}{\includegraphics{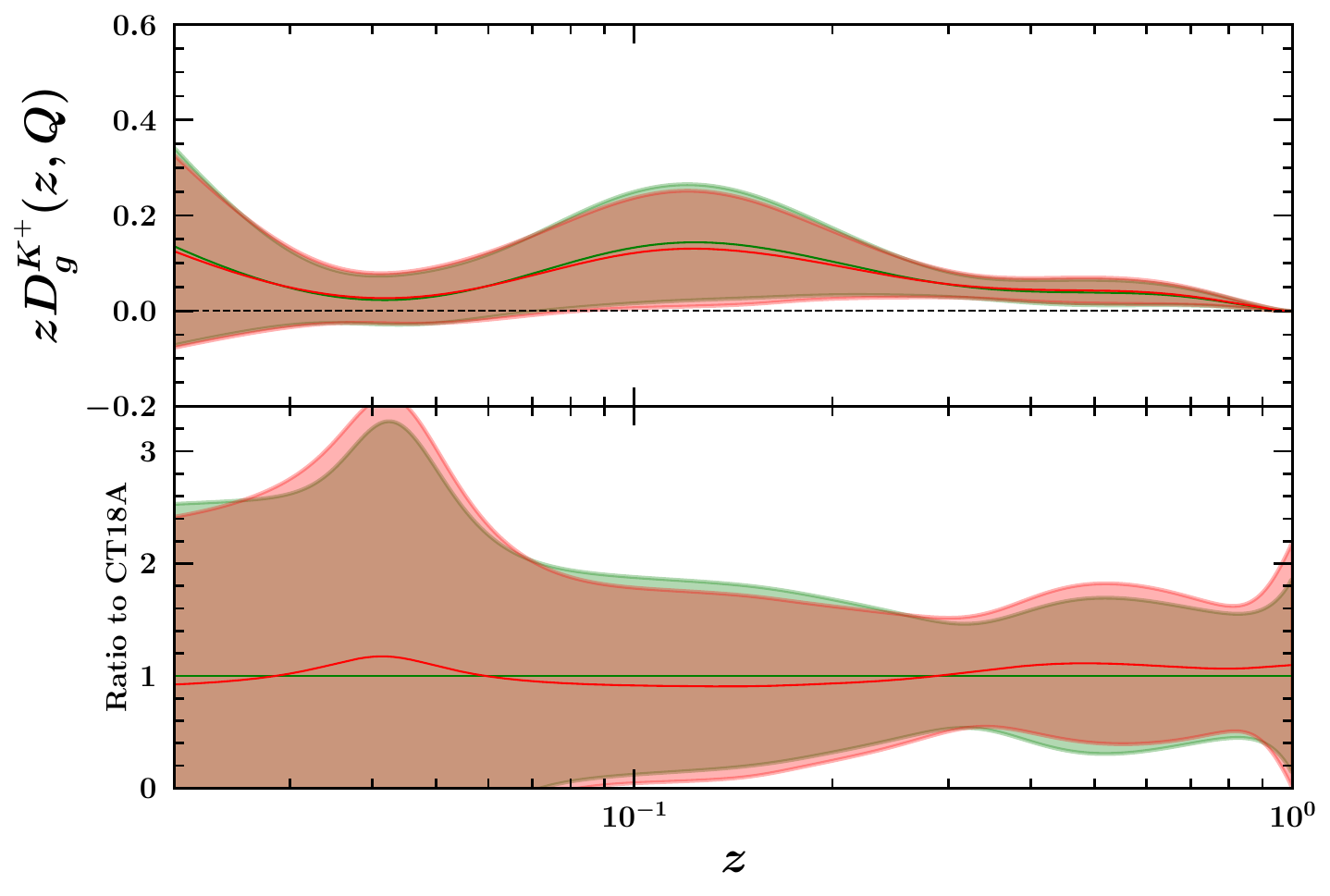}} 	
	\resizebox{0.45\textwidth}{!}{\includegraphics{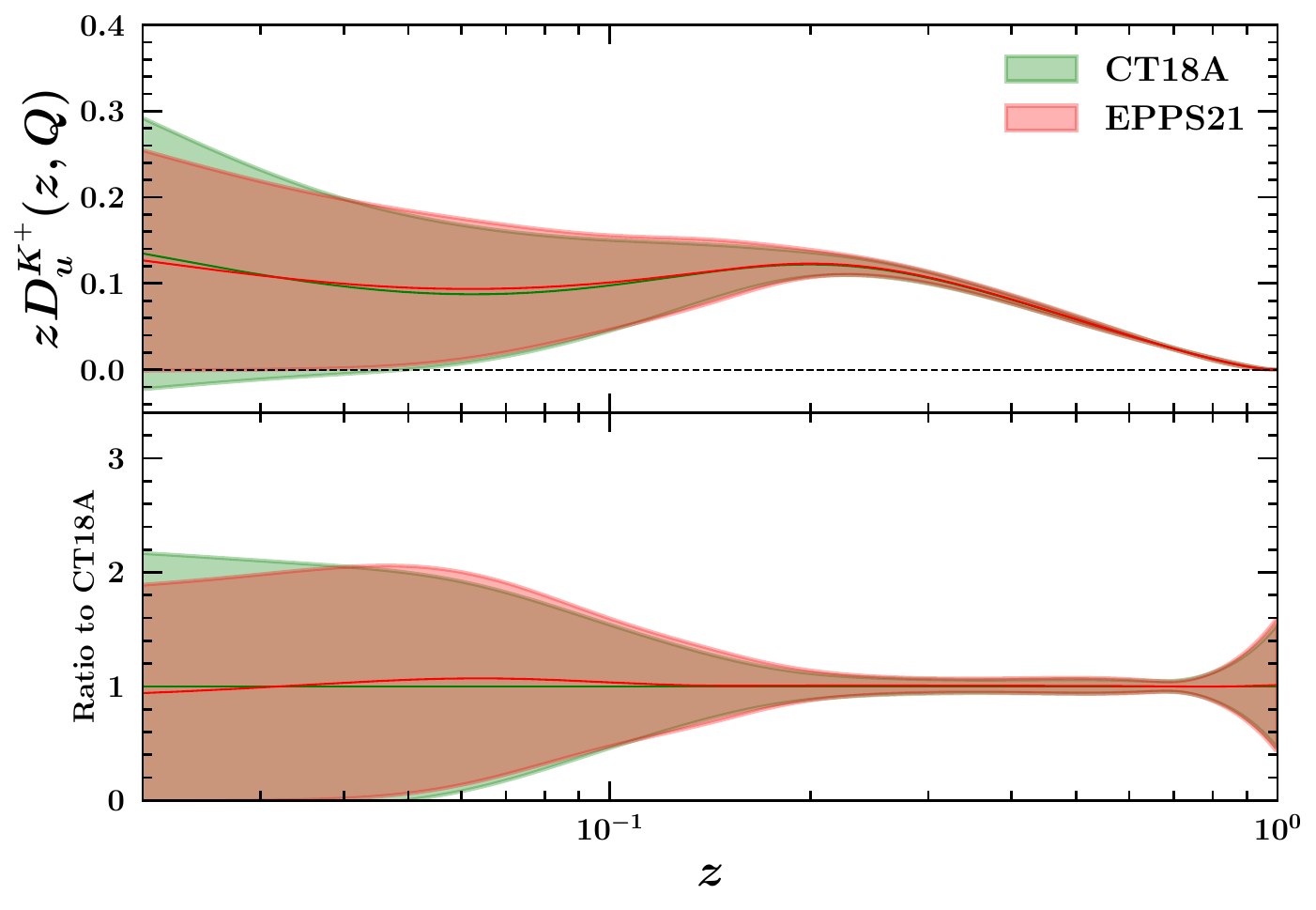}}  	
	\resizebox{0.45\textwidth}{!}{\includegraphics{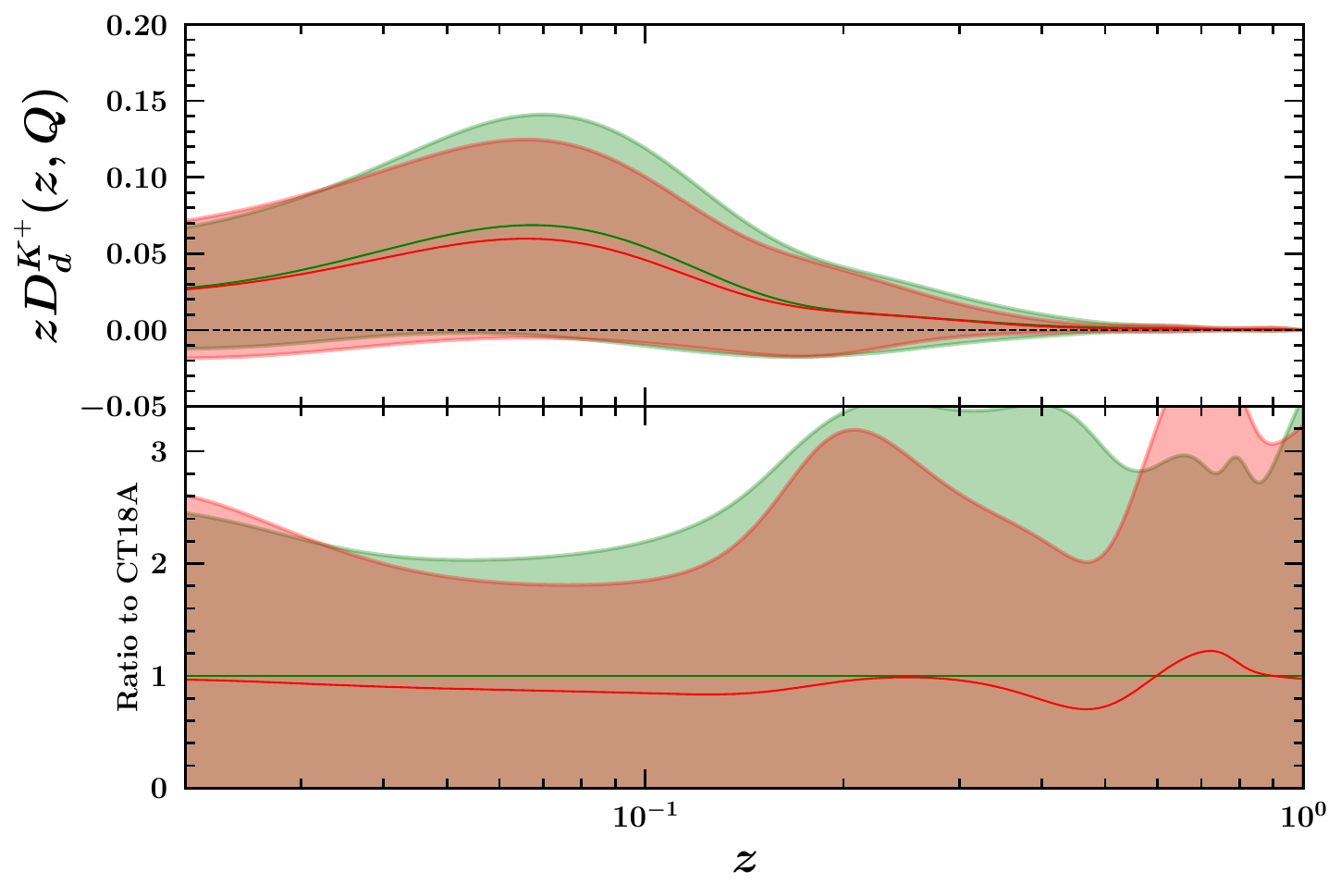}}
	\resizebox{0.45\textwidth}{!}{\includegraphics{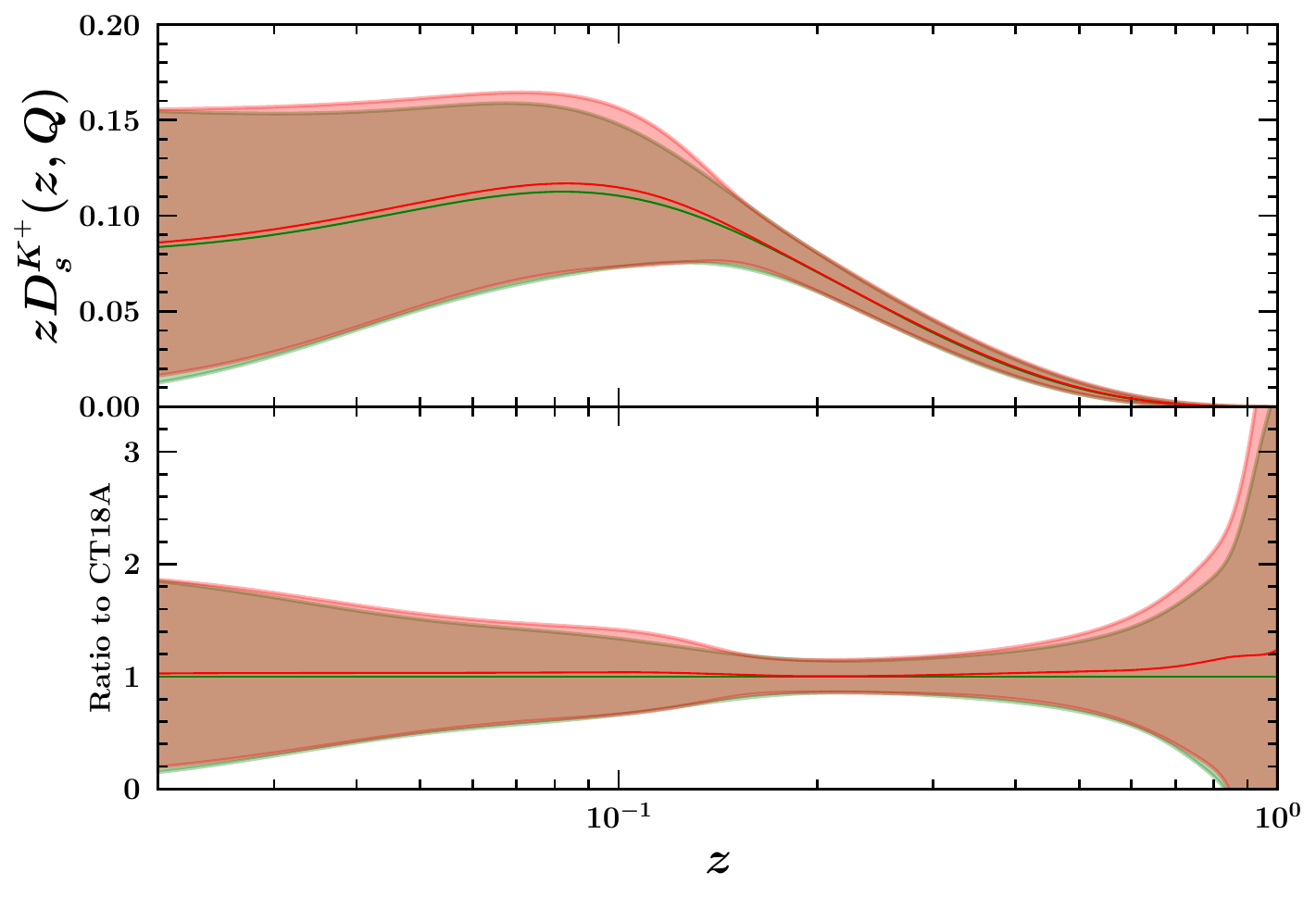}} 
	\resizebox{0.45\textwidth}{!}{\includegraphics{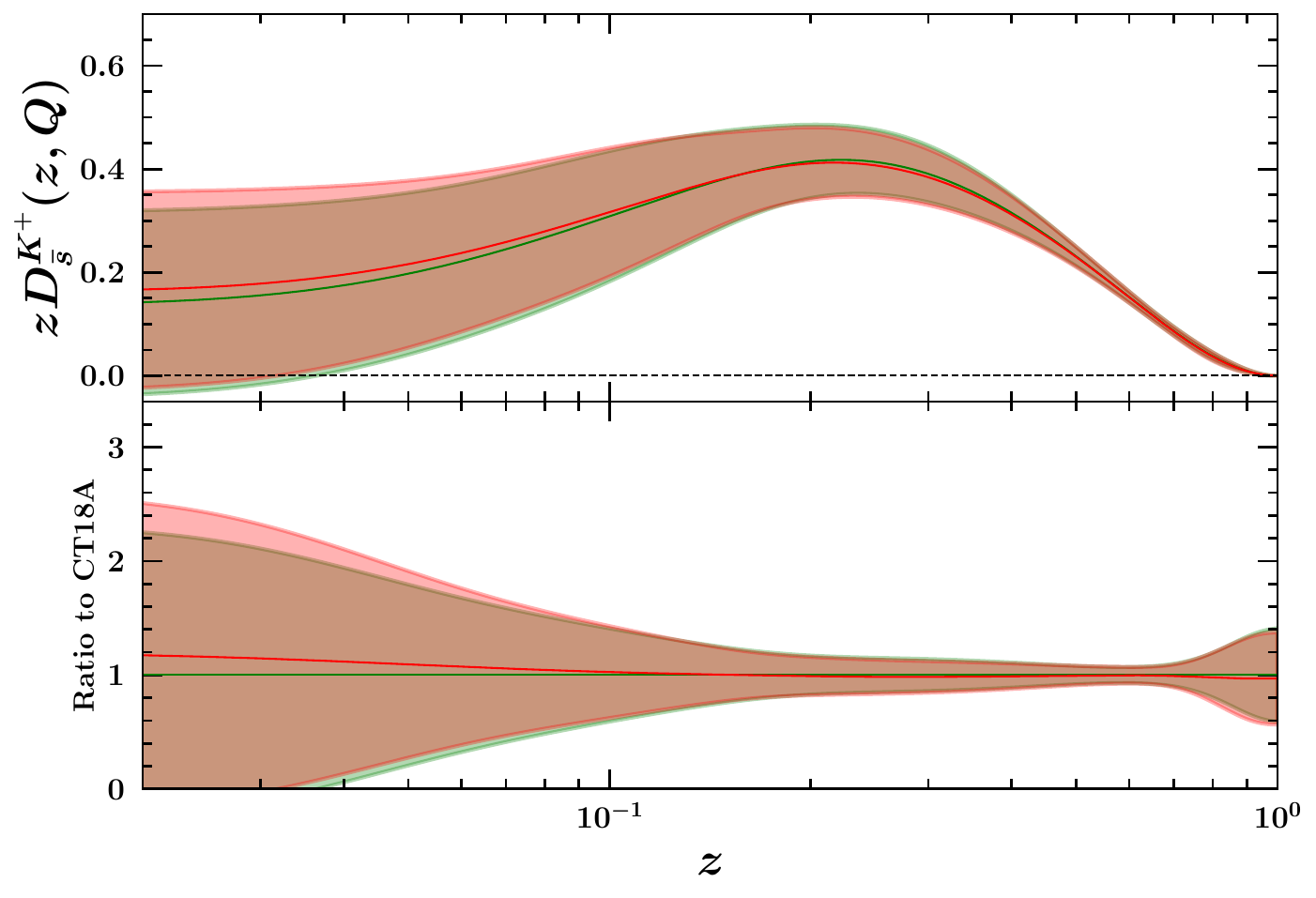}} 
	\resizebox{0.45\textwidth}{!}{\includegraphics{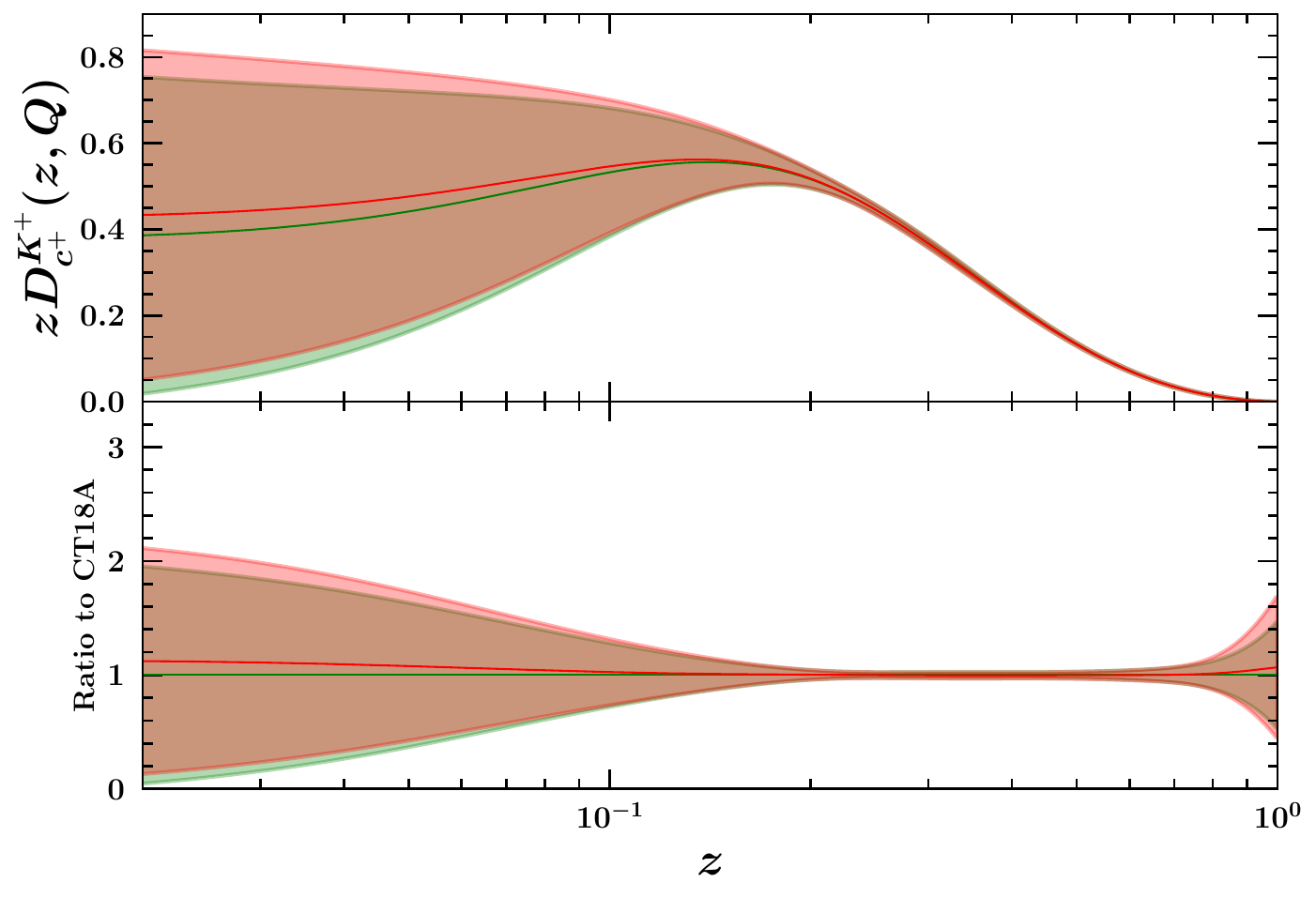}} 
	\resizebox{0.45\textwidth}{!}{\includegraphics{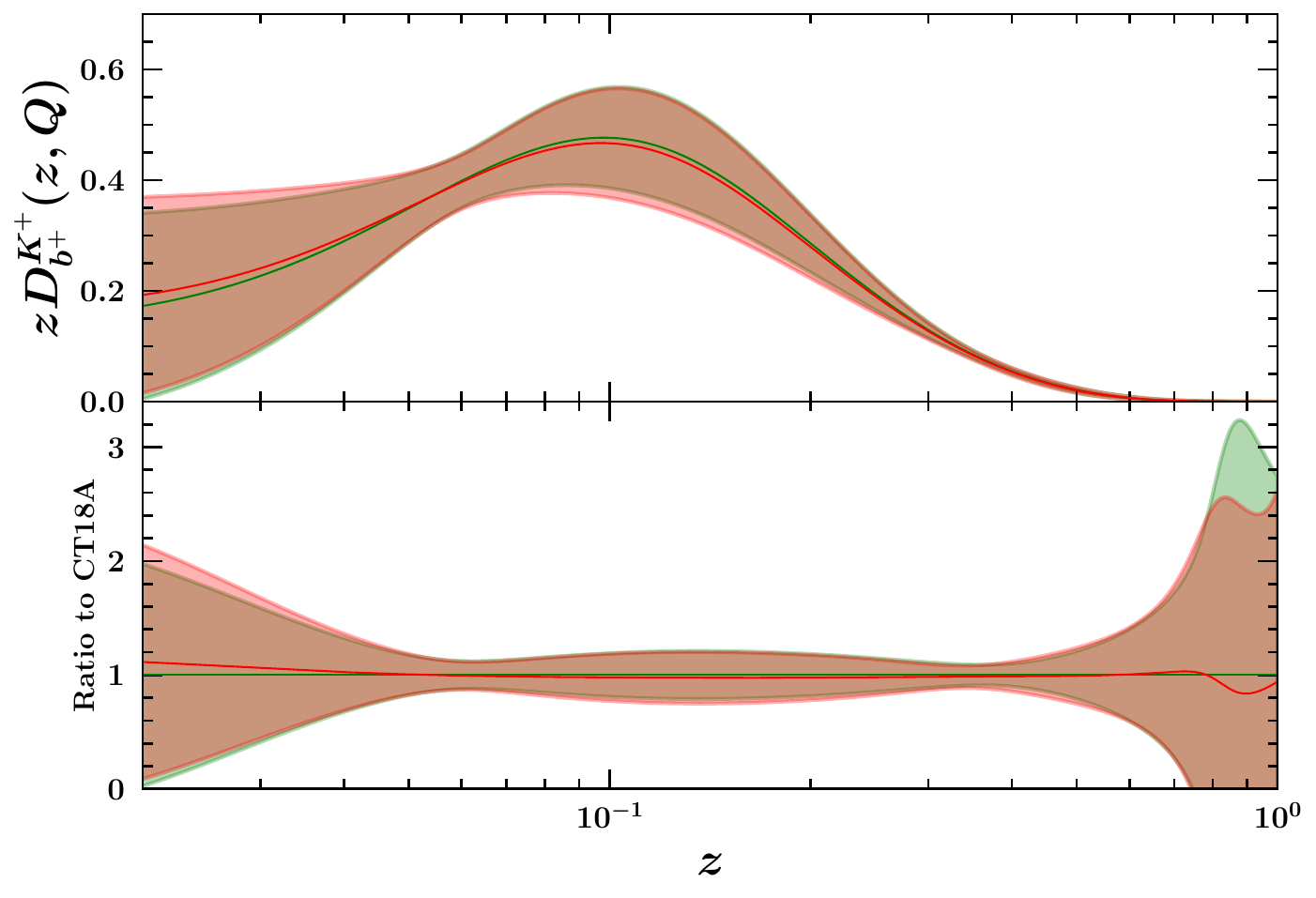}}   	 	
	\begin{center}
		\caption{ 
			\small 
 			Comparison of kaon FFs extracted from {\tt CT18A} proton PDF sets as our baseline, 
			and the nuclear PDF sets from {\tt EPPS21}. 
			We present both the absolute values, and the ratio to the 
			{\tt CT18A} proton PDFs baseline as well.
			The results presented at  $Q=5$ GeV.} 
		\label{fig:kaon-FFs_NLO_EPPS}
	\end{center}
\end{figure*}

\begin{figure*}[htb]
	\vspace{0.50cm}
	\resizebox{0.45\textwidth}{!}{\includegraphics{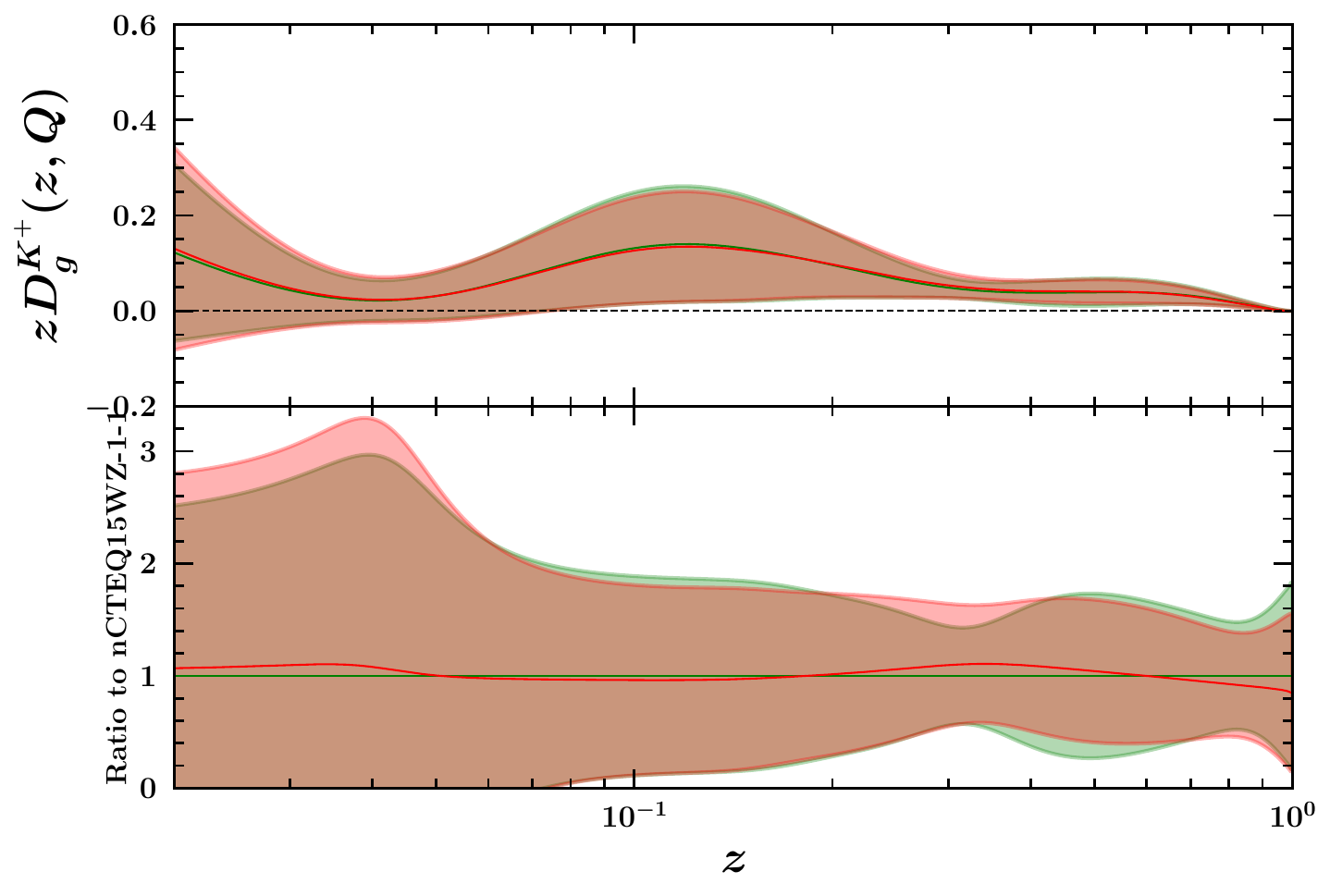}} 	
	\resizebox{0.45\textwidth}{!}{\includegraphics{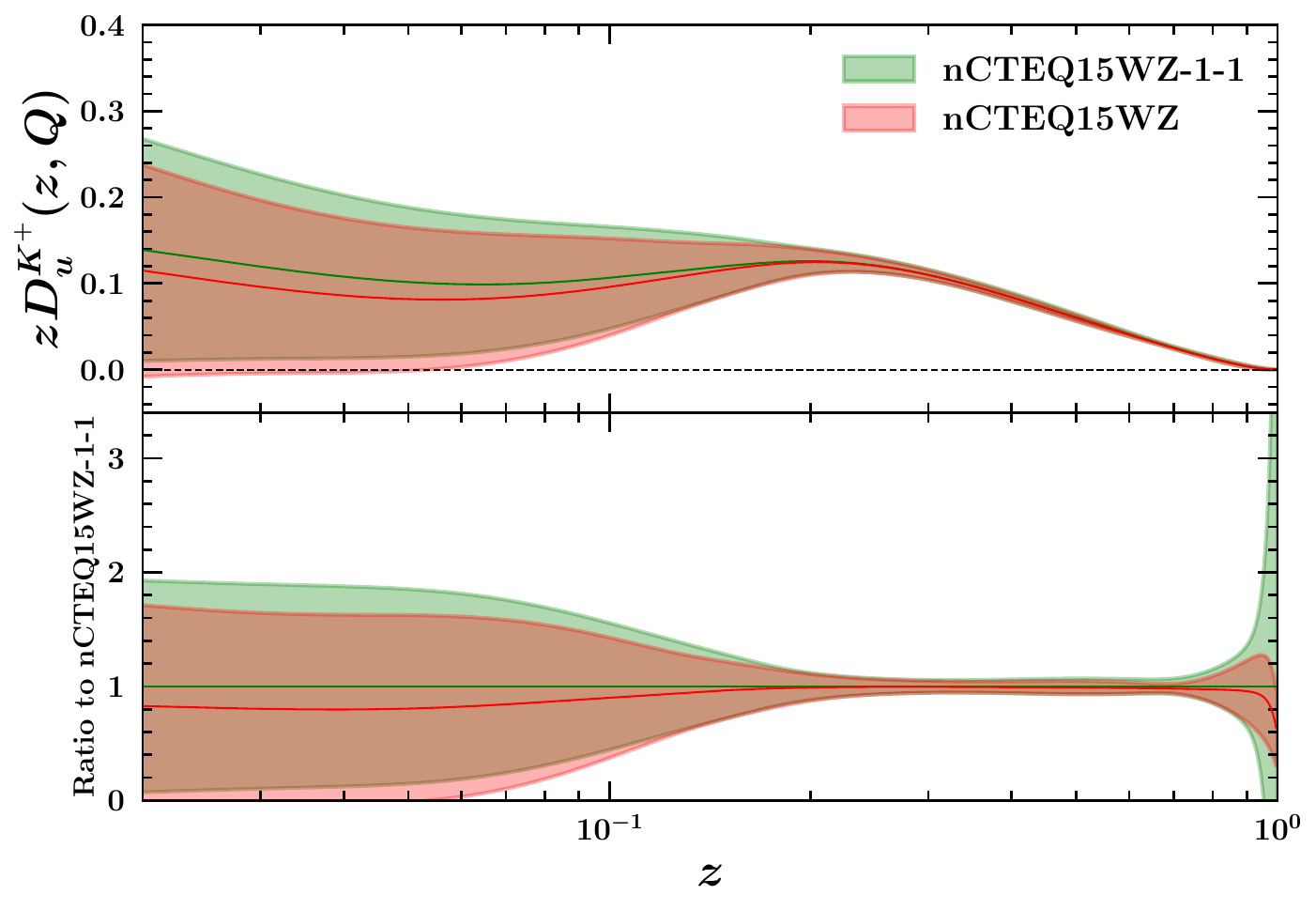}}  	
	\resizebox{0.45\textwidth}{!}{\includegraphics{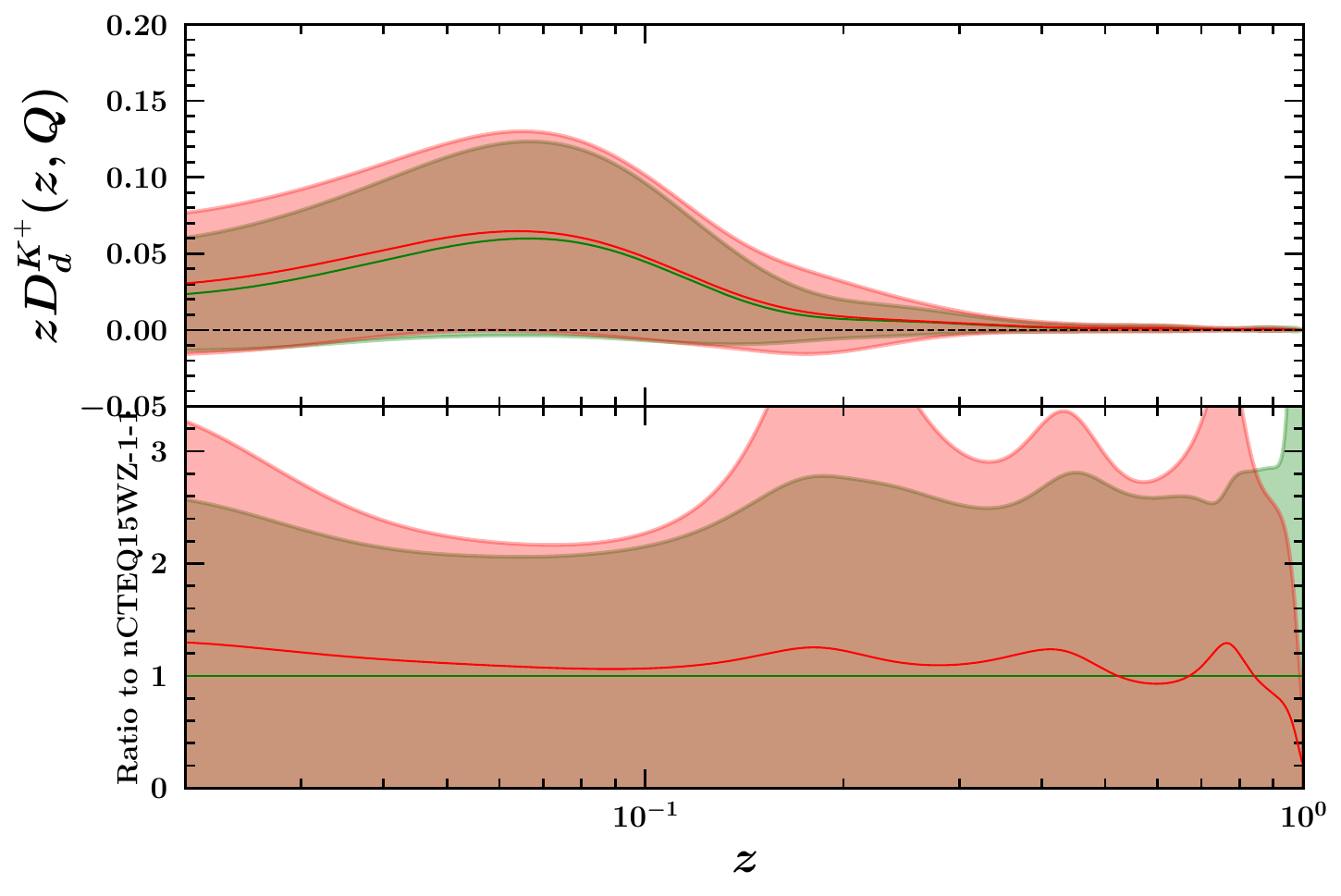}}
	\resizebox{0.45\textwidth}{!}{\includegraphics{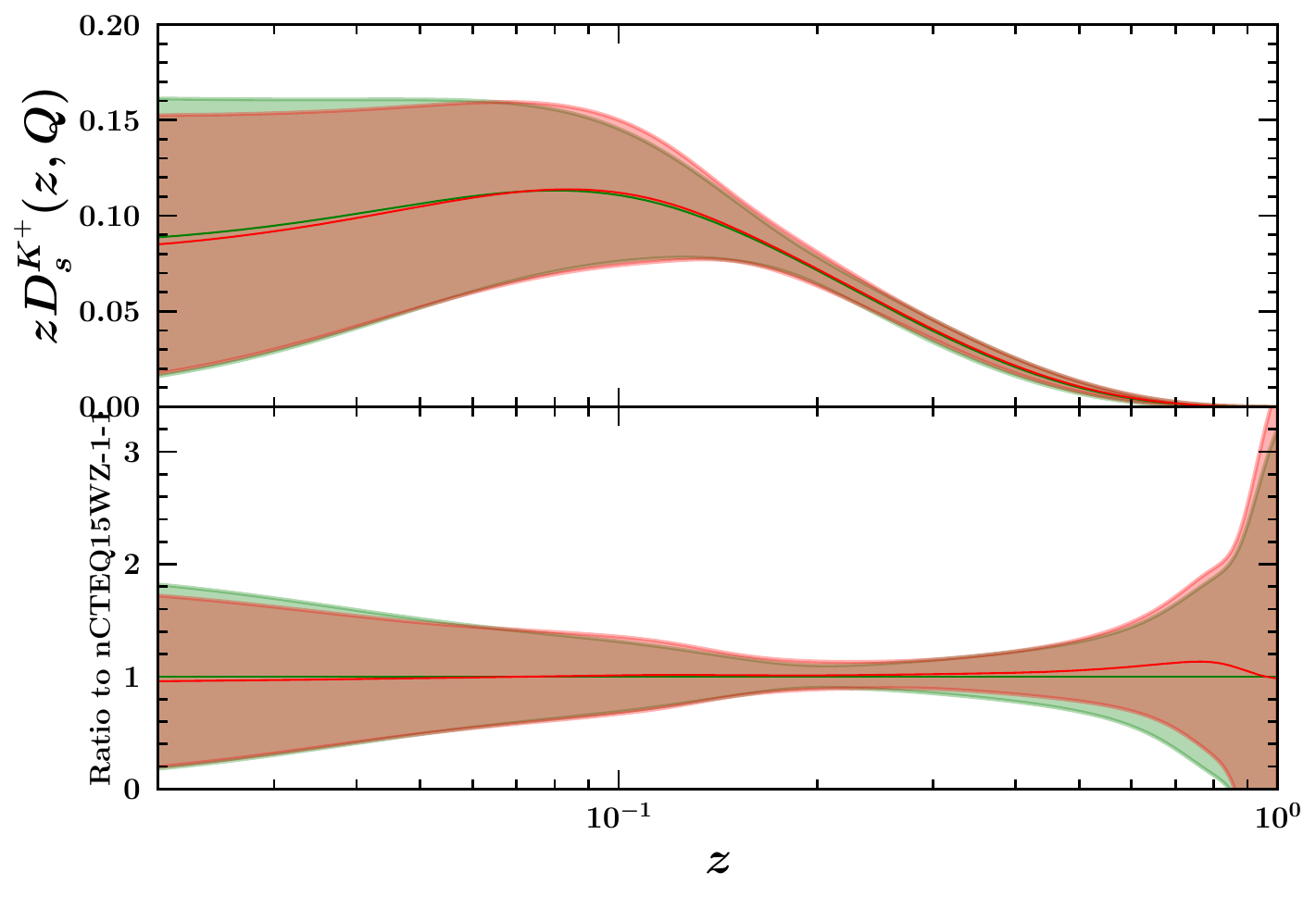}} 
	\resizebox{0.45\textwidth}{!}{\includegraphics{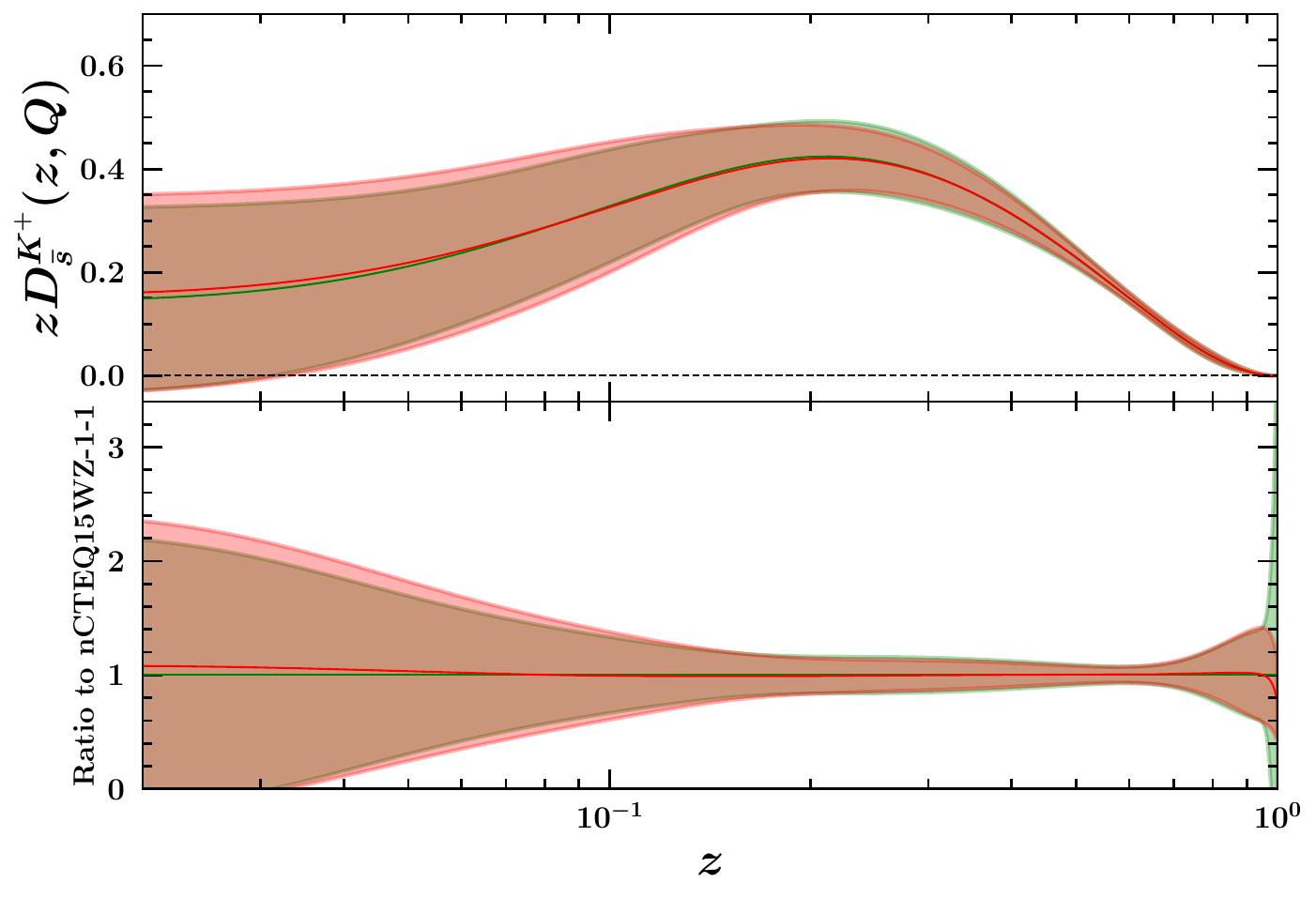}} 
	\resizebox{0.45\textwidth}{!}{\includegraphics{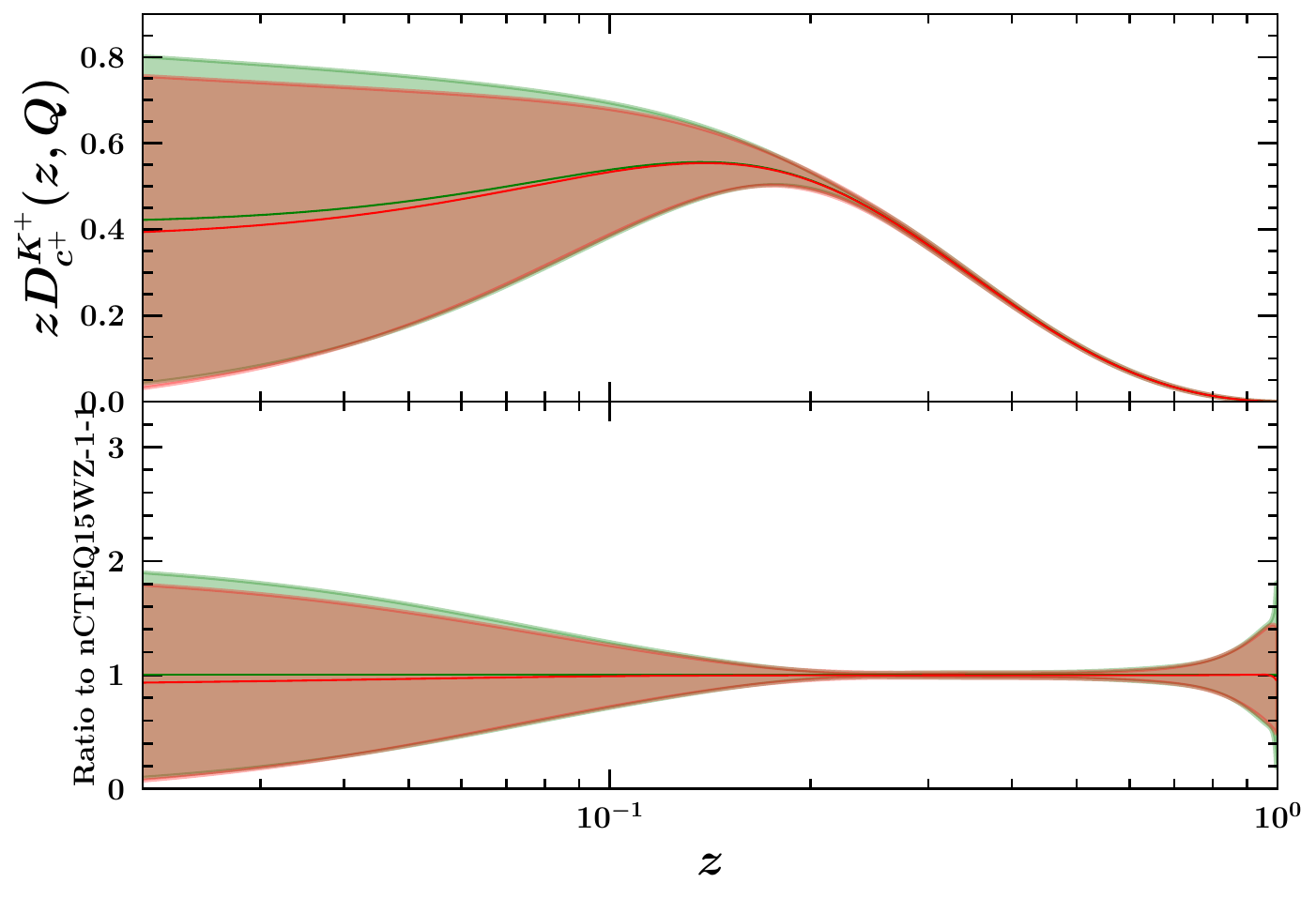}} 
	\resizebox{0.45\textwidth}{!}{\includegraphics{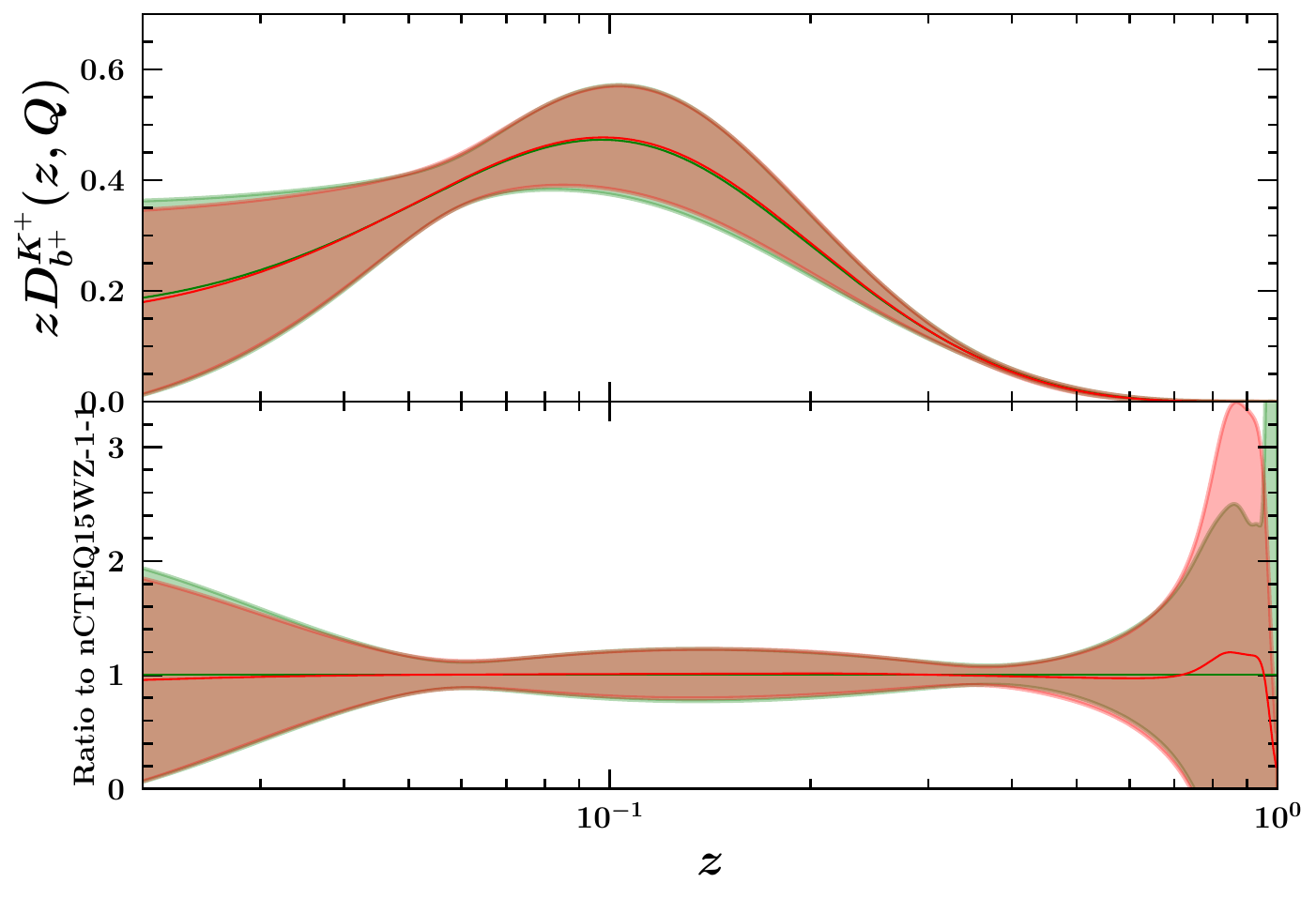}}   	 	
	\begin{center}
		\caption{ 
			\small 
 			Comparison of kaon FFs extracted from {\tt nCTEQ15WZ-1-1} proton PDF sets as our baseline, 
			and the nuclear PDF sets from {\tt nCTEQ15WZ}. 
			We present both the absolute values, and the ratio to the 
			{\tt nCTEQ15WZ-1-1} proton PDFs baseline as well.
			The results presented at  $Q=5$ GeV.} 
		\label{fig:kaon-FFs_NLO_nCTEQ}
	\end{center}
\end{figure*}

Now, we can proceed to discuss the extracted kaon FFs in terms of their 
central values and uncertainty bands, considering the inclusion of 
nuclear effects.
In Figs.~\ref{fig:kaon-FFs_NLO_nNNPDF}, \ref{fig:kaon-FFs_NLO_EPPS} and 
\ref{fig:kaon-FFs_NLO_nCTEQ}, we present 
the kaon FFs for different parton species.
As previously discussed, the nuclear PDF sets from {\tt nNNPDF3.0}, 
{\tt EPPS21}, and {\tt nCTEQ15WZ} are considered to 
investigate the nuclear effects in the extracted kaon FFs. 
We present both the absolute values of the extracted kaon FFs and 
their ratios to the corresponding proton PDFs baseline. 
Overall, it is evident that incorporating the nuclear PDFs brings slight  
changes in both the shape of the central values and the uncertainty 
bands of the extracted kaon FFs. For the case of {\tt nNNPDF3.0}, as 
can be seen from Fig.~\ref{fig:kaon-FFs_NLO_nNNPDF}, such changes are 
more pronounced for the gluon, $d$, and $s$ FFs.

In the context of $d$-quark FFs, the incorporation of nuclear effects not only 
influences the shape of the central values but also results in a narrowing of 
the uncertainty bands when compared to the {\tt nNNPDF3.0-p} proton PDFs
across the majority of $z$ regions. Conversely, for gluon FFs, the inclusion of 
nuclear effects slightly widens the uncertainty bands. In the case of $s$-quarks, a 
small reduction in uncertainty is observed, particularly in the transition from medium 
to small values of $z$. 
Furthermore, the reduction in uncertainty for $u$-quarks is more pronounced 
in the smaller $z$ region.
For the $\bar{s}$, $c^+$, and $b^+$ FFs, the results are 
generally similar, with the inclusion of nuclear effects 
leading to reductions in uncertainties for large 
regions of $z$ for the $b^+$ FF.

In the context of {\tt EPPS21}, as illustrated in Fig.~\ref{fig:kaon-FFs_NLO_EPPS}, light changes are seen in the $d$-quark FFs across the entire range of $z$, mirroring the results obtained with {\tt nNNPDF3.0}. These changes encompass not only alterations in the central values but also a small narrowing of uncertainty bands in the medium $z$ region.
For the gluon FFs, marginal central value modifications are observed across all $z$ regions, and fluctuations in the error bands are apparent across various $z$ ranges.
Furthermore, the $u$ and $c^+$ FFs display notable alterations, especially in the lower $z$ range. A distinct reduction in uncertainty bands is observed in specific cases, such as the $u$-quark FFs at smaller values of $z$.

As can be seen, for the case of {\tt nCTEQ15WZ} nuclear PDFs, changes can 
be observed in both the central values and error bands of 
the gluon, $u$- and $d$-quark FFs. 
For the $s$, $\bar{s}$, $c^+$ and $b^+$ quark FFs with 
the inclusion of {\tt nCTEQ15WZ} nuclear 
PDFs, slight changes in both the shape of the central values and the 
uncertainty bands can also be observed. 
The smaller error bands are achieved 
for $u$, $s$ and $c^+$ FFs.
This reduction in the uncertainty bands is particularly 
notable for the $u$ FF, 
at all values of $z$.

In general, for the kaon, the most significant changes in the central values 
and uncertainty bands can be observed for the gluon and 
$d$ components. 
However, the overall impact on the kaon FFs analysis appears to be small.

\clearpage

%
\subsection{Nuclear effects of the unidentified light charged hadron FFs}\label{sec:hadron_result}
%

In this section, we will present our results for the unidentified 
light charged hadron FFs in the presence of nuclear effects.
The individual and total $\chi^2$ per data points for our analysis 
of the charged hadron FFs are presented 
in Table~\ref{tab:datasets_hadron_NLO}. 
The $\chi^2$ values allow for the quantification of the goodness-of-fit 
for the charged hadron FFs analysis.

As can be seen, in terms of total $\chi^2$, the inclusion of 
the {\tt nNNPDF3.0} and {\tt EPPS21} nuclear PDFs reduces 
the values of $\chi^2$ from 1.059 to 1.030 and from 1.060 
to 1.056, respectively, with respect to their 
baseline proton PDFs fits. 
However, in the case of {\tt nCTEQ15WZ} nuclear PDFs, 
the value of the total $\chi^2$ is increased. 
This finding is also consistent with the individual $\chi^2$ 
values per data point presented in Table~\ref{tab:datasets_hadron_NLO}, 
particularly for the {\tt COMPASS} $H^+$ and {\tt COMPASS} $H^-$ data sets. 
As can be seen, for the {\tt nCTEQ15WZ} fit, the 
individual $\chi^2$ value for the {\tt COMPASS} $H^+$ dataset 
increased from 1.230 to 1.379 when the 
nuclear correction is considered.

\begin{table*}[htb]
	\renewcommand{\arraystretch}{2}
	\centering 	\scriptsize
	\begin{tabular}{|l|c|cc|cc|cr|}				\hline
	
	        ~        & ~ &\multicolumn{6}{|c|}{  $\frac{\chi^2}{N_{\rm dat}}$:}
	        \\
	        
		     Experiment &  $N_{\rm dat}$&~  {\tt nNNPDF3.0-p}    ~&~   {\tt nNNPDF3.0} ~&~  {\tt CT18A} ~&~ {\tt EPPS21} ~&~ {\tt nCTEQ15WZ-1-1} ~&~ {\tt nCTEQ15WZ} 
		\rule[-3mm]{0mm}{5mm}
		\\
		\hline \hline
		{\tt COMPASS} $H^+$\cite{COMPASS:2016xvm}   & 157  & 1.425& 1.316&1.232&1.225&1.230& 1.379  \\
		{\tt COMPASS} $H^-$~\cite{COMPASS:2016xvm} &   157 & 0.784 & 0.801&0.957&0.941&0.939&0.907 \\ 
		{\tt TASSO14}~\cite{TASSO:1982bkc}  & 14 & 1.835& 1.820 &1.809&1.805&1.816&1.734 \\      					
		{\tt TASSO22}~\cite{TASSO:1982bkc}   & 14 & 1.213& 1.211&1.226&1.238&1.256& 1.187\\
		{\tt TPC}~\cite{Aihara:1988su}  & 21 & 0.600&0.605&0.606&0.613&0.609&0.508 \\
		{\tt TASSO44}~\cite{TASSO:1988jma}&14 & 2.605&2.591&2.724&2.778 &2.881&2.629\\
		{\tt ALEPH}~\cite{Buskulic:1995aw}  & 32 & 0.818 &0.822&0.825&0.830&0.823&0.817\\
		{\tt DELPHI} (incl.)~\cite{Abreu:1998vq}   & 21 & 0.606 &0.602&0.599&0.594&0.595&0.592 \\
		{\tt DELPHI} ($uds$ tag)~\cite{Abreu:1998vq}  & 21 & 0.376 &0.373&0.377&0.378&0.371&0.377  \\
		{\tt DELPHI} ($b$ tag)~\cite{Abreu:1998vq}  &21 & 1.014  & 1.009 &1.010& 1.008&1.019& 0.998\\
		{\tt OPAL} (incl.)~\cite{Ackerstaff:1998hz} & 19& 1.824& 1.824&1.808&1.774&1.775& 1.785\\
		{\tt OPAL} ($uds$ tag)~\cite{Ackerstaff:1998hz}   & 19 & 0.791 &0.788&0.796&0.798&0.797&0.787\\
		{\tt OPAL} ($c$ tag)~\cite{ Ackerstaff:1998hz}  &19 & 0.592 & 0.591&0.606&0.611&0.609&0.604 \\
		{\tt OPAL} ($b$ tag)~\cite{Ackerstaff:1998hz}  & 19 & 0.279  &0.278&0.276&0.287&0.277&0.281\\
		{\tt SLD} (incl.)~\cite{Abe:2003iy}   & 34 & 1.064&1.044&1.080&1.086&1.066& 1.086\\
		{\tt SLD} ($uds$ tag)~\cite{Abe:2003iy}   &34& 0.966 & 0.952&0.928& 0.929&0.925& 0.939 \\
		{\tt SLD} ($c$ tag)~\cite{Abe:2003iy} &34&0.996  & 0.991&1.136&1.168&1.169&1.109 \\
		{\tt SLD} ($b$ tag)~\cite{Abe:2003iy} &34 & 1.080 &1.060&1.086&1.085&1.090&1.098 \\ 				
		\hline \hline
		Total $\chi^2/{N_{\rm dat}}$ & 684 &1.059 &1.030&1.060& 1.056 &1.058& 1.080   \\
		\hline \hline	
	\end{tabular}
	\caption{ \small 
 		The list of input data sets for our unidentified light charged hadrons production included in our charged hadron FFs analysis. 
		For each data set, we have indicated the experiments, corresponding published reference and the number of data points. 
		In the last four columns, we show the value of $\chi^2/{N_{\rm dat}}$ resulting from the 
		FF fit at NLO order by considering proton PDF sets from {\tt nNNPDF3.0-p}~\cite{AbdulKhalek:2022fyi}, {\tt CT18A} ~\cite{Hou:2019efy}  and {\tt nCTEQ15WZ-1-1} \cite{Kusina:2020lyz} , and 
		nuclear PDF sets available in the literature, namely {\tt nNNPDF3.0}~\cite{AbdulKhalek:2022fyi}, 
		{\tt EPPS21}~\cite{Eskola:2021nhw}, 
		and {\tt nCTEQ15WZ}~\cite{Kusina:2020lyz}. 
		The total value of the total $\chi^2/{N_{\rm dat}}$ also is shown at the bottom of the table. } 
	\label{tab:datasets_hadron_NLO}
\end{table*}

\begin{figure*}[htb]
	\vspace{0.50cm}
	\resizebox{0.45\textwidth}{!}{\includegraphics{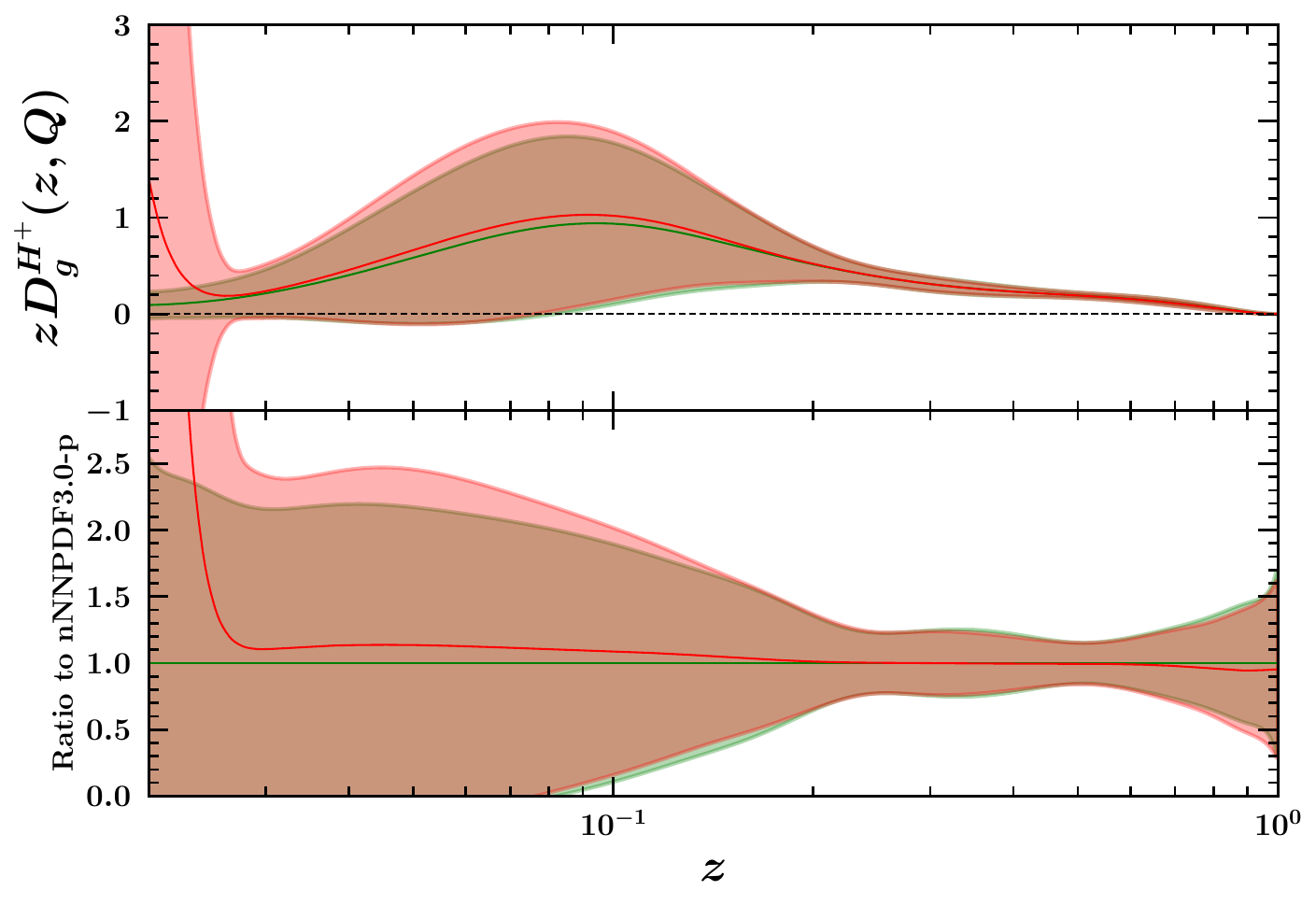}} 	
	\resizebox{0.45\textwidth}{!}{\includegraphics{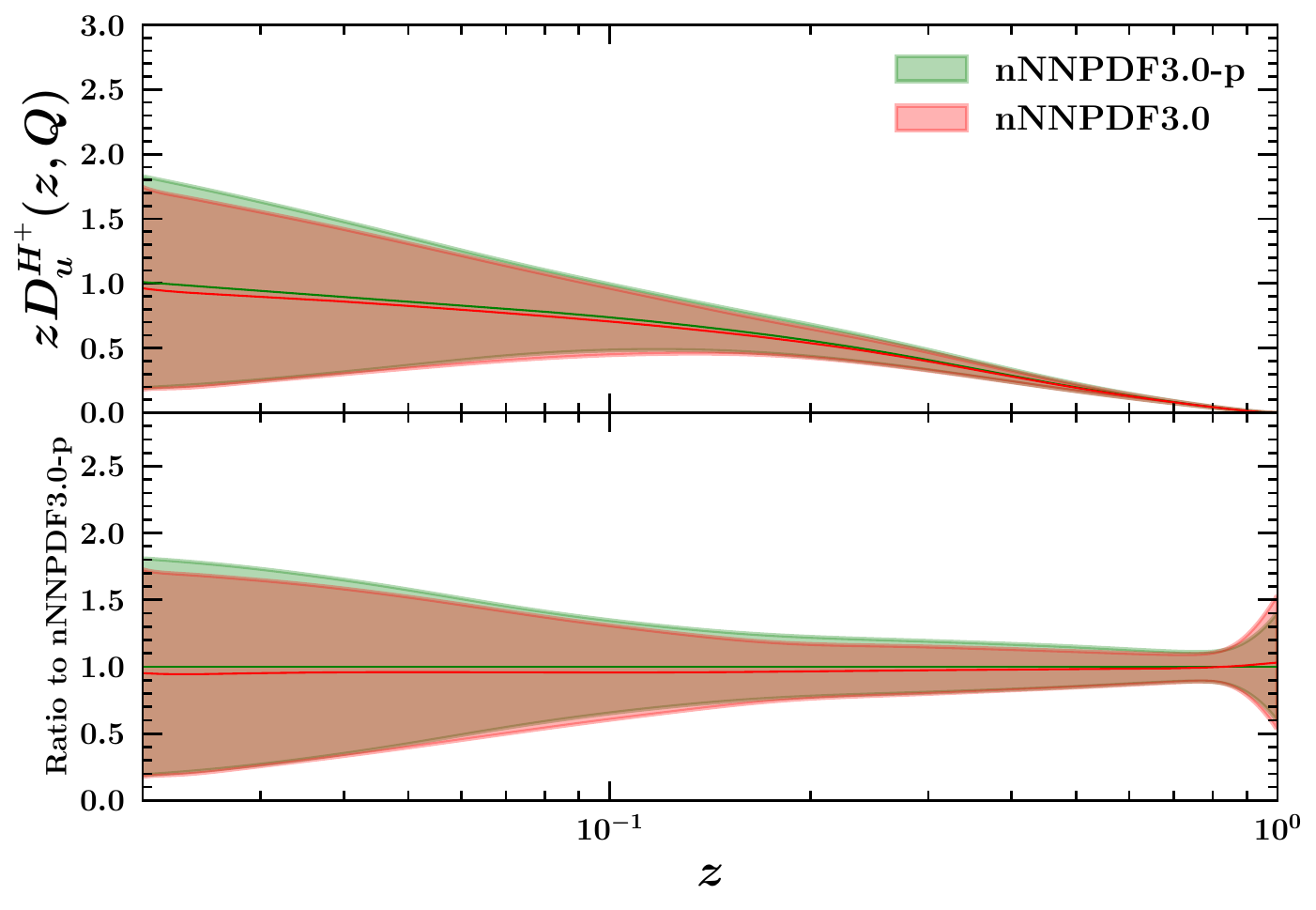}}  	
	\resizebox{0.45\textwidth}{!}{\includegraphics{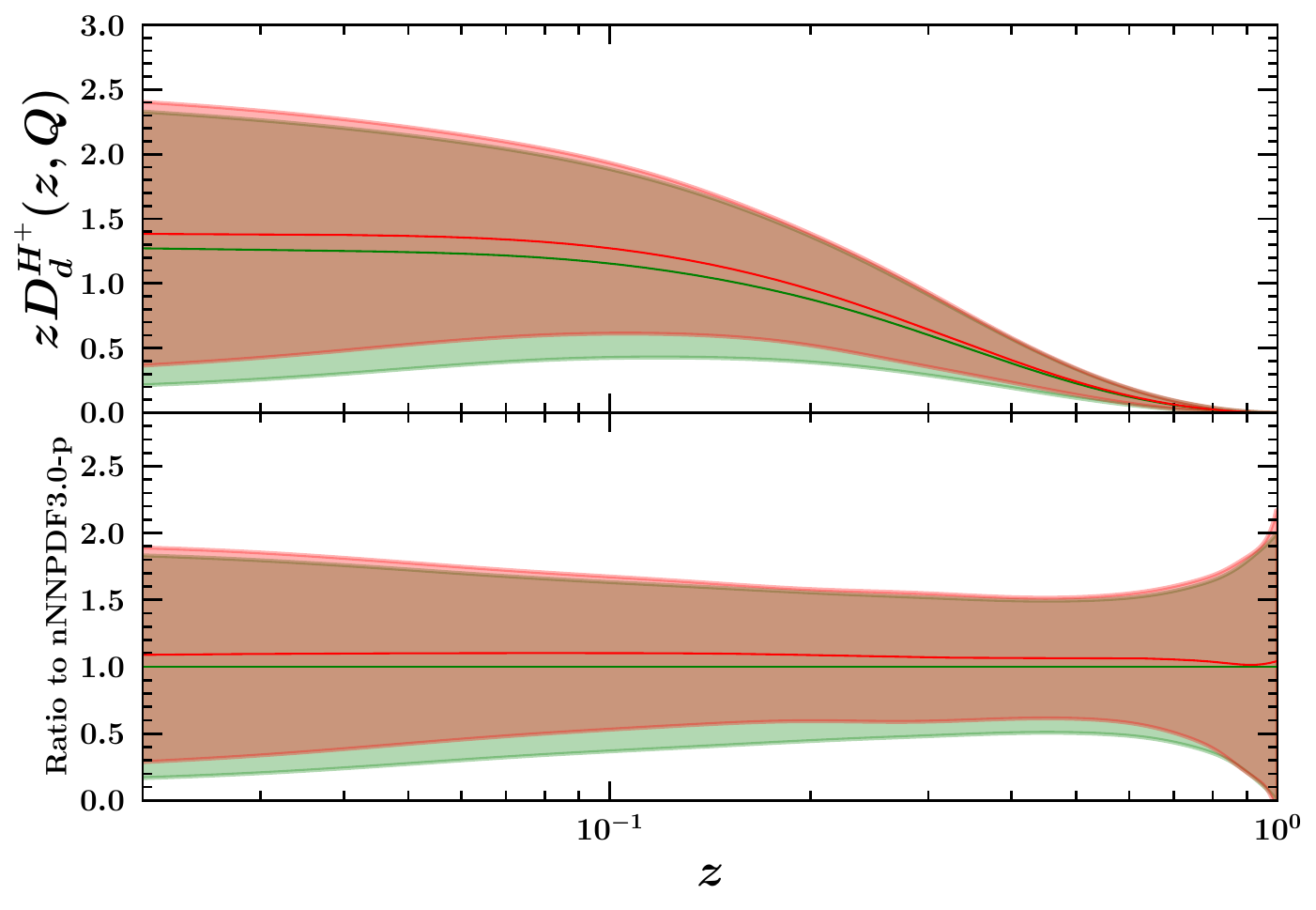}} 
	\resizebox{0.45\textwidth}{!}{\includegraphics{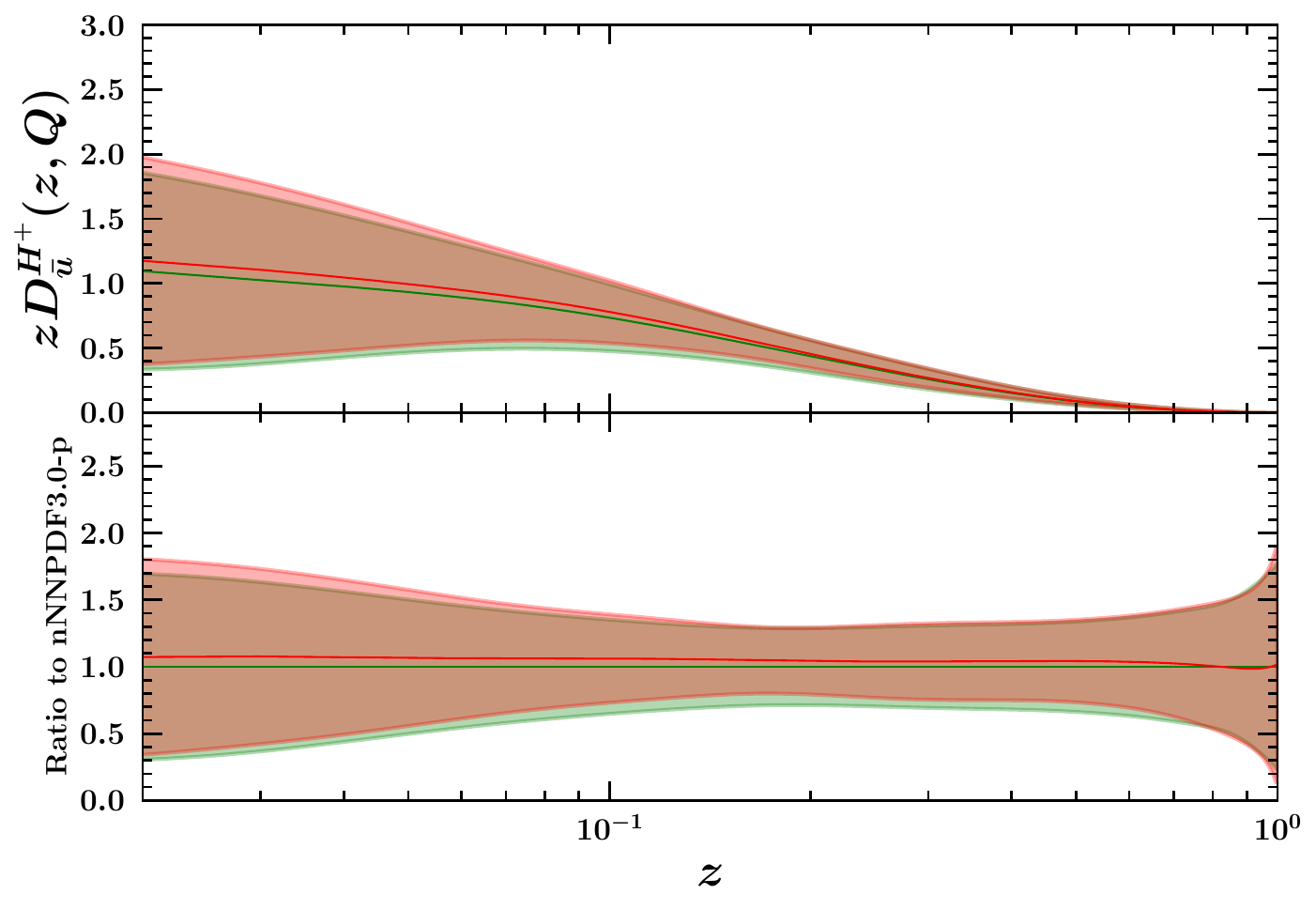}}  
	\resizebox{0.45\textwidth}{!}{\includegraphics{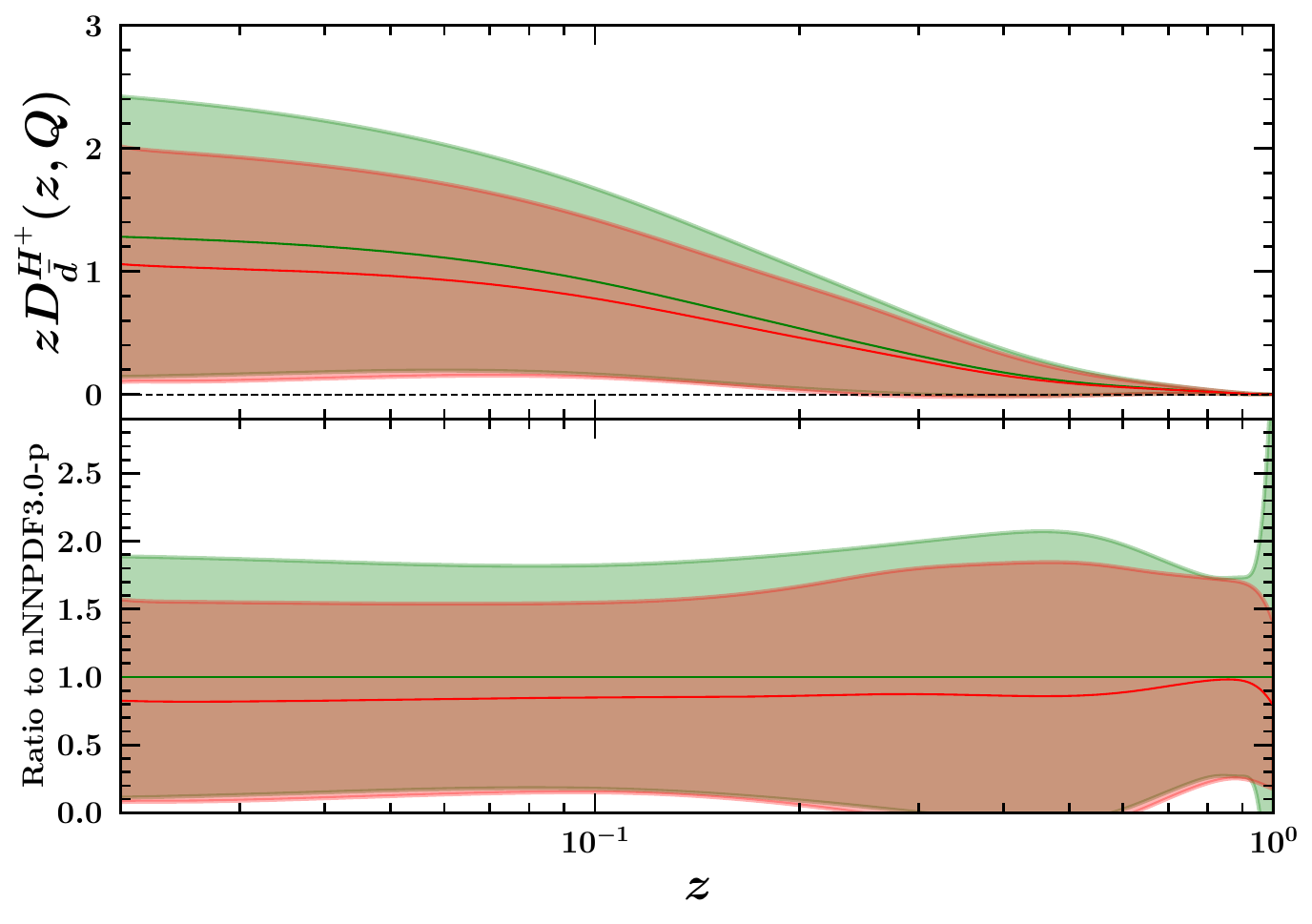}}
	\resizebox{0.45\textwidth}{!}{\includegraphics{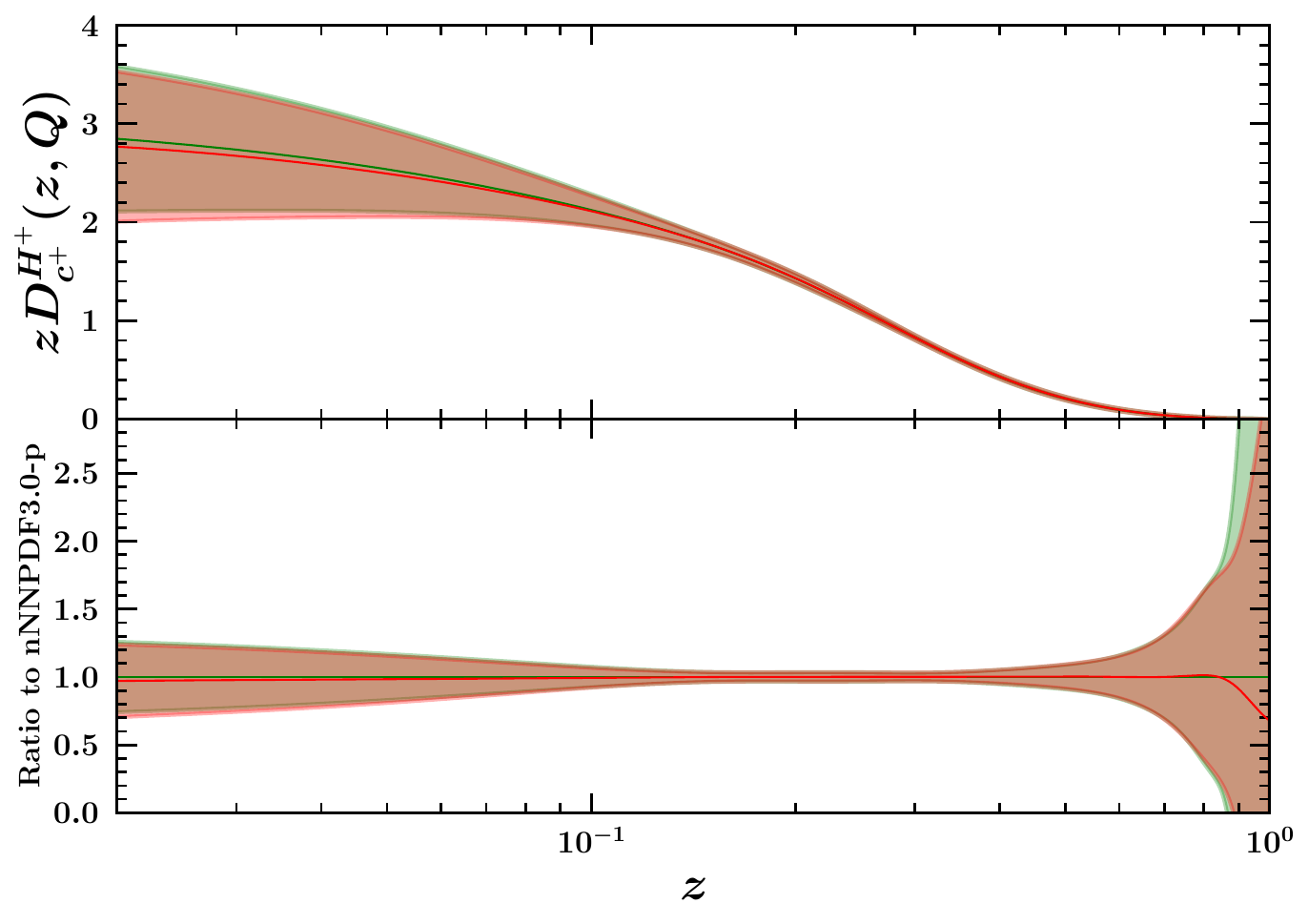}} 
	\resizebox{0.45\textwidth}{!}{\includegraphics{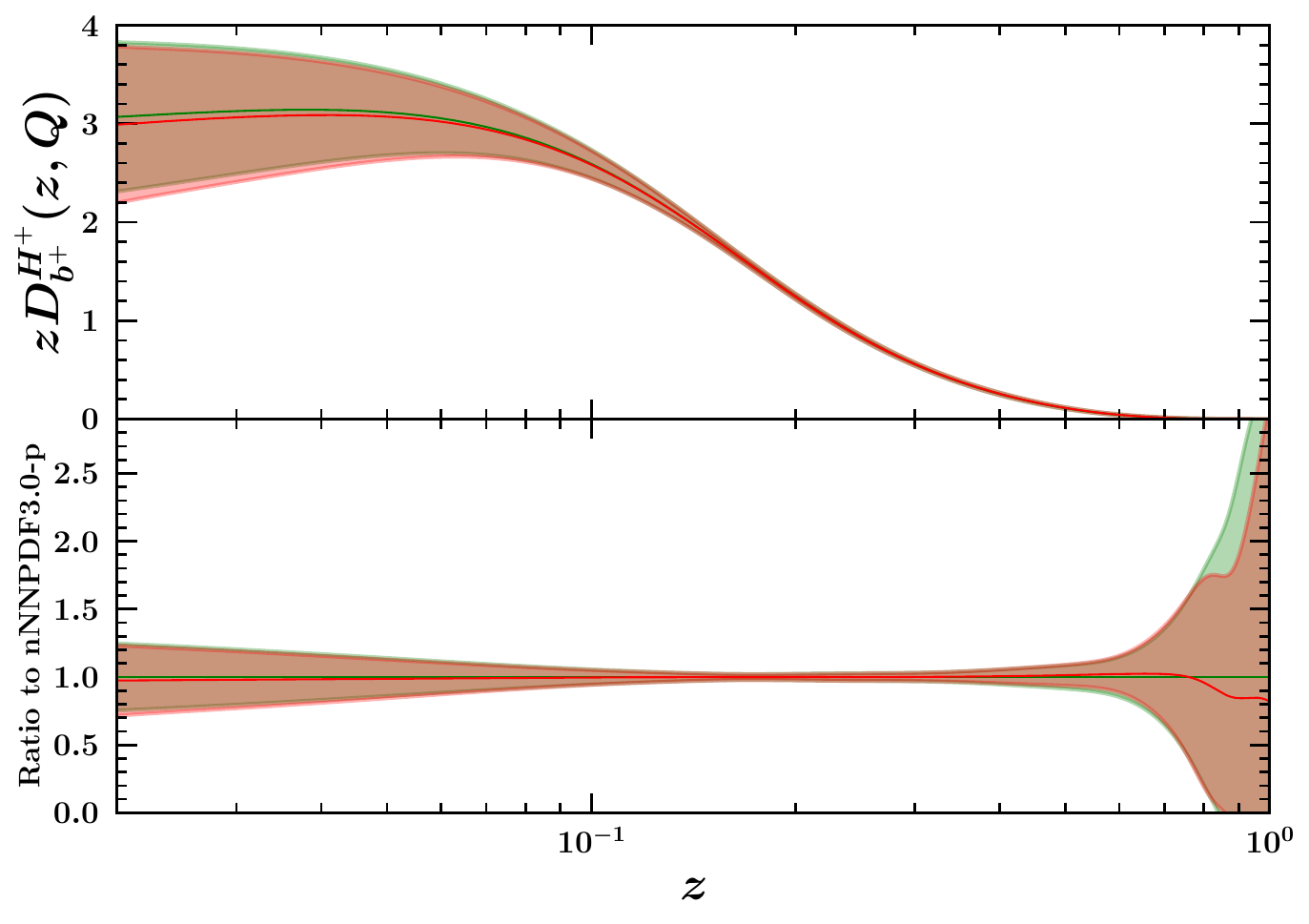}}   	 	
	\begin{center}
		\caption{ 
			\small 
  Comparison of charged hadron FFs extracted from {\tt nNNPDF3.0-p} proton PDF sets as our baseline, 
			and the nuclear PDF sets from {\tt nNNPDF3.0}. 
			We present both the absolute values, and the ratio to the 
			{\tt nNNPDF3.0-p} proton PDFs baseline as well.
			The results presented at  $Q=5$ GeV.} 
		\label{fig:H-FFs_NLO_nNNPDF}
	\end{center}
\end{figure*}

\begin{figure*}[htb]
	\vspace{0.50cm}
	\resizebox{0.45\textwidth}{!}{\includegraphics{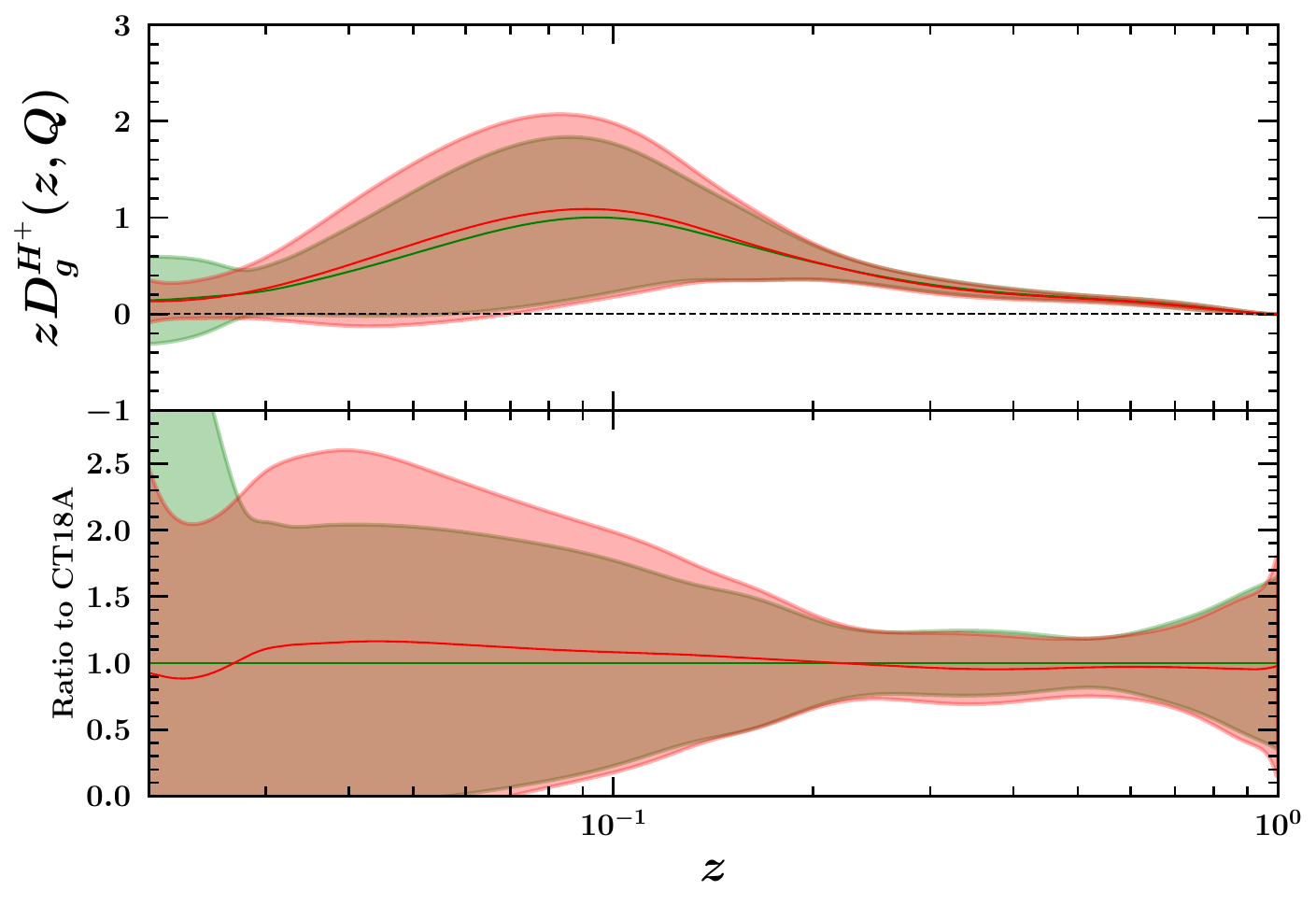}} 	
	\resizebox{0.45\textwidth}{!}{\includegraphics{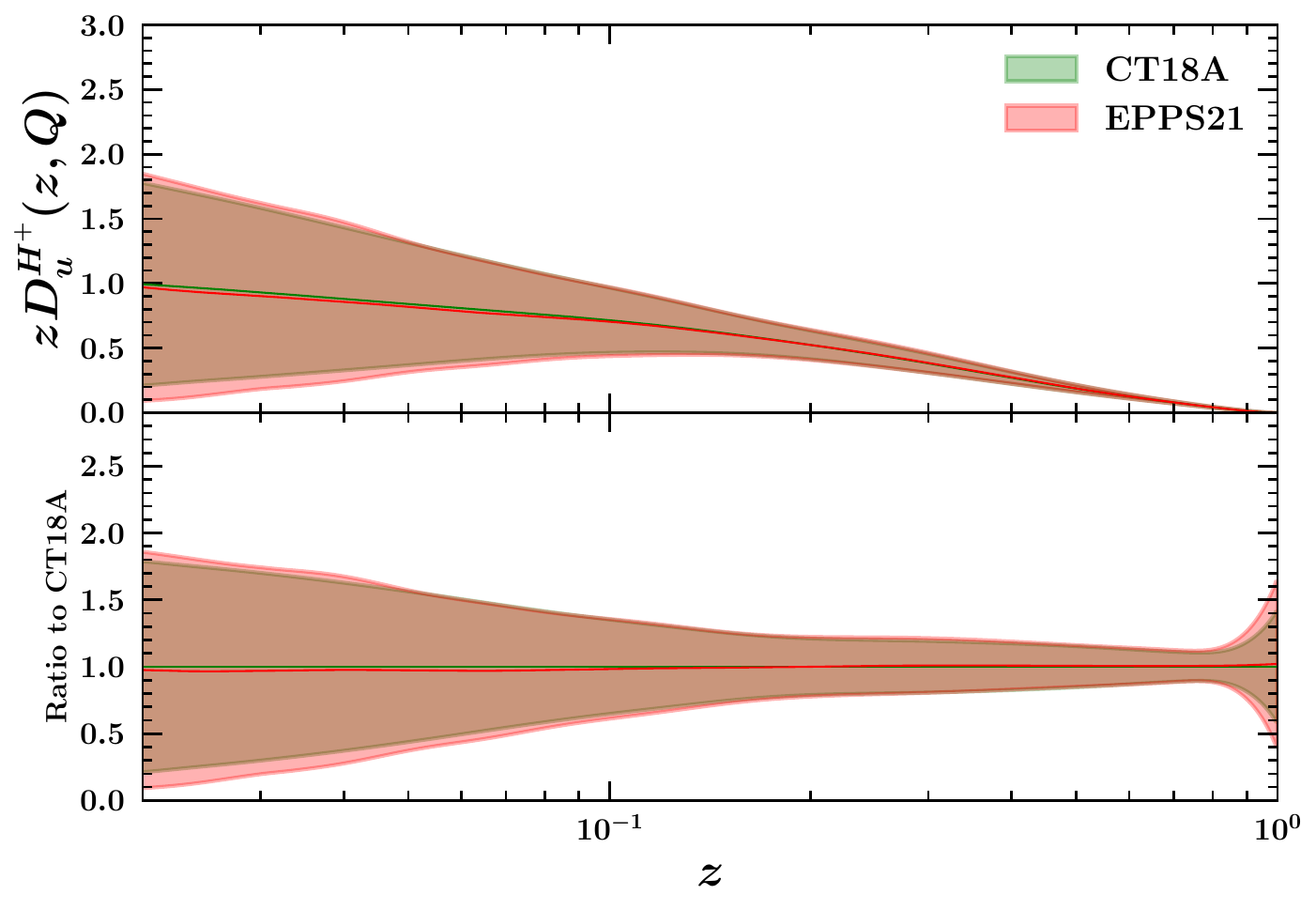}}  	
	\resizebox{0.45\textwidth}{!}{\includegraphics{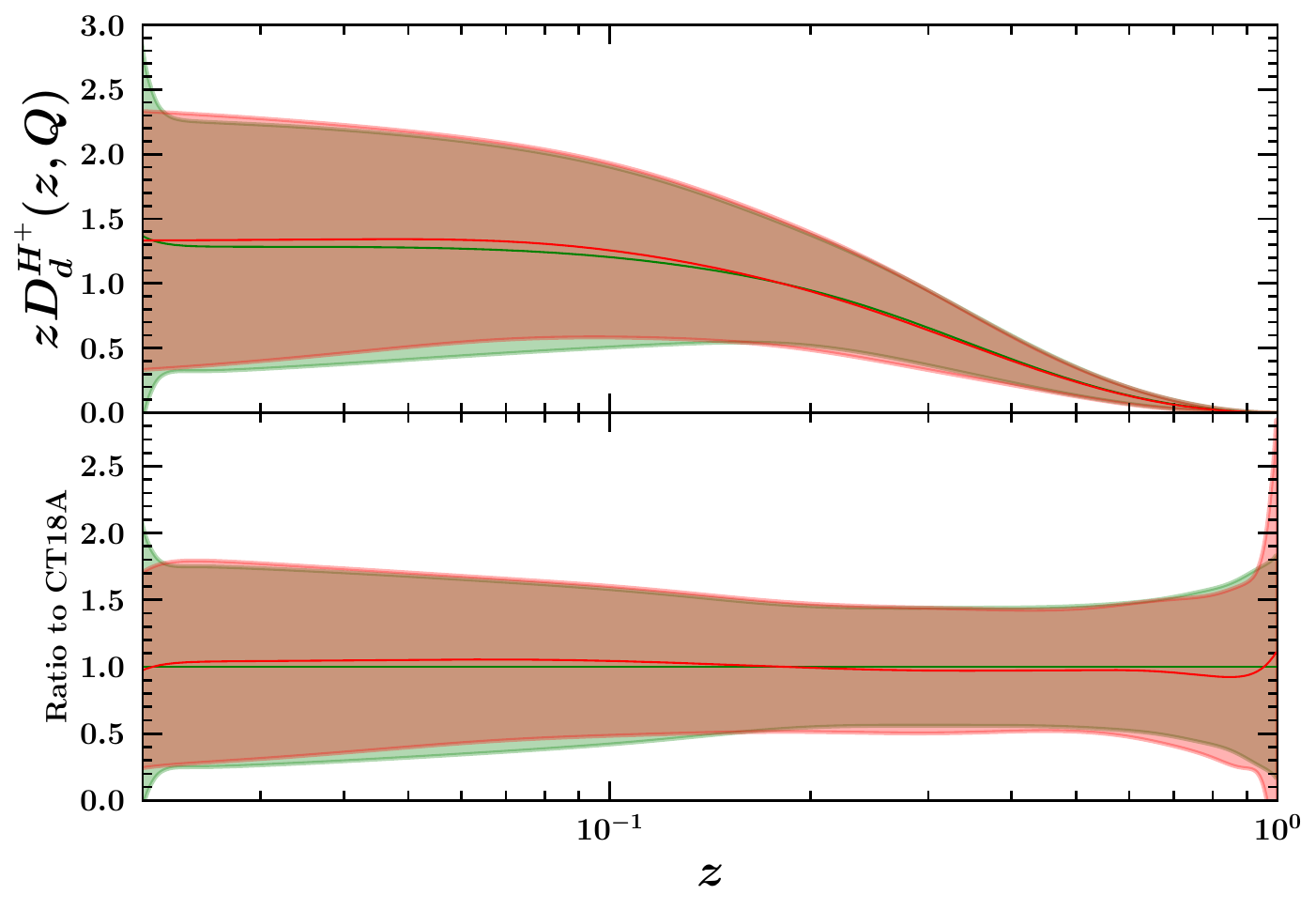}} 
	\resizebox{0.45\textwidth}{!}{\includegraphics{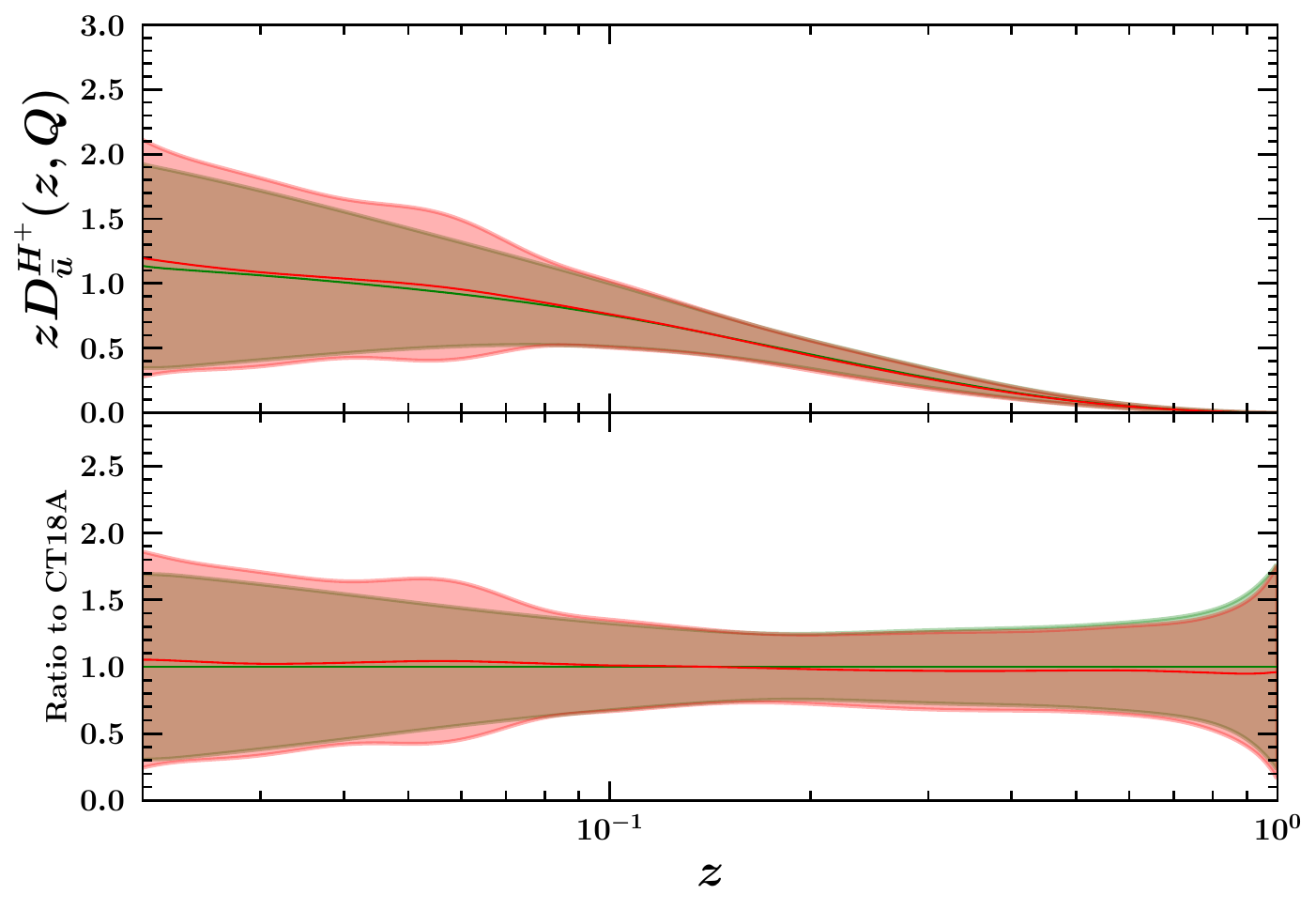}}  
	\resizebox{0.45\textwidth}{!}{\includegraphics{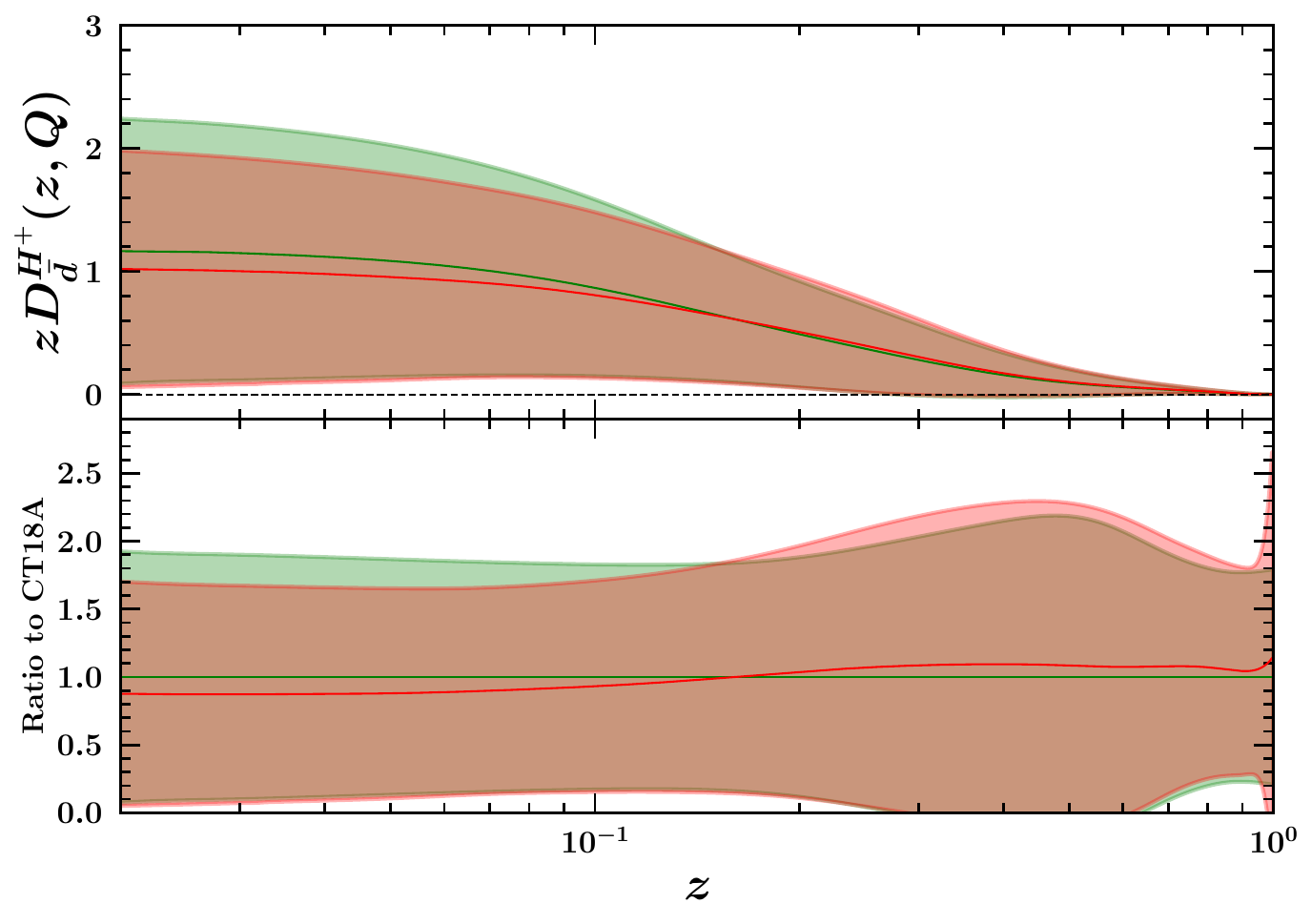}}
	\resizebox{0.45\textwidth}{!}{\includegraphics{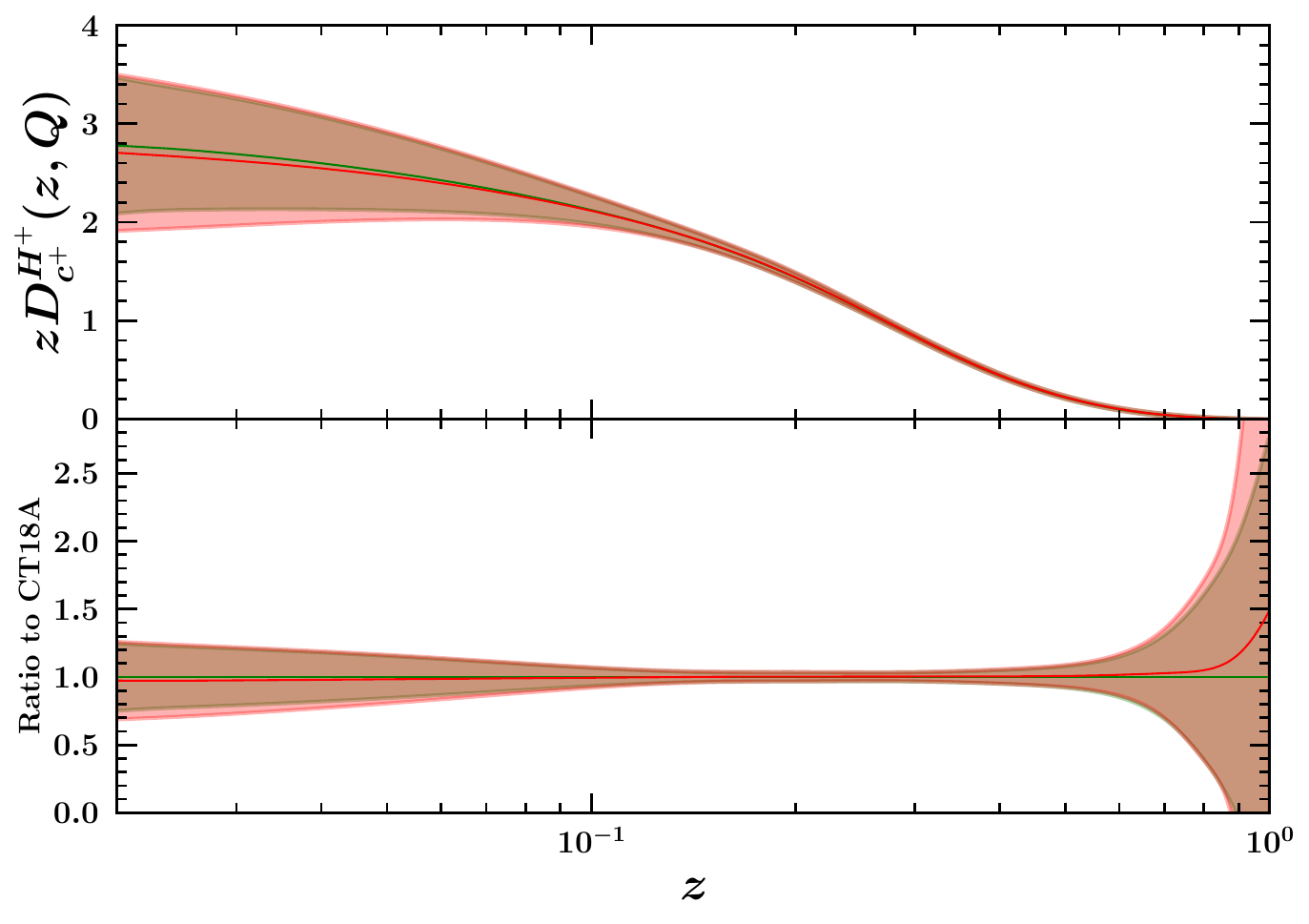}} 
	\resizebox{0.45\textwidth}{!}{\includegraphics{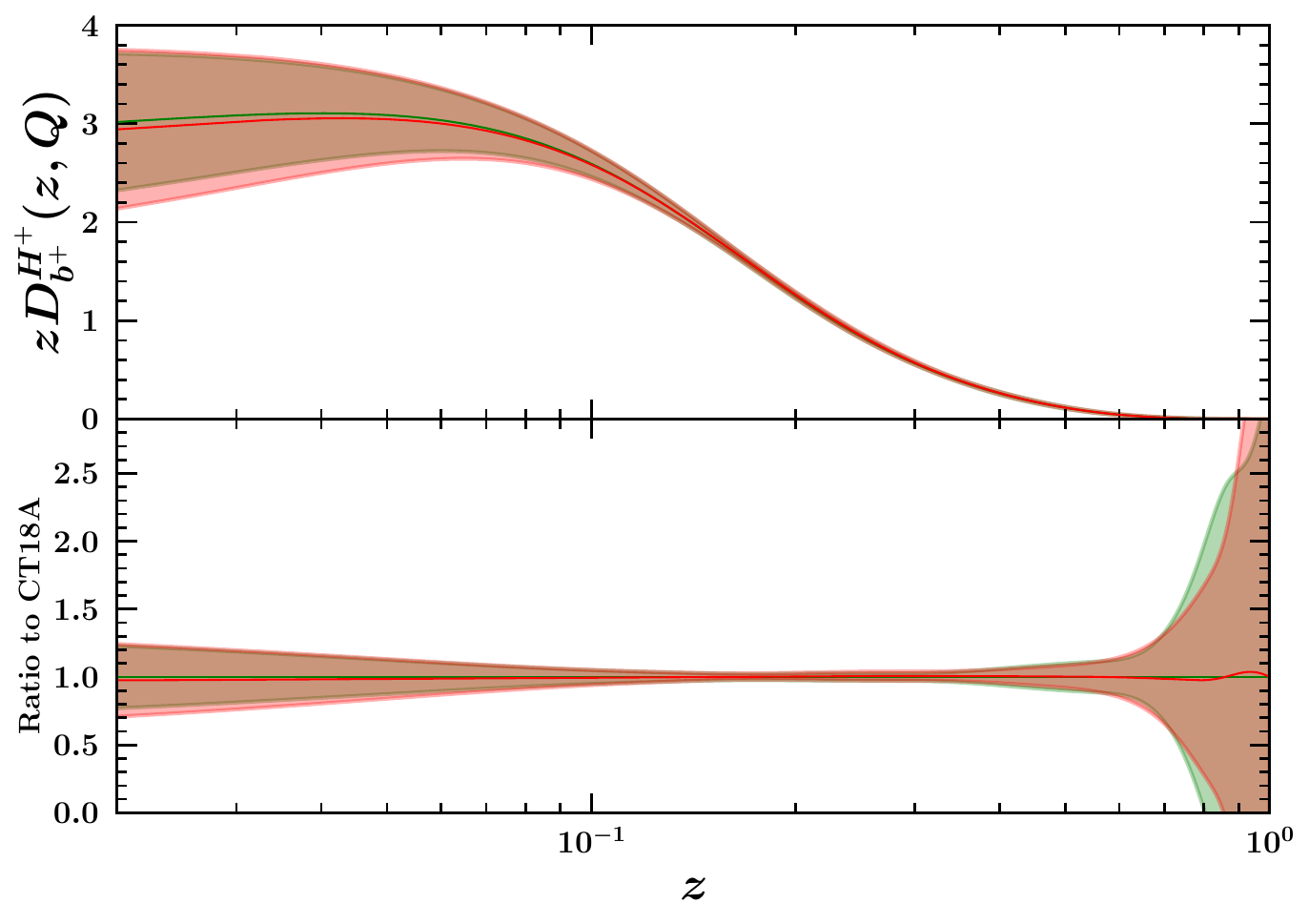}}   	 	
	\begin{center}
		\caption{ 
			\small 
 	Comparison of charged hadron FFs extracted from {\tt CT18A} proton PDF sets as our baseline, 
			and the nuclear PDF sets from {\tt EPPS21}. 
			We present both the absolute values, and the ratio to the 
			{\tt CT18A} proton PDFs baseline as well.
			The results presented at  $Q=5$ GeV.}    
		\label{fig:H-FFs_NLO_EPPS}
	\end{center}
\end{figure*}

\begin{figure*}[htb]
	\vspace{0.50cm}
	\resizebox{0.45\textwidth}{!}{\includegraphics{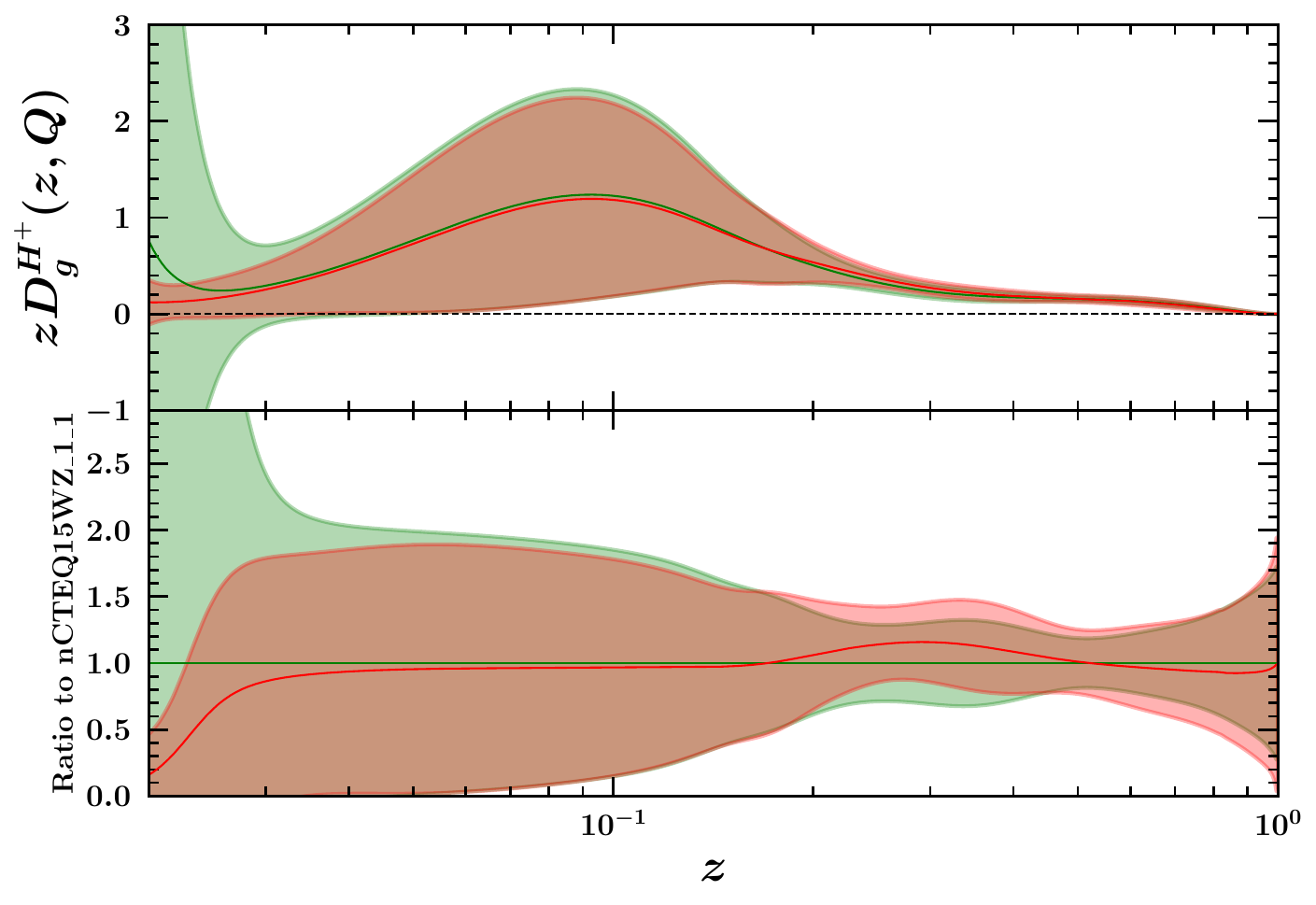}} 	
	\resizebox{0.45\textwidth}{!}{\includegraphics{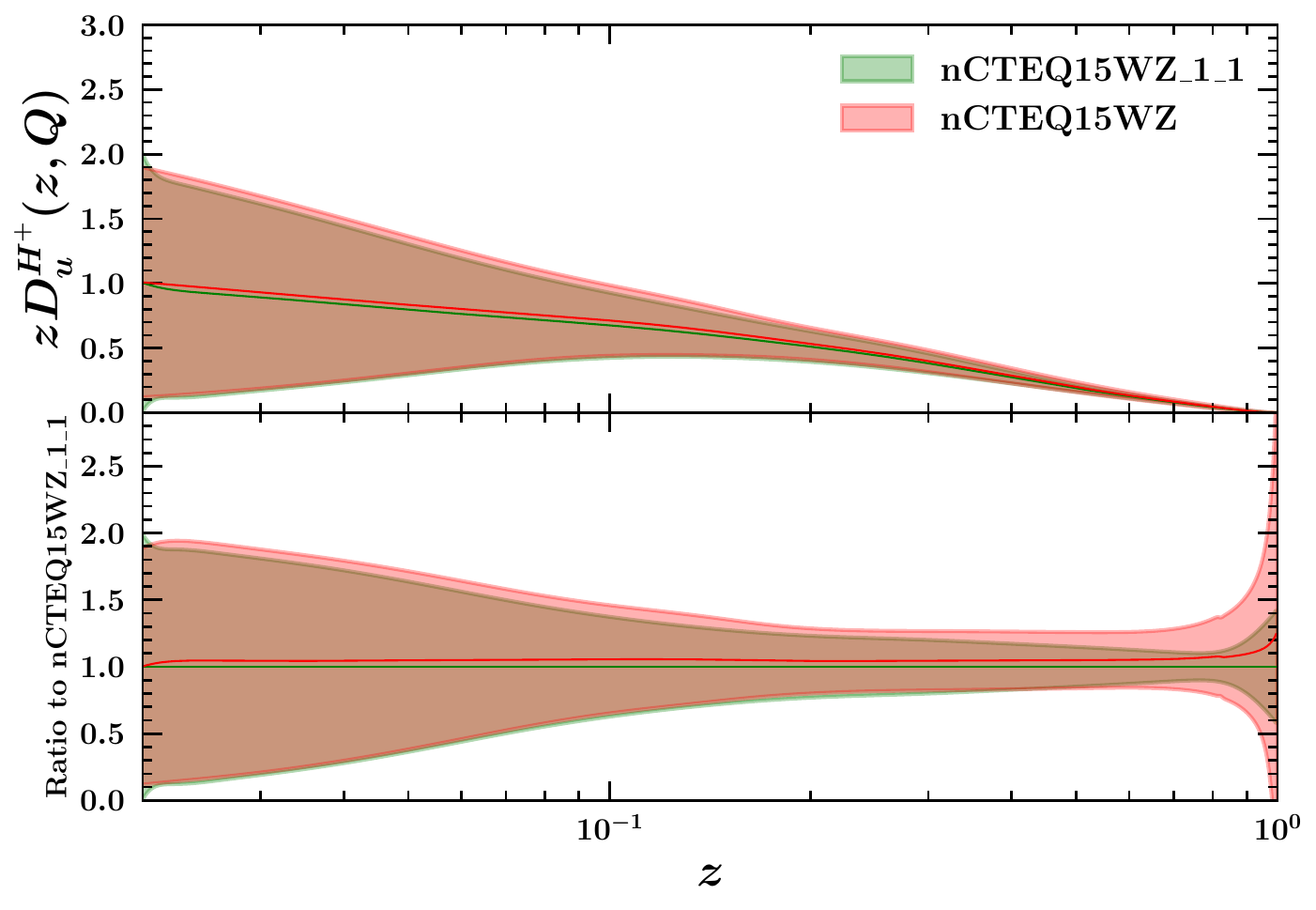}}  	
	\resizebox{0.45\textwidth}{!}{\includegraphics{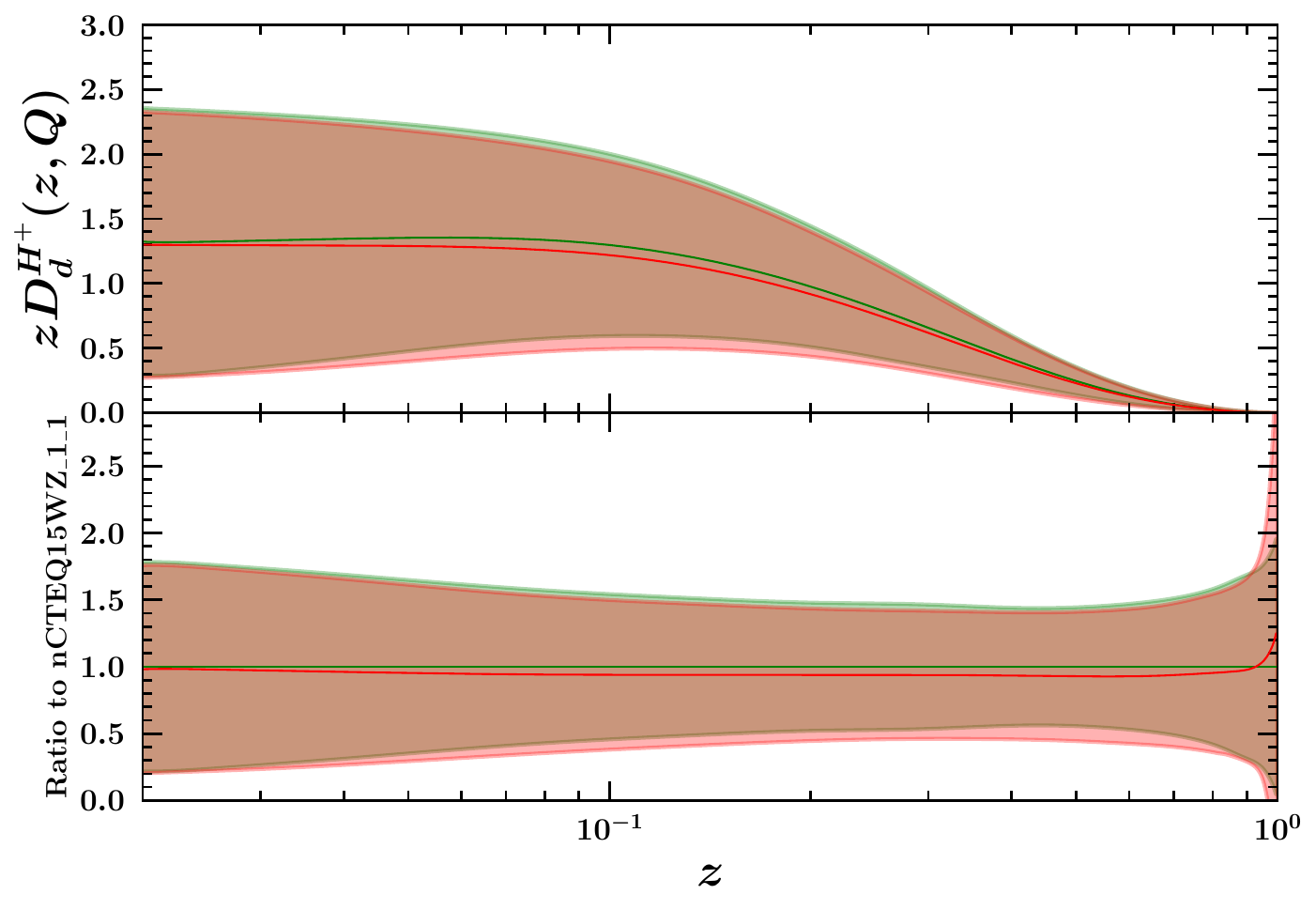}} 
	\resizebox{0.45\textwidth}{!}{\includegraphics{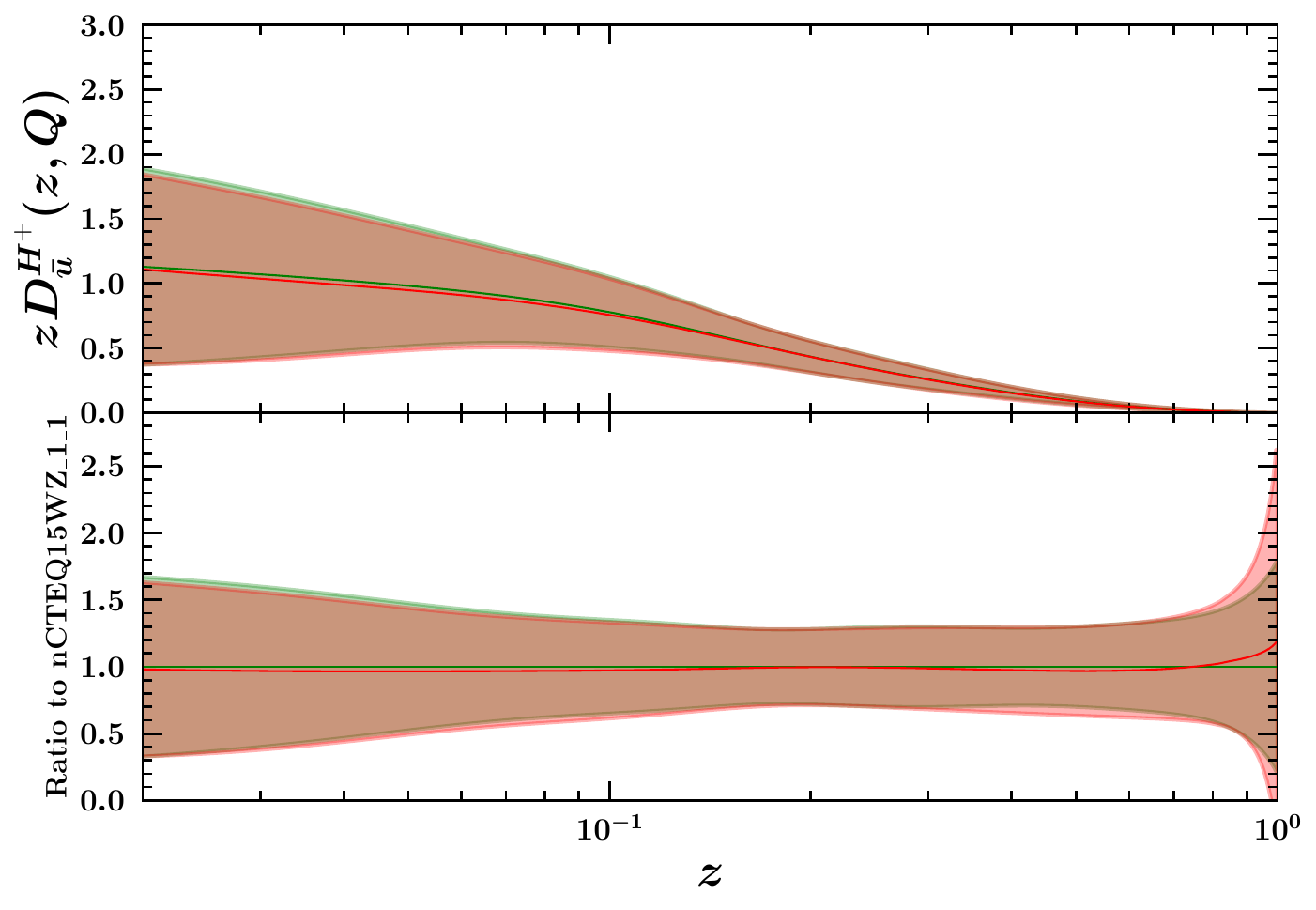}}  
	\resizebox{0.45\textwidth}{!}{\includegraphics{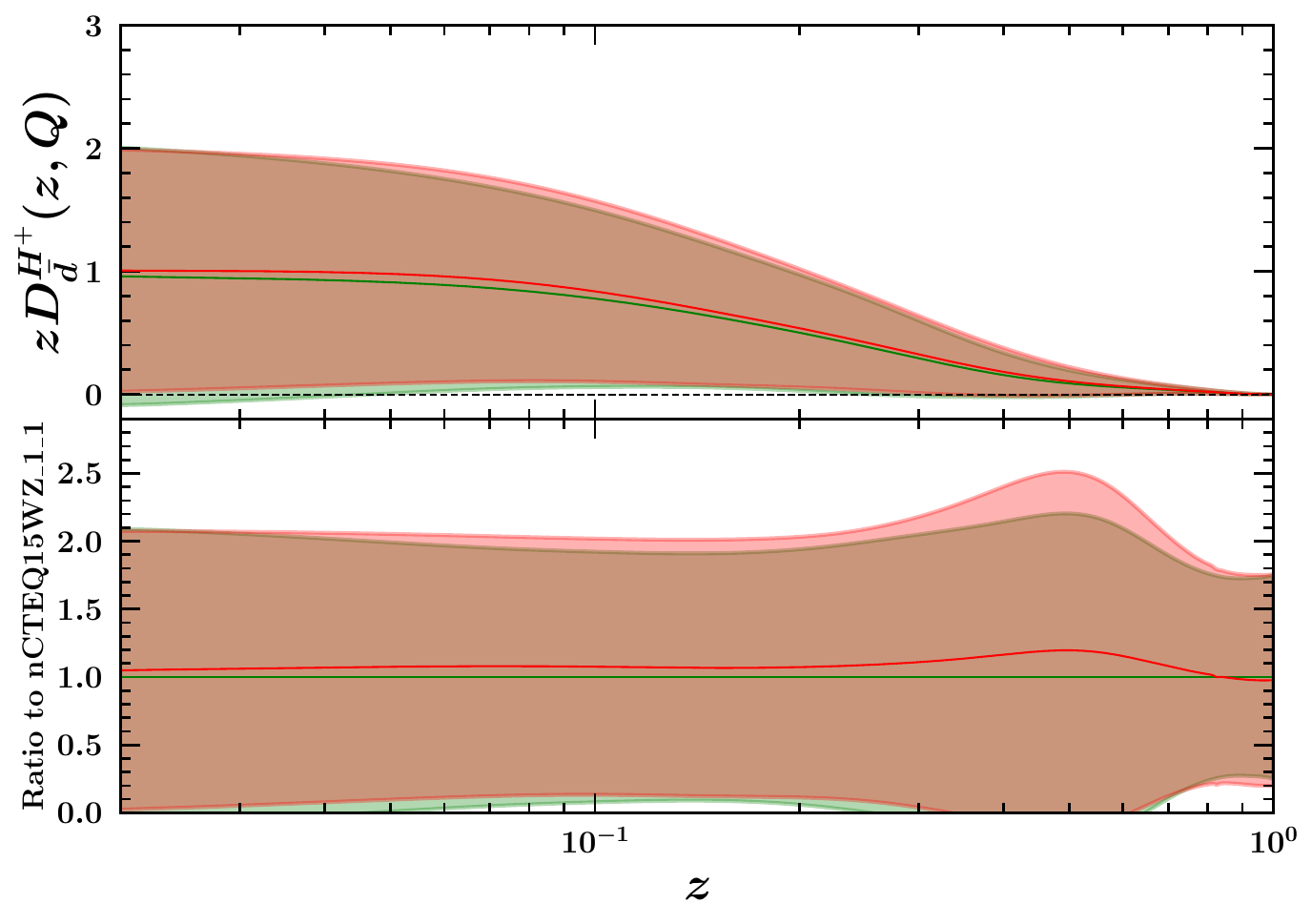}}
	\resizebox{0.45\textwidth}{!}{\includegraphics{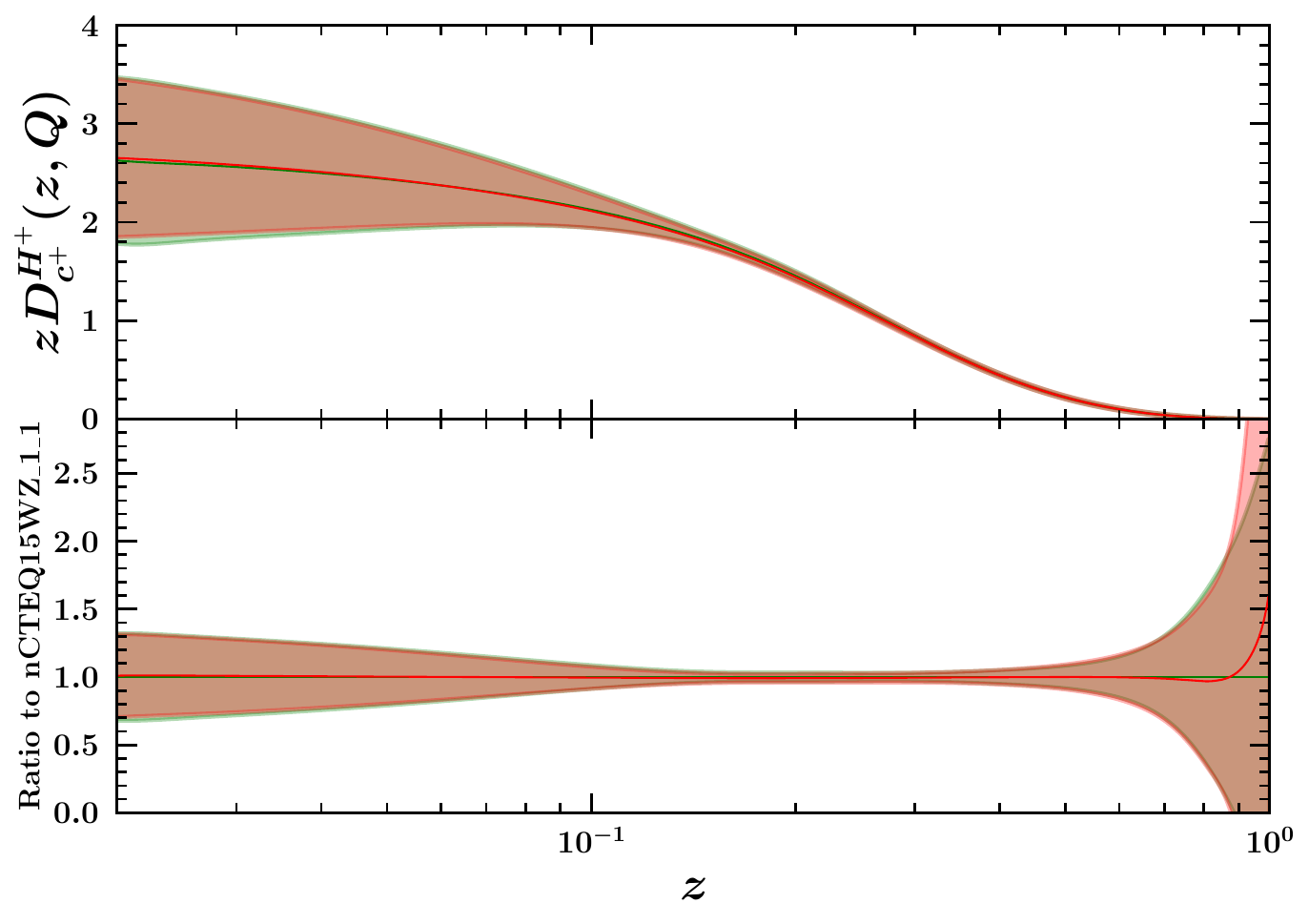}} 
	\resizebox{0.45\textwidth}{!}{\includegraphics{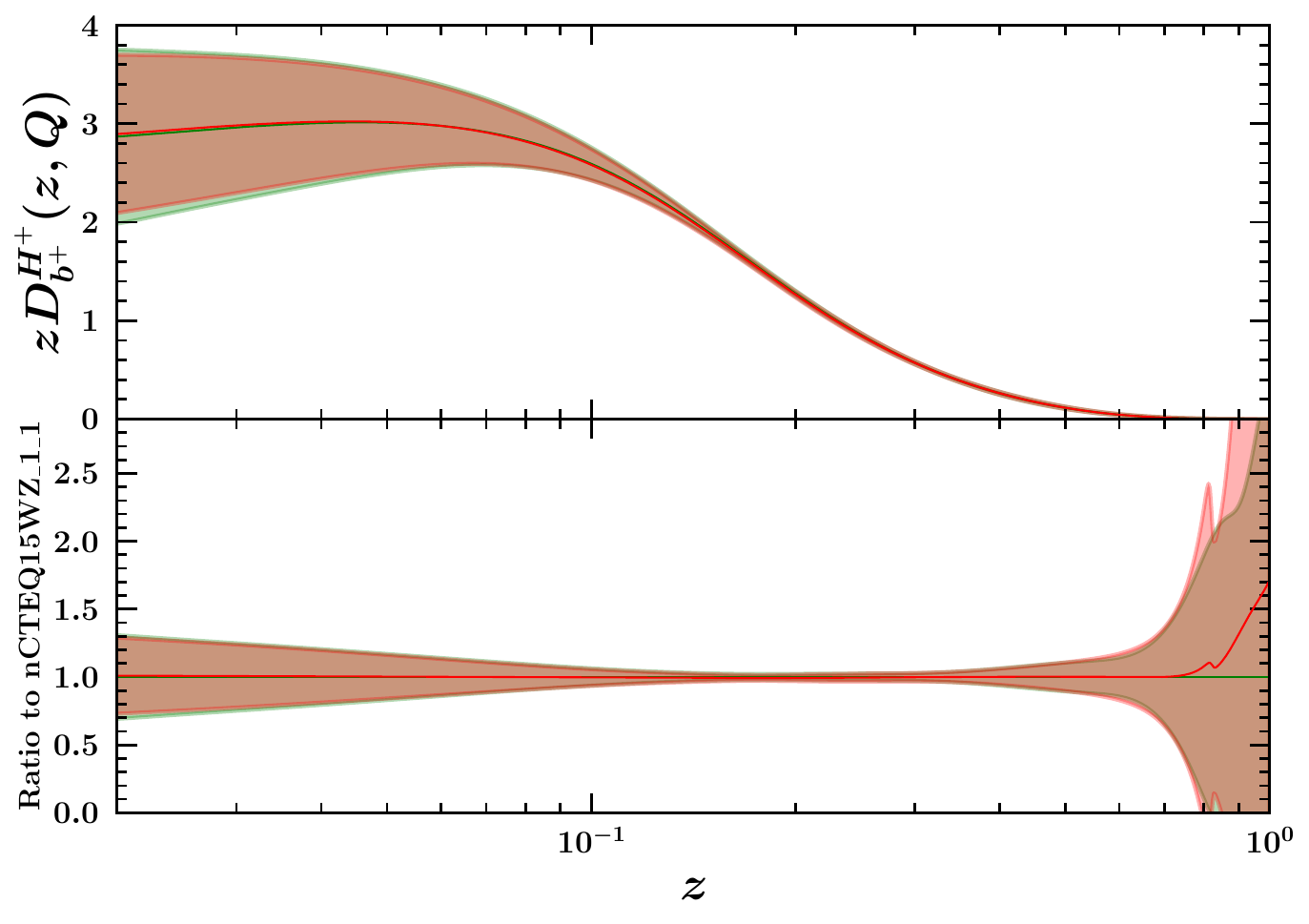}}   	 	
	\begin{center}
		\caption{ 
			\small 
  Comparison of charged hadron FFs extracted from {\tt nCTEQ15WZ-1-1} proton PDF sets as our baseline, 
			and the nuclear PDF sets from {\tt nCTEQ15WZ}. 
			We present both the absolute values, and the ratio to the 
			{\tt nCTEQ15WZ-1-1} proton PDFs baseline as well.
			The results presented at  $Q=5$ GeV.}  
		\label{fig:H-FFs_NLO_nCTEQ}
	\end{center}
\end{figure*}

Now we can proceed to discuss the extracted charged hadron FFs,  which were determined using the available nuclear PDFs.
The results are presented in Figs.~\ref{fig:H-FFs_NLO_nNNPDF}, 
\ref{fig:H-FFs_NLO_EPPS}, \ref{fig:H-FFs_NLO_nCTEQ}.
We present both the absolute values of the charged hadron FFs 
and their ratios to the corresponding proton PDFs baseline.

For the case of unidentified charged hadron FFs 
extracted by {\tt nNNPDF3.0} in Fig.~\ref{fig:H-FFs_NLO_nNNPDF}, the inclusion of 
nuclear effects primarily affected the central values of 
the gluon, $d$, and $\bar{d}$ FFs.
Other FFs such as $u$ and $\bar{u}$ show slight effects, 
particularly over the medium to small values of $z$. However, 
the FFs for $c^+$ and $b^+$ remain unchanged with the 
inclusion of nuclear effects.
As observed, a clear reduction in the error bands can be seen 
for the $u$, $d$ and $\bar{d}$ FFs, 
particularly in different regions of $z$.
For the case of gluon FF, the uncertainty band clearly increased.

For the case of unidentified charged hadron FFs 
extracted using {\tt EPPS21}, as depicted in 
Fig.~\ref{fig:H-FFs_NLO_EPPS}, the nuclear corrections primarily 
affect the gluon and $\bar{d}$ FFs. Notably, a 
light reduction in the uncertainty bands for $\bar{d}$ FFs 
is clearly observed as well.
	
For the case of unidentified charged hadron FFs 
extracted using {\tt nCTEQ15WZ}, as illustrated 
in Fig.~\ref{fig:H-FFs_NLO_nCTEQ}, the nuclear corrections 
once again primarily impact the gluon and $\bar{d}$ FFs. 
Notably, this effect is more prominent for the small 
values of $z$ for the gluon FFs.
The reduction in error bands for $\bar{d}$ is relatively 
small in certain regions of $z$.
In the case of other FFs, both the central values and the 
uncertainty bands remain almost unchanged when 
considering the nuclear PDFs.

In a broader context, when examining charged hadrons FFs, notable changes in both the central 
values and uncertainty bands are often more pronounced in the gluon and 
$\bar{d}$ components. 
Overall, these findings confirm the effects of the nuclear corrections in the discussion of 
charged hadron FFs analysis, while acknowledging that their overall impact on the analysis is modest.


\clearpage


%
\section{Summary and Conclusions}\label{sec:Summary}
%

In this study, we present a new global QCD analysis of FFs for the pion, kaon, and unidentified light charged hadrons. We analyze experimental data from SIA and SIDIS processes. Additionally, recent studies presented in Refs.~\cite{AbdulKhalek:2022laj, Soleymaninia:2022alt} are revisited to investigate the impact of nuclear corrections on SIDIS observables and the newly extracted FFs.

In this work, we utilized a Neural Network parametrization enriched with the Monte Carlo methodology to conduct uncertainty studies. This framework minimizes model bias to the greatest extent possible and ensures that uncertainties from the experimental data and proton PDF sets are appropriately propagated into the extracted FFs.

In our QCD analyses, we have used the {\tt nNNPDF3.0-p}, {\tt CT18A}, and {\tt nCTEQ15WZ-1-1} proton PDF sets as the baseline for our investigation.
To study the nuclear corrections, we compared these baselines with the most recent determinations of nuclear PDF sets available in the literature, namely the {\tt nNNPDF3.0}, {\tt EPPS21}, and {\tt nCTEQ15WZ} at NLO accuracy.

We observed that incorporating nuclear corrections in the analysis of FFs using SIDIS data does not result in an overall reduction in the individual \( \chi^2 \) per data point from \texttt{COMPASS} for all hadrons.
However, this reduction is not universally applicable, as the analysis primarily focuses on targets composed of relatively light nuclei, which exhibit approximately 10-20\% nuclear effects.

Additionally, \texttt{COMPASS} measured the SIDIS cross sections normalized to the inclusive DIS cross sections, where it is expected that some of the nuclear effects would cancel out in this ratio. These factors collectively contribute to minor changes in some instances.
In fact, we have observed that incorporating nuclear corrections slightly affects both the shape of the central values and the uncertainty bands of the extracted charged pion for \( \bar{d} \) FF, charged kaon for \( d \) FF, and unidentified light hadron for \( d \), \( \bar{d} \), and to some extent \( \bar{u} \) quark FFs.
Moreover, the central values and uncertainty bands of the gluon FF for most of our results are marginally impacted by the choice of nuclear PDF sets.

In summary, we investigate that it is not possible to conclude that nuclear PDFs corrections have a significant
and beneficial impact on FFs of pion, kaon, and light charged hadrons in the SIDIS process.
Overall, the observed changes are rather small, and in some cases, they are negligible and can safely be ignored.

The extracted FFs in standard LHAPDF format are available from authors upon request.

%
\begin{acknowledgments}
%

The authors gratefully acknowledge many helpful discussions and comments by 
Valerio Bertone. Authors thank the School of Particles and Accelerators, Institute 
for Research in Fundamental Sciences (IPM) for financial support of 
this project. 
Maryam Soleymaninia is thankful to the Iran Science Elites Federation
for the financial support.
Hamzeh Khanpour appreciates financial support from NAWA under grant 
number BPN/ULM/2023/1/00160, as well as from the IDUB programme at the AGH University.

\end{acknowledgments}

\clearpage

%
\section{Data Availability Statement}
%
The authors declare that the data supporting the findings of this study are 
available within the Durham High-Energy Physics Database (HEPData) 
as an open-access repository for scattering data from experimental 
particle physics (https://www.hepdata.net/).


\end{document}